  \documentclass[12pt,preprint]{aastex}
\newcommand{\ri}{r_{\rm i}}

\newcommand{\cs}{c_{\rm s}}
\newcommand{\css}{c_{\rm s}^2}
\newcommand{\bp}{B_{\rm P}}

\newcommand{\msun}{{\rm M}_{\sun}}
\newcommand{\rsun}{{\rm R}_{\sun}}
\newcommand{\etat}{{\eta}_{\rm t}}
\newcommand{\betat}{{\beta}_{\rm t}}
\newcommand{\betai}{{\beta}_{\rm i}}
\newcommand{\deltai}{{\delta}_{\rm i}}
\newcommand{\rhoi}{{\rho}_{\rm i}}
\newbox\grsign \setbox\grsign=\hbox{$>$}
\newdimen\grdimen \grdimen=\ht\grsign
\newbox\laxbox \newbox\gaxbox
\setbox\gaxbox=\hbox{\raise.5ex\hbox{$>$}\llap
     {\lower.5ex\hbox{$\sim$}}}\ht1=\grdimen\dp1=0pt
\setbox\laxbox=\hbox{\raise.5ex\hbox{$<$}\llap
     {\lower.5ex\hbox{$\sim$}}}\ht2=\grdimen\dp2=0pt

\newcommand{\lax}{$\mathrel{\copy\laxbox}$}

\shorttitle{MHD jet collimation}
\shortauthors{Ch.~Fendt}

\begin{document}

\title{Collimation of astrophysical jets -- the r\^ole of \\
       the accretion disk magnetic field distribution}

\author{Christian Fendt\altaffilmark{1}}
\affil{ Max Planck Institute for Astronomy, K\"onigstuhl 17,
D-69117 Heidelberg, Germany}
\email{fendt@mpia.de}

\altaffiltext{1}{Part of the numerical work has been accomplished 
at the 
Astrophysikalisches Institut Potsdam and at the University of
Potsdam, Germany}

\begin{abstract}
We have applied axisymmetric magnetohydrodynamic (MHD) simulations
in order to investigate the impact of the accretion disk magnetic
flux profile on the collimation of jets.
Using the ZEUS-3D code modified for magnetic diffusivity, 
our simulations evolve from an initial state in hydrostatic equilibrium
and a force-free magnetic field configuration.
Considering a power law for the disk poloidal magnetic field profile
$\bp \sim r^{-\mu}$
and for the density profile of the disk wind $\rho \sim r^{-\mu_{\rho}}$,
we have performed a systematic parameter study over a wide range of 
parameters $\mu$ and $\mu_{\rho}$.
We apply a toy parameterization for the magnetic diffusivity derived 
from the internal turbulent Alfv\'enic pressure. 
We find that the degree of collimation 
(quantified by the ratio of mass flow rates in axial and lateral direction)
decreases for a steeper disk magnetic field profile (increasing $\mu$).
Varying the total magnetic flux does not change the degree of jet collimation
substantially, it only affects the time scale of outflow evolution and the 
terminal jet speed.
As our major result we find a general relation between the
collimation degree with the disk wind magnetization power law exponent. 
Outflows with high degree of collimation resulting from a flat disk magnetic
field profile tend to be unsteady, producing axially propagating knots
as discussed earlier in the literature.
Depending slightly on the inflow density profile this unsteady behavior
sets in for $\mu < 0.4$.
We also performed simulations of jet formation with artificially enhanced
decay of the toroidal magnetic field component in order to investigate
the idea of a purely "poloidal collimation" previously discussed in the
literature. 
These outflows remain only weakly collimated and propagate with lower 
velocity.
Thanks to our large numerical grid size (about $7 \times 14$\,AU for protostars),
we may apply our results to recently observed hints of jet rotation (DG\,Tau)
indicating a relatively flat disk magnetic field profile, $\mu \simeq 0.5$.
In general, our results are applicable to both stellar and extragalactic sources 
of MHD jets.
\end{abstract}

\section{Introduction}

Astrophysical jets as highly collimated high speed beams of matter 
have been observed as common phenomenon in a variety of astronomical
sources, among them young stars, micro-quasars and active galactic
nuclei.
The current understanding of jet formation is that jets are launched
by {\em magnetohydrodynamic} (MHD) processes in the vicinity of the
central jet object -- 
an accretion disk surrounding a protostar or a black hole 
\citep{blan82,pudr83,came90}.
The general properties of jet sources and the ongoing physical 
processes have been investigated for a long time, however,
the principal mechanism which actually {\em launches} a jet from 
the disk at a certain time is not yet known.
%

A number of numerical simulations investigating the MHD jet formation 
have been published. 
We may distinguish between those simulations which take into account
the evolution of the {\em disk structure} and those which consider
the disk as a fixed {\em boundary condition} for the simulation
of the disk wind.
The first approach allows to study directly the 
launch\footnote{By jet {\em formation} and jet {\em launching} we denote 
the acceleration and collimation of a disk wind, 
and the generation of a disk wind out of the accretion flow, respectively}
of the outflow,
i.e the ejection of accreting plasma into an outflow in direction
vertical to the disk.
However, due to the limited time resolution and spatial coverage typical 
for these simulations,
little can be learned concerning the ultimate jet acceleration and 
collimation which takes place at radii beyond the Alfv\'en surface
\citep{uchi85,mill97,kudo98}.
The second approach allows to investigate the {\em long term evolution}
of jet formation on a large spatial scale
\citep{ouye97a,ouye97b,kras99,fend00,fend02}.
The mechanism which launches the outflow out of the 
accretion stream cannot be studied by this approach.

In this paper we study how jet formation  -- i.e. jet collimation and
acceleration -- depends on the magnetic field profile of 
the jet launching accretion disk.
The initial magnetic field is located in an initially hydrostatic 
density distribution.
We prescribe various magnetic field profiles along the disk surface
which is taken as a boundary condition fixed in time.
The coronal (i.e. jet) magnetic field evolves in time from 
the initial state and governed by a prescribed mass inflow from the
disk surface into the corona.
We apply the ZEUS-3D MHD code modified for magnetic diffusivity
\citep{fend02}.
It is clear that the "real" disk-jet magnetic field and the mass
flux into the jet is a result of
dissipative MHD processes in the disk in interaction with the external
field structure.
The disk structure and evolution will also be affected by the
existence of a magnetized disk wind/jet.
Although most of the processes involved are understood to some extent,
the self-consistent dynamical modeling of such systems has just
started (e.g. simulations of a disk dynamo generated jet magnetic
magnetic field \citep{reko04})
and the results have to be considered still as preliminary.
For the purpose of the present paper we have to keep in mind that
some of the parameter runs considered may be more realistic than
others concerning the internal disk physics.

We note that in a contemporaneous and independent study 
\citet{pro06} have undertaken a similar approach.
Our results generally agree with their paper, but also transcend their approach 
in several aspects. 
Among the additional features treated in our paper are a much larger grid, 
the description of a turbulent magnetic diffusivity, 
a quantitative measure of the collimation degree,
and a broader range of parameters studied.

Our paper is organized as follows.
Section 2 briefly summarizes the issue of MHD jet collimation.
Section 3 introduces our model approach.
In Sect.~4 we present our results before concluding the paper with the
summary.
Preliminary results of the present investigation have been published
earlier (see Fig.~2 in \citet{rued02}).

\section{Self-collimation of rotating MHD outflows}
%
Theoretical studies considering analytical solutions of the stationary,
axisymmetric MHD force-balance in the asymptotic jet flow (i.e. far from
the source) have clearly shown that such outflows {\em must} collimate into
a narrow beam when carrying a {\em net poloidal electric current}
\citep{heyv89,chiu91}. 
This fundamental statement has been recently confirmed and generalized
\citep{okam03,heyv03}.

\subsection{Stationary studies of MHD jets}
A self-consistent treatment of the collimation {\em process} - i.e.~the
question how collimation is achieved along the initial outflow - has to
consider at least a 2.5-dimensional (i.e. axisymmetric three-dimensional)
MHD problem.
Until about a decade ago, 
MHD jet theory has been mostly limited to the stationary approach.
Time-independent studies, however, 
cannot really prove whether a MHD solution derived will actually
be realized during the time evolution of a forming jet.
Hence, the final answer concerning jet self-collimation can only 
be given by time-dependent MHD simulations.
Although stationary {\em self-similar} models gave clear indication for 
collimated MHD jets 
\citep{blan82,saut94,li93,cont94} 
one has to keep in mind that such a constraint has implications
which are critical for collimation, but are not really feasible for jets.
For example, self-similarity considers an {\em infinite jet radius}
and does exclude the symmetry axis of the outflow.
Truly 2.5-dimensional (axisymmetric 3D) solutions considering the
local force-balance and global boundary conditions have been obtained
only in the force-free limit \citep{fend95,lery98,fend01},
and provide the shape of the collimating jet as determined by the 
internal force-equilibrium.
So far, 
the only self-consistent two-dimensional stationary MHD solution 
of a collimating wind flow has been published decades ago
\citep{saku85,saku87}.
These solutions collimate on logarithmic scales only - a result of the
low rotation rate applied.

\subsection{Time-dependent MHD simulations}
The first numerical evidence of the MHD jet self-collimation process 
has been provided by \citet{ouye97a} following a pioneering approach 
for disk winds by \citet{usty95}.
The simulations by Ouyed \& Pudritz were been performed from an initially 
force-free setup and fixed boundary conditions for the mass inflow into
the jet from an underlying Keplerian disk surface.
After a few hundred of inner disk rotations the MHD disk wind collimates 
into a narrow beam and saturates into a stationary state outflow.
For a recent review on numerical progress in simulating disk winds and jets
we refer to \citep{ppv06}.

It is clear that different inflow conditions for the disk wind
(mass flow rate or magnetic field profile)
will result in a different dynamical evolution of the jet, 
as there are different time scales, velocities or a different 
degree of collimation.
It is therefore interesting to investigate a wide parameter range for the
leading jet parameters.
This has partly been done in a number of papers discussing 
the variation of the jet mass load \citep{ouye99,kato02},
a time-dependent disk magnetic field inclination \citep{kras99},
magnetic diffusivity in the jet \citep{kuwa00,fend02,kuwa05},
time-dependent mass loading \citep{vito03},
the extension of the simulation box on collimation \citep{usty99},
or,
in difference to these cases of a monotonously distributed disk magnetic field,
the long-term evolution of a stellar dipolar magnetosphere in connection
with a Keplerian disk \citep{fend00}.

%
The next step will be to include the disk structure in the numerical
simulations \citep{haya96,mill97,cass02,kudo02} and follow the jet
launching process over many hundreds of disk rotations.
Substantial progress have been made in this respect in simulations
already in particular concerning the accretion process from disk
down to the star \citep{roma02, roma03, roma04, roma05} but collimated 
disk jets have not yet been found.

\subsection{``Toroidal or poloidal collimation''?}  \label{sec_tor_pol}

Concerning jet self-collimation
it is interesting to mention a proposal by \citet{spru97} pointing out 
that the toroidally dominated jet magnetic field is {\em kink-instable}.
Spruit et al. suggested that jets should be rather collimated by the
{\em poloidal} disk magnetic field pressure and less by the toroidal 
pinching force.
The authors find that ``{\em poloidal collimation}'' works best for 
a disk magnetic field profile $|\bp|\sim r^{-\mu}$ with $\mu \le 1.3$.
Indication for ``poloidal collimation'' of outflows has been reported
also by MHD simulations \citep{matt03}, launching the outflow
in a dipolar stellar magnetosphere and collimating it 
by the surrounding disk magnetic field.

The kink instability of jets can be investigated by non-axisymmetric
simulations only.
It is therefore important to note that 3D-simulations of MHD jet
formation have indeed proven the feasibility of
MHD self-collimation under the influence of non-axisymmetric
perturbations in the jet launching region \citep{ouye03,kigu05}.
Kelvin-Helmholtz modes are usually fastest growing and are particularly
dangerous for the jet in the super-Alfv\'{e}nic regime.
Essentially, it is the ``backbone'' of the jet flow, i.e. the axial region of
high field strength and low density which stabilizes the jet against
the instabilities as this highly sub-Alfv\'{e}nic region of \citep{ouye03,ppv06}.
In general, the 3D simulations show that the jets starts as a stable outflow,
then destabilizes after about 100 inner disk rotations before it stabilizes
again at the time of 200 inner disk rotations. 
The detailed behavior of the different instability modes depend of course
on the flow parameters and the general setup of e.g. the non-axisymmetric
boundary conditions (see \citet{ouye03} for details).

As an interesting fact, we note that all studies of MHD jet formation so
far (time-dependent or stationary) which lead to a collimated jet,
the disk magnetic field profile satisfies the Spruit et al. criterion
$\mu \le 1.3$.
Examples are the stationary \citep{blan82} solution of a self-similar, cold
jet with $\mu = 5/4$, simulations of a collimating jet by \cite{ouye97a} 
applying $\mu = 1$,
or by \cite{fend00} treating an initially dipolar field 
distribution $\mu = 3$.

%

\section{Model setup}

We perform axisymmetric MHD simulations of jet formation for a set
of different boundary conditions for the accretion disk magnetic
field.
We apply the ZEUS-3D MHD code modified for magnetic diffusivity
\citep{fend02}.
As initial state we prescribe a force-free magnetic field within
a gas distribution in hydrostatic equilibrium.
This is essential in order to avoid artificial relaxation processes 
caused by a non-equilibrium initial condition.
The gas is "cold" and supported by additional turbulent Alfv\'enic 
pressure.

A Keplerian disk is taken as a boundary condition for the mass
inflow from the disk surface into the corona and the magnetic flux.
The poloidal magnetic field lines are anchored in the rotating disk.
A gap extends between the central body and the inner disk radius $\ri$ 
implying a vanishing mass load for the jet from this region.
Compared to our previous studies 
\citep{fend00,fend02},
several new features are included in the present approach.
\begin{itemize}
\item[-] 
Our major goal is to investigate a broad parameter range of different accretion disk 
         magnetic flux profiles and disk wind density (i.e. mass flux) profiles.
\item[-] 
We apply a non-equidistant numerical grid covering a large spatial
         domain of $150 \times 300$ inner disk radii corresponding to 
         e.g. about $(6.7\times 13.3)$\,AU assuming an inner disk radius of
         $10\,\rsun$ for protostars.
\item[-] 
We consider a simple parameterization of turbulent magnetic diffusivity, 
         variable in space and time, in agreement with also considering 
         turbulent Alfv\'enic pressure.
         However, compared to our previous studies the magnitude of 
         magnetic diffusivity is low, thus not affecting the degree of
         jet collimation.
\end{itemize}
In the following we give some details for the new features of our the model setup.

\subsection{Resistive MHD equations}

We numerically model the time-dependent evolution of jet formation 
considering the following set of resistive MHD equations,
\begin{equation}
{\frac{\partial \rho}{\partial t}} + \nabla \cdot (\rho \vec{v} ) = 0,\quad 
\nabla \cdot\vec{B} = 0,\quad 
\frac{4\pi}{c} \vec{j} = \nabla \times \vec{B},
\end{equation}
\begin{equation}
\rho \left[ {\frac{\partial\vec{u} }{\partial t}}
+ \left(\vec{v} \cdot \nabla\right)\vec{v} \right]
+ \nabla (p+p_A) +
 \rho\nabla\Phi - \frac{\vec{j} \times \vec{B}}{c} = 0,
\end{equation}
\begin{equation}
{\frac{\partial\vec{B} }{\partial t}}
- \nabla \times \left(\vec{v} \times \vec{B}
-{\frac{4\pi}{c}} \eta\vec{j}\right)= 0,
\end{equation}
\begin{equation}
\rho \left[ {\frac{\partial e}{\partial t}}
+ \left(\vec{v} \cdot \nabla\right)e \right]
+ p (\nabla \cdot\vec{v} )
- {\frac{4\pi}{c^2}}\eta \vec{j}^2= 0,
\end{equation}
with the usual notation for the variables 
(see \citet{fend00}).

We apply a polytropic equation of state for the gas, $p=K\rho^\gamma$
with the polytropic index $\gamma$.
Thus, we do not solve the energy equation (4) and the internal energy 
of the gas is reduced 
\footnote{Note, however, that 
ongoing work by Clarke and collaborators 
seems to indicate that relaxation of the polytropy assumption may
affect the dynamical evolution in certain domains of the jet,
in particular regions with shocks or contact discontinuities
\citep{clar04}}
to $ e=p/(\gamma-1)$.
The Ohmic heating term in energy equation (4) has shown to be 
generally negligible for the dynamics of the system 
(e.g. \citet{mill97}).
This is usually due to the low levels of dissipation presumed in
astrophysical cases.
Comparison of the compression and dissipation terms shows that this 
holds also for our study 
(plasma beta $\betai \simeq 1$ and magnetic diffusivity $\eta \simeq 0.01$,
see below).
On the other hand, simulations treating violent reconnection processes
in stellar magnetospheres and/or current sheets have shown that
Ohmic heating might establish regions of hot plasma (e.g. \citet{haya96}).
As we are primarily interested in the flow dynamics we neglect Ohmic
heating.

Applying a polytropic equation of state simplifies the numerical
challenge considerably.
We do not believe that this simplification affects the general
results of our simulations.
Clearly, certain number values numerically derived in our study 
may well change (as e.g. the exact degree of collimation)
and their interpretation must be strictly limited to a comparison
{\em within} our parameter range applied.
Future studies on this topic will have to consider and to compare
to a non-polytropic equation of state as well.
This is, however, beyond the scope of the present paper.

In addition to the hydrostatic gas pressure $p$ we consider 
{\em Alfv\'{e}nic turbulent pressure} assuming 
$p_{\rm A}\equiv p/\betat$ with $\betat = {\rm const.}$
This approach has been applied in a series of papers by
various groups 
(\citet{ouye97a}, \citet{ouye97b}, \citet{fend00}, \citet{fend02},
\citet{vito03}, \citet{ouye03}, \citet{pro06}).
As we intend to complement these studies, we have followed once
again the same approach.
In general, turbulent Alfv\'{e}nic pressure may explain the
presence of a {\em cold} corona above protostellar accretion 
disks as observationally suggested.
We have previously argued that the existence of such an additional 
pressure component is inevitable as turbulent Alfv\'{e}n waves will
be launched naturally in the highly turbulent accretion disk and 
then propagate into the disk wind and outflow \citep{fend02}.

For the polytropic index we apply $\gamma=5/3$.
This is a value generally assumed for non-relativistic astrophysical
polytropic gas dynamics.
However, note that with such $\gamma$, 
Ohmic heating (see above) and also turbulent Alfv\'{e}nic
pressure is potentially neglected for the thermodynamics. 
Detailed (linear) analysis of magnetohydrodynamic turbulence in media 
under various conditions
has shown that sub-Alfv\'{e}nic pressure exerted by Alfv\'{e}n waves 
generally obeys a polytropic equation of state with polytropic index 
$\gamma_{\rm t} = 1/2$.
This has been derived analytically for well defined model 
assumptions\footnote{A thermodynamic system undergoing 
temporal changes the {\em adiabatic} index of (compressible) Alfv\'{e}n 
waves is instead $\gamma_{\rm wawe} = 3/2$ \citep{mckee95}.
$P_{\rm wave} \sim \rho^{1/2}$ applies in steady state media or 
Alfv\'{e}n waves propagating in a density gradient.}
which are applicable for a variety of physical states (see \citet{mckee95})
and has also proven numerically \citep{gamm96,pass03}.
However, the exact number value of the polytropic index for turbulent Alfv\'{e}nic 
pressure in the case of expanding and accelerating magnetohydrodynamic outflows is 
not known (if it exists at all).
Therefore, for simplicity, we apply a turbulent Alfv\'{e}nic pressure polytropic 
index equal to that of the gas, $\gamma_{\rm t} = \gamma = 5/3$.
The effective polytropic index for multi-pressure polytropes 
($N$ pressure components $p_n$) would be given by
\begin{equation}
\gamma_{\rm eff} = \Sigma_{n=1}^N \left(\frac{p_n}{p}\right) \gamma_n =
                   \frac{\betat\gamma + \gamma_{\rm t}}{\betat + 1}
\end{equation}
(see \citet{curr00} and references therein) 
indicating that for low $\betat$ the polytrope would be dominated by the
turbulent Alfv\'{e}nic pressure term. For $\gamma_{\rm t} = \gamma = 5/3$ and
$\betat \simeq 0.03$, also $\gamma_{\rm eff} \simeq 5/3$.

We solve the MHD equations applying the ZEUS-3D code 
\citep{stone92a,stone92b,hawl95}
in the axisymmetry option for cylindrical coordinates $(r,\phi,z)$.
We added physical magnetic resistivity to the original ZEUS-3D ideal MHD code 
(see description and extensive tests in \citet{fend02}).
We normalize all variables to their value measured at the inner disk radius
$r_{\rm i}$.
For example, time is measured in units of the Keplerian period at the inner disk 
radius, and for the density $\rho \rightarrow \rho /\rho_{\rm i}$.
The point mass for the gravitational potential $\Phi = - 1/\sqrt{r^2 + z^2}$ 
is located at the origin.
Finally, the normalized equation of motion treated numerically is
\begin{equation}
\frac{\partial\vec{v}' }{\partial t'}
+ \left(\vec{v}' \cdot \nabla'\right)\vec{v}' =
\frac{2 \,\vec{j}' \times \vec{B}' }{\deltai\,\betai\,\rho'}
- \frac{\nabla' (p' + p_{\rm A}')}{\deltai \,\rho'} - \nabla'\Phi'\,.
\end{equation}
The coefficients
$\betai \equiv 8 \pi p_i / B_i^2 $ and
$\deltai \equiv \rho_i v^2_{K,i} / p_i $ with
the Keplerian speed $v_{K,i} \equiv (GM/r_i)^{1/2} $,
correspond to the plasma beta and the Mach number of the
gas at the inner disk radius \citep{ouye97a}.
For a ``cold'' corona with $p_{\rm A}' > 0$ we have 
$\betat = 1 / (\deltai (\gamma - 1) / \gamma - 1)$.
In the following primes are omitted
and
only normalized variables are used if not explicitly declared otherwise.

\subsection{Turbulent magnetic diffusivity}

Magnetic diffusivity caused by turbulent motion in the jet material affects both
collimation and acceleration of the jet \citep{fend02}.
Magnetic diffusivity also plays an essential r\^ole in the jet launching
process in the accretion disk \citep{li95,ferr97}.
The magnetic diffusivity $\eta$ we apply in Eq.~(3) is ``anomalous'' and 
is considered to be provided by macroscopic MHD instabilities.
As the jet launching accretion disk is highly turbulent itself, 
we naturally expect that the turbulent pattern will propagate into the jet
when the disk material is lifted up from the disk surface.
The turbulence evolution along the jet will certainly be affected
by advection and stretching and may vary in different regimes.
However, the exact dynamical evolution of turbulence in jets has
not yet been investigated and is therefore poorly understood.

Supposing that the magnetic diffusivity is primarily due by the same 
turbulent 
Alfv\'{e}nic waves which are responsible for the turbulent Alfv\'{e}nic
pressure applied in the simulations,
we have derived a toy parameterization relating both effects \citep{fend02}.
Such an approach extends and generalizes the studies by \citet{ouye97a},
\citet{fend00}, and \citet{pro06}.
We achieve this by parameterizing the {\em turbulent magnetic diffusivity} 
similar to the Shakura-Sunyaev parameterization of turbulent viscosity,
$\etat = \alpha_{\rm m} v l$, where $\alpha_{\rm m}\leq\!1$
and $l$ and $v$ are the characteristic dynamical length scale and velocity 
of the system, respectively.

The turbulent magnetic diffusivity $\etat$ can be related to the Alfv\'{e}nic
turbulent pressure $p_{\rm A}$ 
by first selecting the turbulent velocity field $v_{\rm t}$
(i.e. the {\em turbulent} Alfv\'{e}nic velocity) as characteristic velocity. 
This gives $\etat = \alpha_{\rm m} v_{\rm t} l$ and with
$\betat = ( \cs / v_{\rm t})^2$ and $\css = \gamma\,p/\rho$, it follows
\begin{equation}
v_{\rm t}^2 = \frac{\gamma}{\betat} \frac{p}{\rho},
\quad {\rm or} \quad
{v'}_{\rm t}^2 = \frac{\gamma}{\deltai\betat} \rho'^{\gamma-1}.
\end{equation}
For the chosen polytropic index this implies a normalized
magnetic diffusivity $\eta\sim \rho^{1/3}$ if $l$ is constant.
The power index for $\eta(\rho)$ would be affected by a change in the
polytropic index
for the turbulent Alfv\'{e}nic pressure (see discussion above).
The characteristic length scale $l$ of turbulent magnetic diffusivity in the
jet is not a known quantity.
If we assume the turbulence being generated within the disk and launched 
from there into the outflow, 
we may assume an initial length scale $l$ corresponding to the Shakura-Sunyaev
parameterization of the disk turbulent viscosity, 
i.e. limited by the disk height $l < h(r)$. 
In our normalization, this would imply $l \simeq 0.1$ for regions close 
to the star as typically $h/r \simeq 0.1$ and $l \simeq 1.0$ at $r \simeq 10$.
For regions at greater distance to the star, $l$ could be correspondingly
larger.
A fully self-consistent treatment of Alfv\'{e}nic turbulence would have to
consider also its temporal and spatial expansion as well as damping or decay 
as the turbulence cells propagate along the collimating jet.
This is clearly beyond the scope of our purpose and we instead {\em assume}
a constant characteristic length scale $l$.

In our simulations we apply sub-Alfv\'{e}nic turbulence, 
$\betat = 0.03$ and $\deltai = 100$ \citep{ouye97a,fend02}
resulting in a magnetic diffusivity 
$\eta= \eta_0 \rho^{1/3}$ with $\eta_0 = 0.03$.
This would correspond to a number value $(\alpha_{\rm m} l) = 0.04$.

We finally note that the magnitude of normalized magnetic diffusivity 
is in the same range as typically applied in the literature 
\citep{haya96,mill97,reko04}. 
In a previous study investigating the relation between magnitude
of magnetic diffusivity and degree of jet collimation we have shown 
that such (low) level of diffusivity does not affect the degree of
jet collimation - jet de-collimation as due to magnetic diffusivity 
applies above a critical diffusivity  about a factor of 100 larger 
as used in the present paper \citep{fend02}.
In fact, as our goal is to study the relation between jet collimation 
and disk magnetic field distribution, the magnitude of diffusivity 
was chosen such that is will not affect the degree of collimation.

\subsection{Numerical grid}

We use the ``scaled grid'' option by ZEUS with the grid element size decreasing 
inwards by a factor of 0.99.
The numerical grid size is $(256\times 256)$ resulting in a 
a spatial resolution from $(0.13\times 0.26)$ close to the origin
to $(1.6\times 3.2)$ at the opposite corner.
The physical grid size for all simulations is $(r\times z) = (150\times 300)\ri$
corresponding to $(6.7\times 13.3)$\,AU for $\ri \simeq 10\,\rsun$ if
applied for a protostar.
This is several times larger than in previous studies 
\citep{ouye97a,fend02,pro06}
and comparable to the observational resolution for young 
stars \citep{bacc02}.

\subsection{Initial conditions}

Prescribing a stable initial condition is essential for any time-dependent
simulation.
Otherwise, the early evolution of the system will be dominated by artificial
relaxation processes of the instable initial setup.
This is particularly important if only few evolutionary time steps are
computed.

We start our simulations from a density stratification in hydrostatic equilibrium 
with an initially force-free (and also current-free) magnetic field.
Such a configuration would remain in its initial state if not disturbed by the given
boundary conditions.
The initial density distribution is $\rho(r,z) = (r^2 + z^2)^{-3/4}$ 
\citep{ouye97a,fend02}.

The initial magnetic field distribution is calculated using a stationary,
axisymmetric finite element code described elsewhere \citep{fend95}.
Essentially, this code calculates the axisymmetric force-free magnetic flux
distribution for {\em any} given boundary value problem.
We compute the $\phi$-component of the vector potential\footnote{
The $\phi$-component of the vector potential $r A_{\phi}$ is identical 
to the magnetic flux 
$\Psi(r,z) = (1/2\pi) \int {\vec {B}}_{\rm P} \cdot d{\vec{A}}$
}
$A_{\phi}$ 
using a numerical grid with twice the resolution of the grid applied
in the ZEUS code.
Having obtained the vector potential of the initial field distribution
by the finite element code,
we derive the magnetic field distribution for the time-dependent 
simulation with respect to the ZEUS-3D {\em staggered mesh}
(see \citet{fend00}.
Our approach guarantees $|\nabla \cdot\vec{B}| < 10^{-15}$, respectively 
$|\vec{j} \times \vec{B}|= 0.01 |\vec{B}|$.

For the finite element grid, we prescribe the disk magnetic 
flux $\Psi(r,z=0) = r^{\mu_{\Psi}}$ as Dirichlet boundary condition,
corresponding to a poloidal magnetic field
$\bp(r,0)\!\sim\! r^{-\mu}$ with $\mu = 2 - \mu_{\Psi}$.
Along the axial boundary $\Psi(r=0,z) = 0$.
Along the outer boundaries we prescribe a magnetic flux decreasing
towards the axes (here the exact flux profile is marginally important 
as the initial field will be immediately perturbed by the outflow). 

\subsection{Boundary conditions}

Along the $r$-boundary we distinguish between the gap region extending 
from $r=0$ to $r=r_i = 1.0$ and the disk region from $r=1.0$ to $r=r_{\rm out}$.
The disk region governs the mass inflow from the disk surface into corona
(denoted also as "disk wind" in the following).
The poloidal magnetic field along the $r$-boundary is fixed in time and is,
hence, determined by the choice of the initial magnetic flux distribution.
Thus, the magnetic flux across the equatorial plane is conserved,
$\Psi(r,z=0;t) = \Psi(r,z=0;t=0)$.
The poloidal magnetic field profile along the disk is chosen as
a power law, $\bp(r,0)\!\sim\! r^{-\mu}$.
Equivalent to \citet{ouye97a}
the disk toroidal magnetic field component is force free with
$B_{\phi} = {\mu}_i / r $ for $r\geq r_{\rm i}$, $\mu_i=B_{\phi,{\rm i}}/B_{\rm i}$.

\subsubsection{Hydrodynamic boundary conditions}
The hydrodynamic boundary conditions are ``inflow'' along the $r$-axis,
``reflecting'' along the symmetry axis and ``outflow'' along the outer boundaries.
Matter is ``injected'' from the disk into the corona parallel to the poloidal
magnetic field with low velocity
$\vec{v}_{\rm inj}(r,0) = \nu_{\rm i} v_{\rm K}(r) \vec{B}_{\rm P}/B_{\rm P}$
and with a density $\rho_{\rm inj}(r,0) = \rhoi\,\rho(r,0)$.
In some simulations we applied a slightly enhanced mass flow rate from the outermost
grid elements of the disk by increasing the inflow velocity by a certain factor.
This choice helped to deal with some problems at the intersection of inflow and outflow 
boundary conditions just in the corner at $(r_{\rm max}, z=0)$, and does not affect the 
flow on larger scales.
Along the gap, the mass flux into the corona is set to zero.
We chose a power law for the inflow density profile,
\begin{equation}
\rho_{\rm inj}(r,0) \equiv \rho_0(r) = \rhoi r^{-\mu_{\rho}}.
\end{equation}
As a first choice, it is natural to assume $\mu_{\rho} \sim 3/2$ as this 
is the disk density profile both for the Shakura-Sunyaev accretion disk
and for advection dominated disks
(see e.g. the self-similar Blandford \& Payne solution, or the
\citet{ouye97a} model).
This is, however, the average density profile of a thin disk and one
may expect deviations from that profile for several reasons.
Those may arise 
(i) as the structure of jet-driving disks can be altered by the outflow, 
(ii) due to magnetic fields close to equipartition, 
or may be simply suggested by the fact that 
(iii) we do not yet fully understand the mass 
      loading of the jet / disk wind out of the accreting matter.
We have therefore applied different density profiles 
$\mu_{\rho} = 0.3, ..., 2.0$
for the disk wind mass flux in our simulations.

For the injection velocity, one may assume that the initial disk wind speed is
in the range of the sound speed in the disk,
$v_{\rm inj}(r) \simeq c_{\rm s}(r) \simeq v_{\rm kep} \sim r^{-1/2}$.
However, the actual velocity along the disk surface will be different from
the thin disk estimate and will, moreover, depend in detail from the launching
mechanism.
For simplicity, all simulations presented in the paper assume the same injection
velocity. 
Therefore, the total mass flux profile is varied only by the density profile.

\subsubsection{The disk magnetic field profile}
The disk magnetic flux profile is less constrained, or better, less understood.
A thin disk equipartition magnetic field will follow a profile $\sim r^{-5/4}$
\citep{blan82}. 
However, it is not clear to date where the disk/jet magnetic field originates, 
how it is generated and how it evolves.
We find it therefore valuable to explore a wider range of disk magnetic flux
profiles. This is the main goal of the present paper.
The choice of having a certain disk magnetic field distribution fixed in 
time should be understood as this field structure being
{\em determined} by the internal disk physics
(implying that different disk internal conditions may lead to a different
disk flux distribution).
In this respect, we do not care about how the magnetic field has been
{\em built up} in time. 
Instead, we assume that internal disk processes like diffusion, advection,
or turbulence work together in generating and supporting a large scale
stationary field structure with the chosen disk surface field distribution.
In the introduction we have already stressed the point that concerning
the internal disk physics and the interaction between disk and outflow
(in both directions)
some of the parameter runs considered may finally be proven to be more
realistic than others if compared to a fully self-consistent simulation 
including the disk internal disk physics and dynamics in relation 
with the outflow itself.
So far, this problem has not been solved in generality, although there 
are
first simulations treating the disk-jet system over many hundreds of 
rotational periods applying simplified disk models 
\citep{cass02,kudo02,ceme04,reko04}.
In particular we mention the work by \citep{reko04} who in their simulations
did not constrain the disk magnetic field but let it evolve by a dynamo
process from an initial stellar magnetosphere. 
Indeed the disk magnetosphere evolves into an open field structure, however
the radial profile of the disk field is not explicitly mentioned.
The outflows from disk and star are pressure driven but a magneto-centrifugally
driven outflow region co-exists originating in the inner part of the disk.
However, no collimated jet flows have been observed in these simulations
covering a area of $0.1\times0.2$\,AU. 

Some insight into the parameter space investigated in our simulations
may be gained by comparing them to stationary models of accretion-ejection 
structure. 
As an example, in the following we discuss the solutions presented 
by \citet{ferr97}.
Ferreira et al. apply self-similar and stationary MHD for the disk-outflow
system and derive the so-called ejection (or mass loading) index 
$\xi \equiv d \ln \dot{M}_{\rm acc} / d \ln r $ 
as the leading parameter governing the structure of the outflow.
They find this parameter being constrained by a lower limit of $0.004$ 
due to the vertical disk equilibrium, 
and a maximum value $\xi < 0.15$ due to the condition of having a 
trans-Alfv\'{e}nic jet\footnote{Note that with
$ d\dot{M}_{\rm acc} / dr \sim d\dot{M}_{\rm jet} / dr$
we have $\mu_{\rho} = 3/2 - \xi$}.
Since by definition $\xi>0$, it follows that $\mu_{\rho}< 3/2$.
In fact, for $\mu_{\rho} = - 3/2$ the Ferreira et al. approach gives
the Shakura-Sunyaev disk density distribution and $\xi=0$, 
i.e. a constant accretion rate $ d\dot{M}_{\rm acc}(r)$ and no
outflow.
Their upper limit $\xi = 0.15$ would imply a very narrow range
for the radial density profile, $1.35 < \mu_{\rho} <1.5$.
However, in our simulations we obtain trans-Alfv\'{e}nic outflows 
also for lower $\mu_{\rho}$.
Our interpretation is that this constraint might be an artifact of
the self-similar approach undertaken by Ferreira et al.
Without considering the disk structure self-consistently in a 
time-dependent simulation this question cannot be answered.

The self-similar study further provides a constraint for the profile
of the magnetic flux configuration $0 < \mu_{\Psi} < 2$ 
(note that $\beta$ in Ferreira's notation corresponds to $\mu_{\Psi}$ in our's).
This is in agreement with all parameter runs in our study.
Another relation found by Ferreira et al. is that $\mu_{\Psi}$ can be 
related to $\xi$ by $\mu_{\Psi} = 3/4 + \xi /2$, 
or
$\mu = 1/2 + \mu_{\rho}/2$, respectively.
(the number value $3/4$ indicates the Blandford-Payne solution).
This quantifies the relation between the profile of the disk mass loss 
and that of the disk magnetic field.
The above mentioned constraint $\mu_{\rho} <3/2$ implies that $\mu < 5/4$,
which tells us that our simulations in the upper part of Tab.~\ref{tab_all}
are inconsistent with the self-similar disk-jets discussed by Ferreira et al.
However, considering the fact that 
(i) the steepness of the field profile is not so different from the other runs
and that
(ii) the analytical constraint is limited by the self-similarity assumption, 
we still think that these runs provide valuable scientific information.

Finally, we can only stress the point once more that both the the mass 
loading from disk into the outflow and the field structure will eventually
defined by the disk physics and the constraints to it set by the existence 
of an outflow.

\subsection{Jet flow magnetization}  \label{sec_sigma}

In ideal MHD the magnetic field is ``frozen'' into the matter which itself slides 
along (but not parallel to) the field lines.
In an axisymmetric and stationary configuration the magnetic field lines are 
located on {\em magnetic flux surfaces} $\Psi(r,z)$ and the velocity vector can be 
expressed as
$\vec{v} =  k(\Psi)\vec{B}/\rho - r\Omega_{\rm F}(\Psi)$,
where 
$ k(\Psi) = (d\dot{M}/d\Psi) = (\rho v_{\rm p}/B_{\rm p}) $
is the mass flow rate per magnetic flux surface
and $\Omega_{\rm F}(\Psi)$ the iso-rotation parameter\footnote{
For simplicity, the iso-rotation parameter $\Omega_{\rm F}$ can be 
interpreted as the angular velocity of a magnetic field line.
Both $k(\Psi)$ and $\Omega_{\rm F}(\Psi)$ are conserved quantities
along the magnetic flux surface.
}.
One of the most important parameters for MHD wind theory is the 
{\em magnetization} parameter \citet{mich69},
\begin{equation}
\sigma \equiv \frac{B^2_{\rm p} r^4 \Omega^2_{\rm F}}{4 \dot{M} c^3},
\label{eq_sigdef}
\end{equation}
where $\dot{M}(\Psi) $ is the mass flow rate within the magnetic flux surface $\Psi $. 
The magnetization is inverse to the mass load $\sigma \sim k^{-1}$.
For outflows expanding in spherically {\em radial direction,} \citet{mich69} derived
an analytical relation between the asymptotic outflow velocity and the magnetization,
$v_{\infty} \sim \sigma^{1/3}$ (Michel scaling).
The outflow collimation results in a variation in the power index of the Michel 
scaling \citep{fend96}.
Essentially, the magnetization summarizes the important parameters for launching a high
velocity outflow -- rapid rotation, strong magnetic field and/or comparatively low mass 
load.
It will therefore be interesting to study also the influence of the magnetization 
{\em profile} across the disk wind on the collimation of this wind into a jet.
Having prescribed a power law for the profiles of the disk wind initial poloidal velocity
$v_{\rm inj}(r) = v_{\rm inj} v_{\rm Kep}(r)$, 
the rotation $\Omega_{\rm F}(r) = \Omega_{\rm Kep} \sim r^{-3/2}$, 
the density and magnetic field profile, 
the disk wind magnetization profile $\sigma_0(r)\equiv \sigma(r,z\!=\!0)$
follows a power law as well,
\begin{equation}
\sigma_0(r)  \sim
 \frac{B_{\rm p}^2(r)}{\rho(r)} r^{-1/2} \sim r^{-(2\mu - \mu_{\rho} +1/2)} \equiv r^{\mu_{\sigma}}
\end{equation}
(hence, $\mu_{\sigma} = \mu_{\rho} - 2\mu - 1/2$).
A choice of parameters as 
$\rho_{\rm inj}(r) \sim r^{-3/2}$, $B_{\rm p}(r,z\!=\!0) \sim r^{-1}$
\citep{ouye97a,fend02} will
result in a radial disk wind magnetization profile $\sigma_0(r)\sim r^{-1}$.
The self-similar Blandford \& Payne model gives $\mu_{\sigma} = -3/2$.
In a recent paper \citet{pro06} have followed an approach similar to
the present paper, varying the mass load $k(r)$ at the disk surface
as $k(r) \sim r^{-1}, r^{-3/4}, r^{-1/2}, r^{-1/4}$.
This would correspond to magnetization profiles as $\mu_{\sigma} = 1, 0.75, 0.5, 0.25$,
respectively, a parameter space which is also covered in our paper.

\section{Results and discussion}

We have run a large number of MHD jet formation simulations covering a wide parameter
range concerning both the disk wind poloidal magnetic field profile and the
"injection" density profile along the disk surface (see Tab.~\ref{tab_all}).
The results presented here are preliminary in the sense that not all simulation runs
could be performed over time scales sufficiently long enough for the MHD outflow 
{\em as a whole} to reach the grid boundaries or to establish a stationary state.
This is due to the large grid size in combination with steep gradients in the disk
wind parameters which in general allow only for a weak outflow from large disk radii.
Nevertheless, our results allow for a firm interpretation of the overall outflow
structure and evolution in the MHD jet formation region.
In most cases we have obtained a quasi-stationary state for the outflow within the
inner region (over at least $50\%$ of the grid).
In some cases the nature of the outflow prevents that a stationary state can be 
reached.

\clearpage
\begin{figure}
\centering

\includegraphics[height=0.7cm]{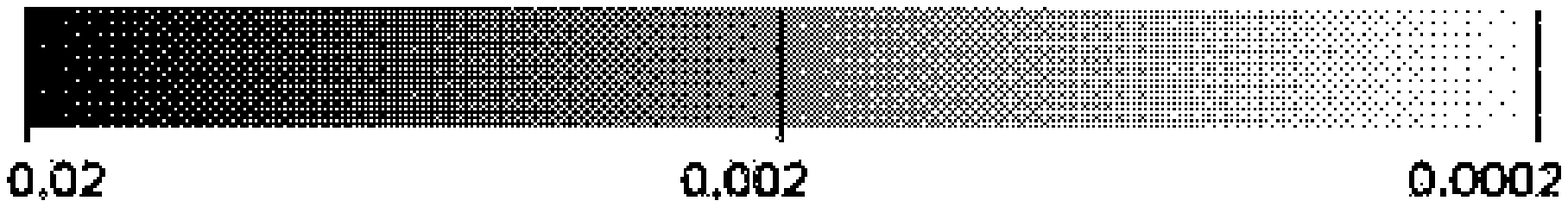}

\includegraphics[height=3.5cm]{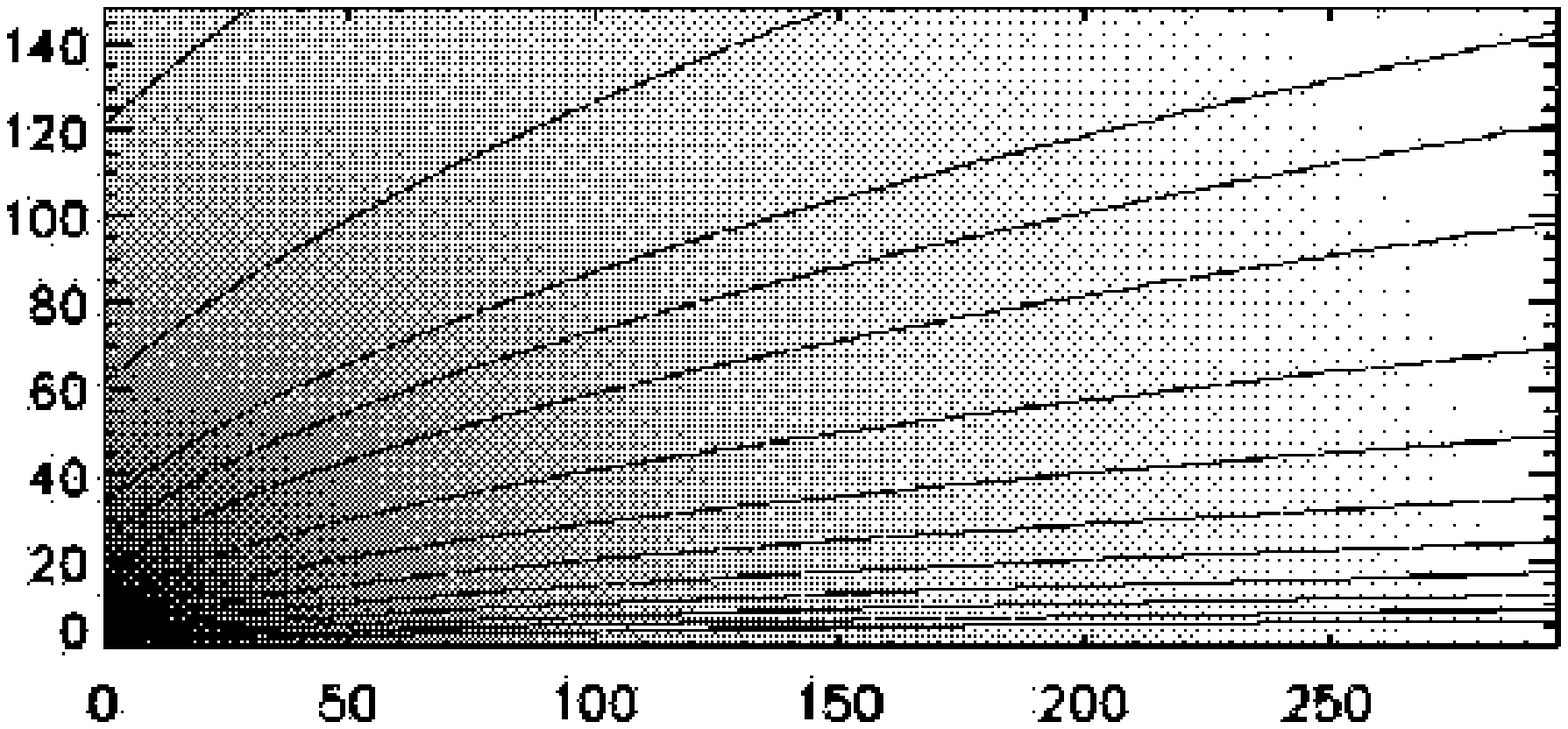}

\includegraphics[height=3.5cm]{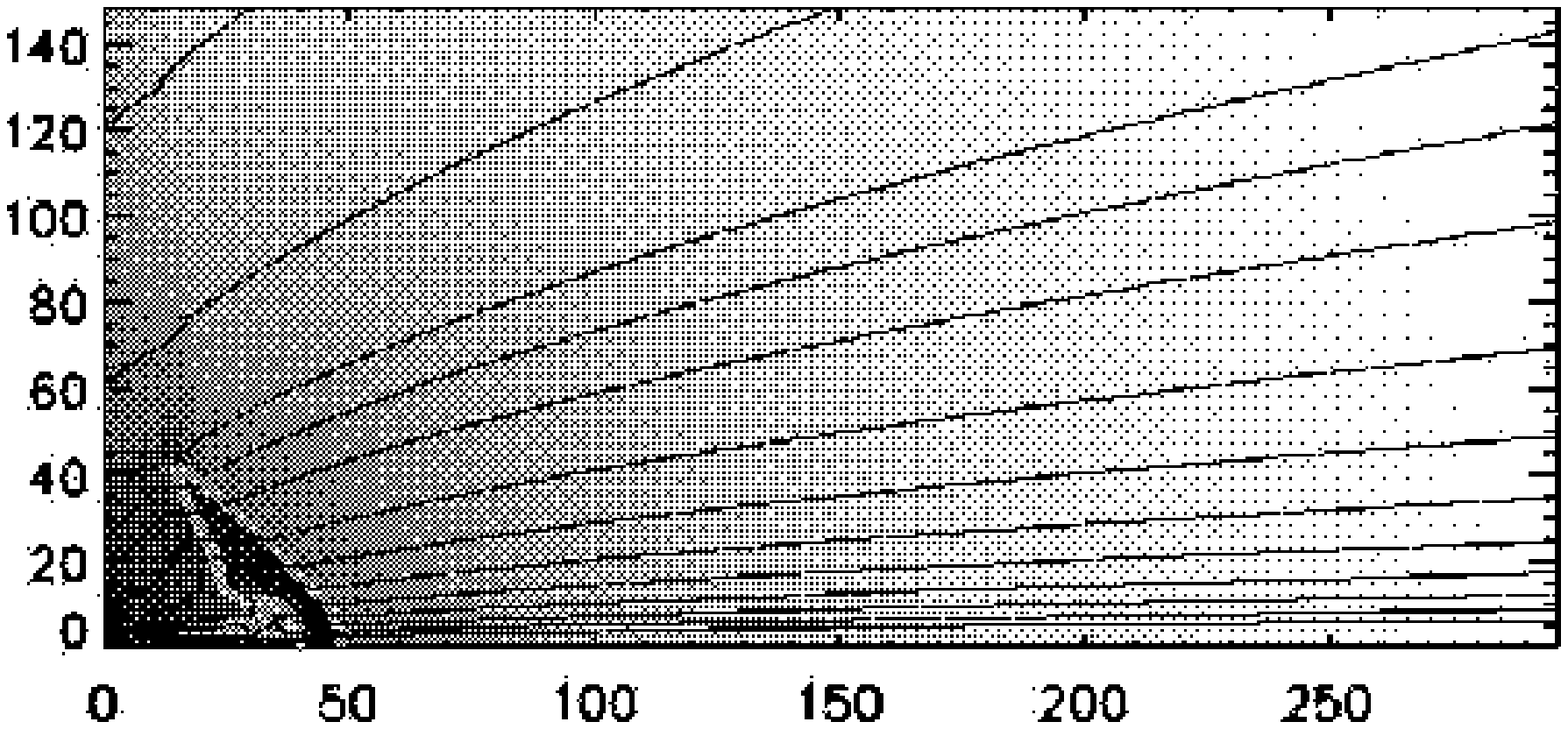}

\includegraphics[height=3.5cm]{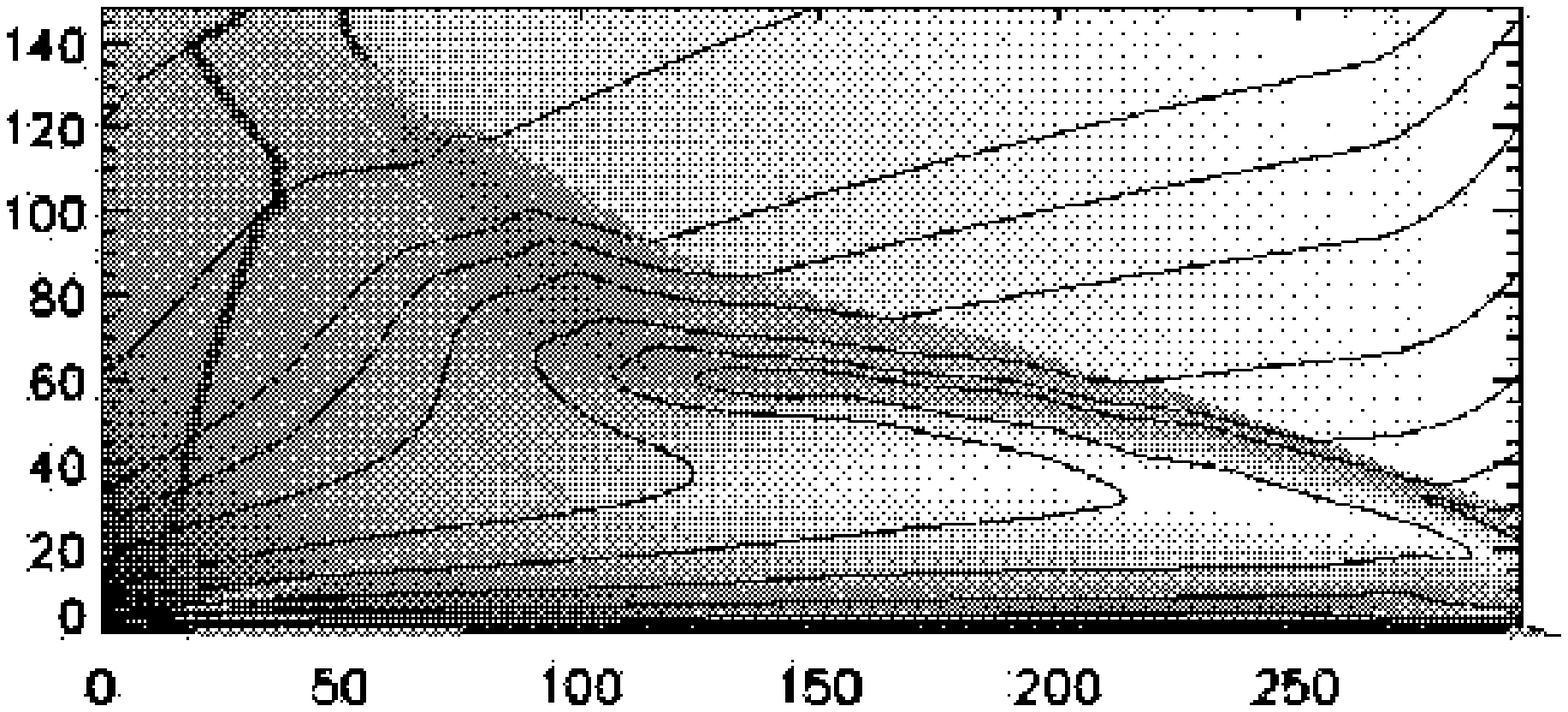}

\includegraphics[height=3.5cm]{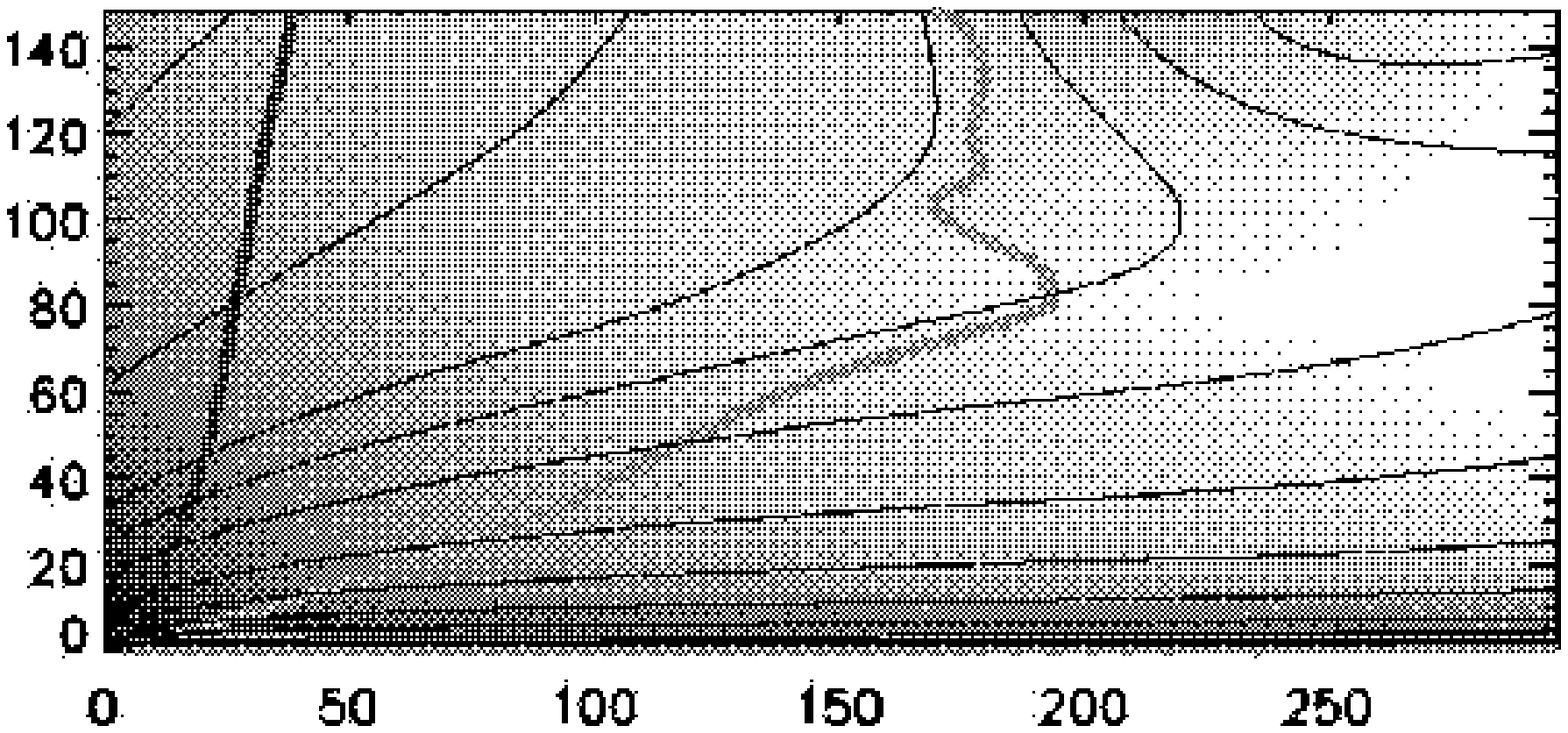}

\includegraphics[height=3.5cm]{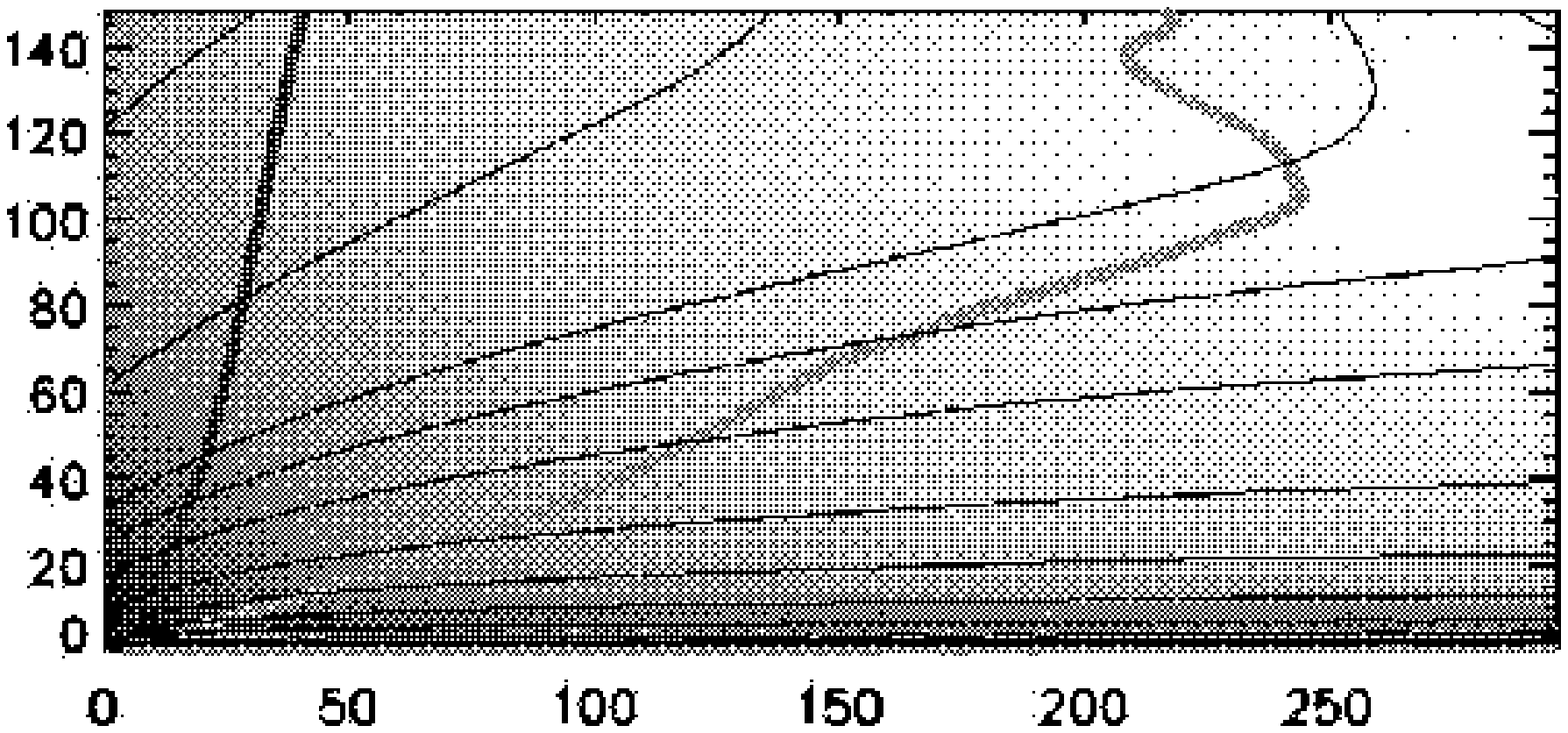}

\caption{Simulation run i11. 
Mass density in {\em grey colors} 
(logarithmic scale between between the density values shown at the 
{\em color bar} on the very top) and 
poloidal magnetic field lines ({\em black lines}) at 
times $t = 0, 100, 500, 1500, 2090$ (from {\em top {\rm to} bottom}).
The $r$ and $z$-axis point in vertical and horizontal direction, respectively.
The dynamic range of the color coding is increased by setting densities above 
(below) the values given at the color bar to black (white).
The thick dark grey and light grey contours indicate the run of the Alfv\'{e}n 
surface and the fast magnetosonic surface, respectively.
\label{fig_long}
}
\end{figure}

\begin{figure}
\centering

\includegraphics[height=0.5cm]{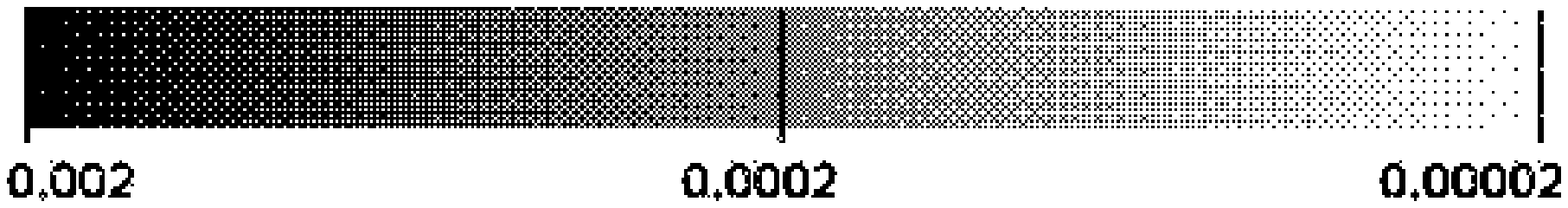}

\includegraphics[height=3.5cm]{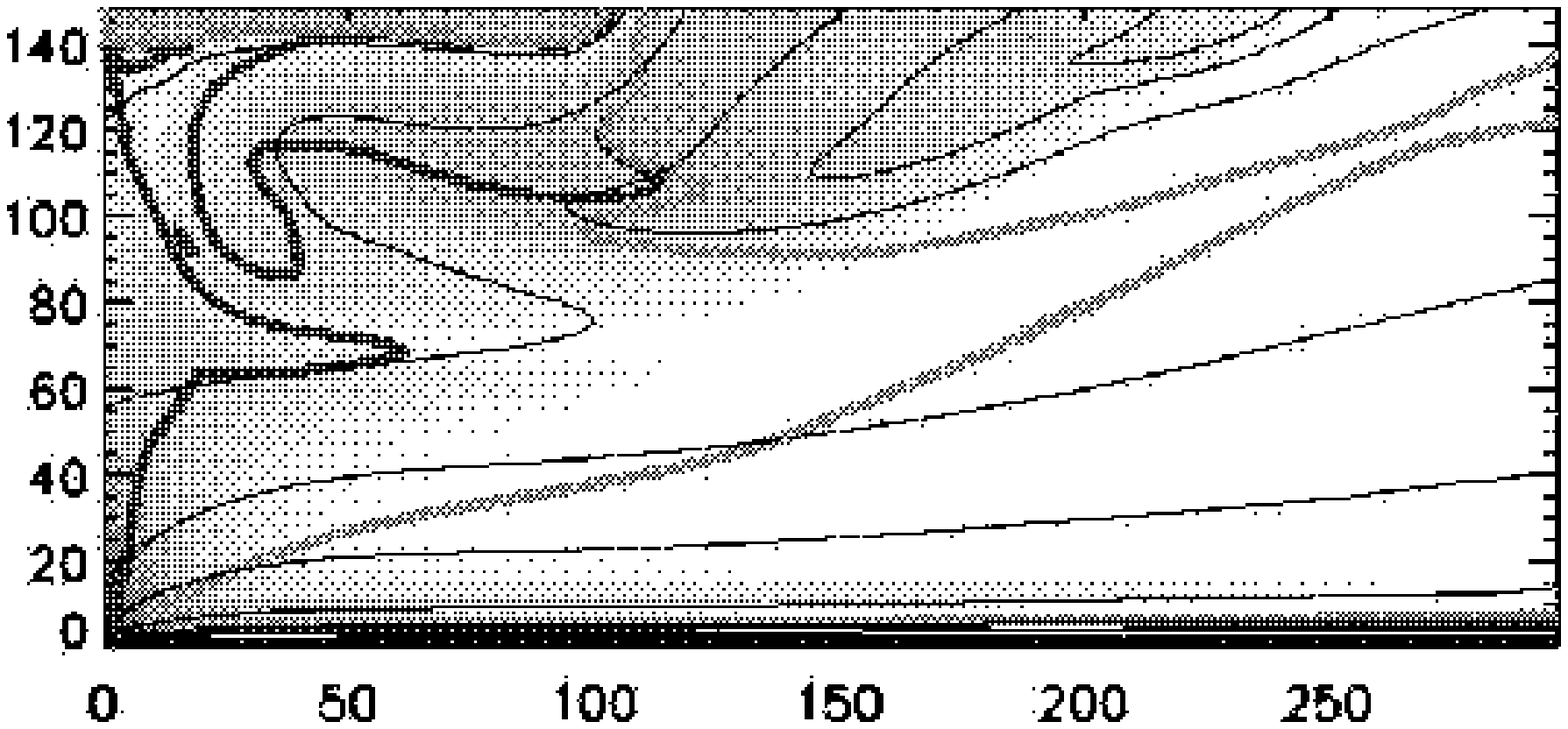}

\includegraphics[height=0.5cm]{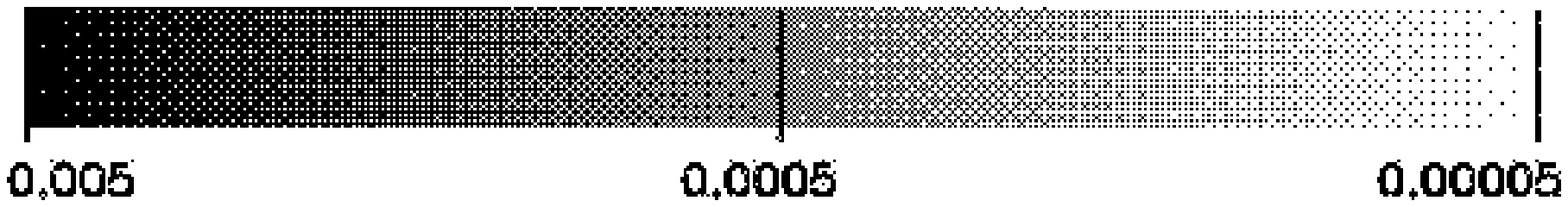}

\includegraphics[height=3.5cm]{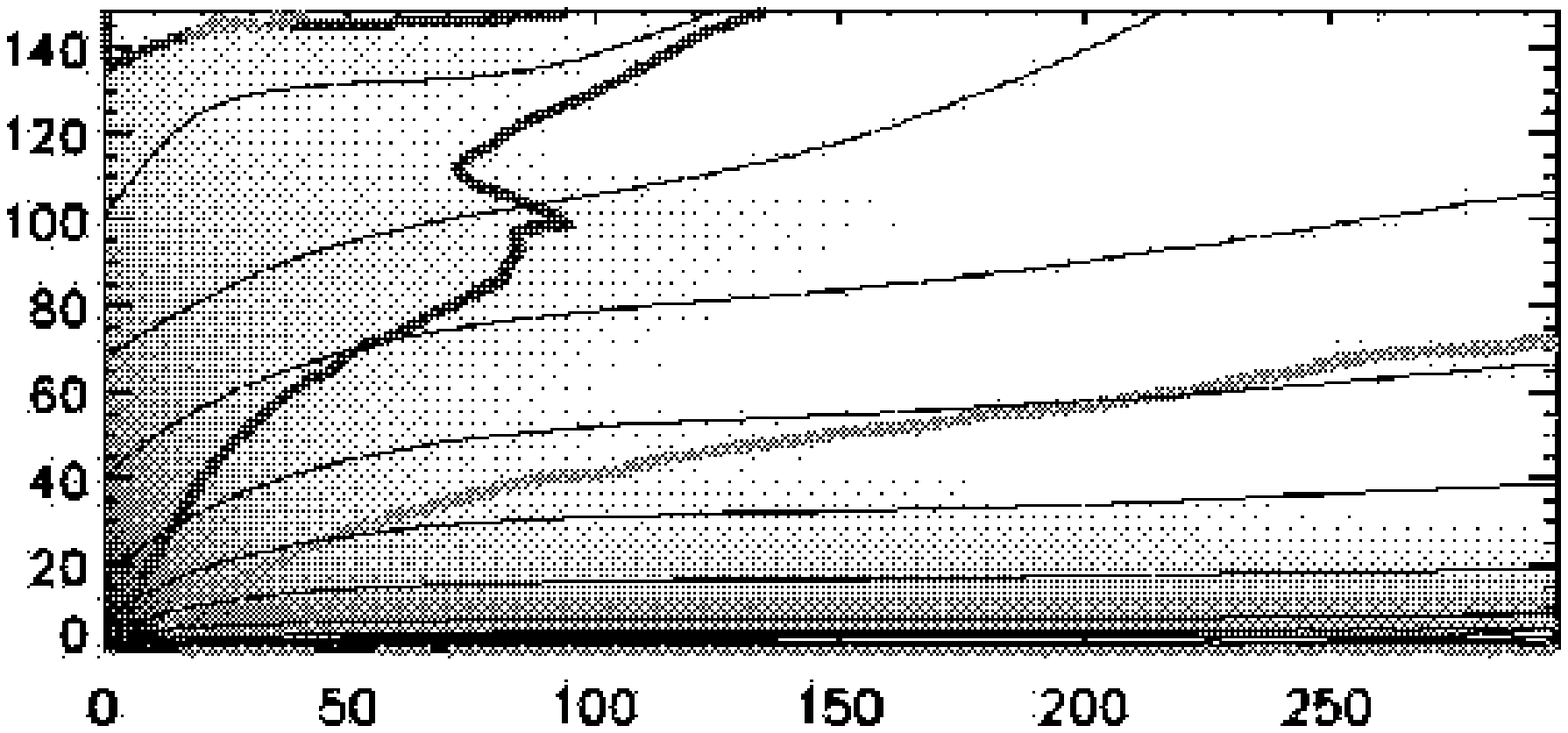}

\includegraphics[height=0.5cm]{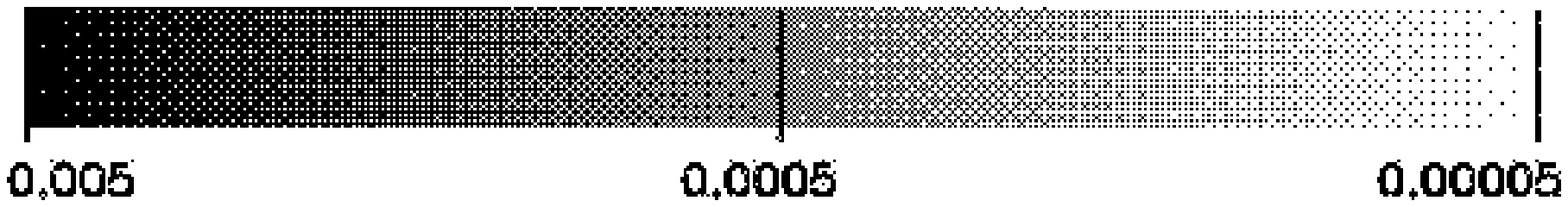}

\includegraphics[height=3.5cm]{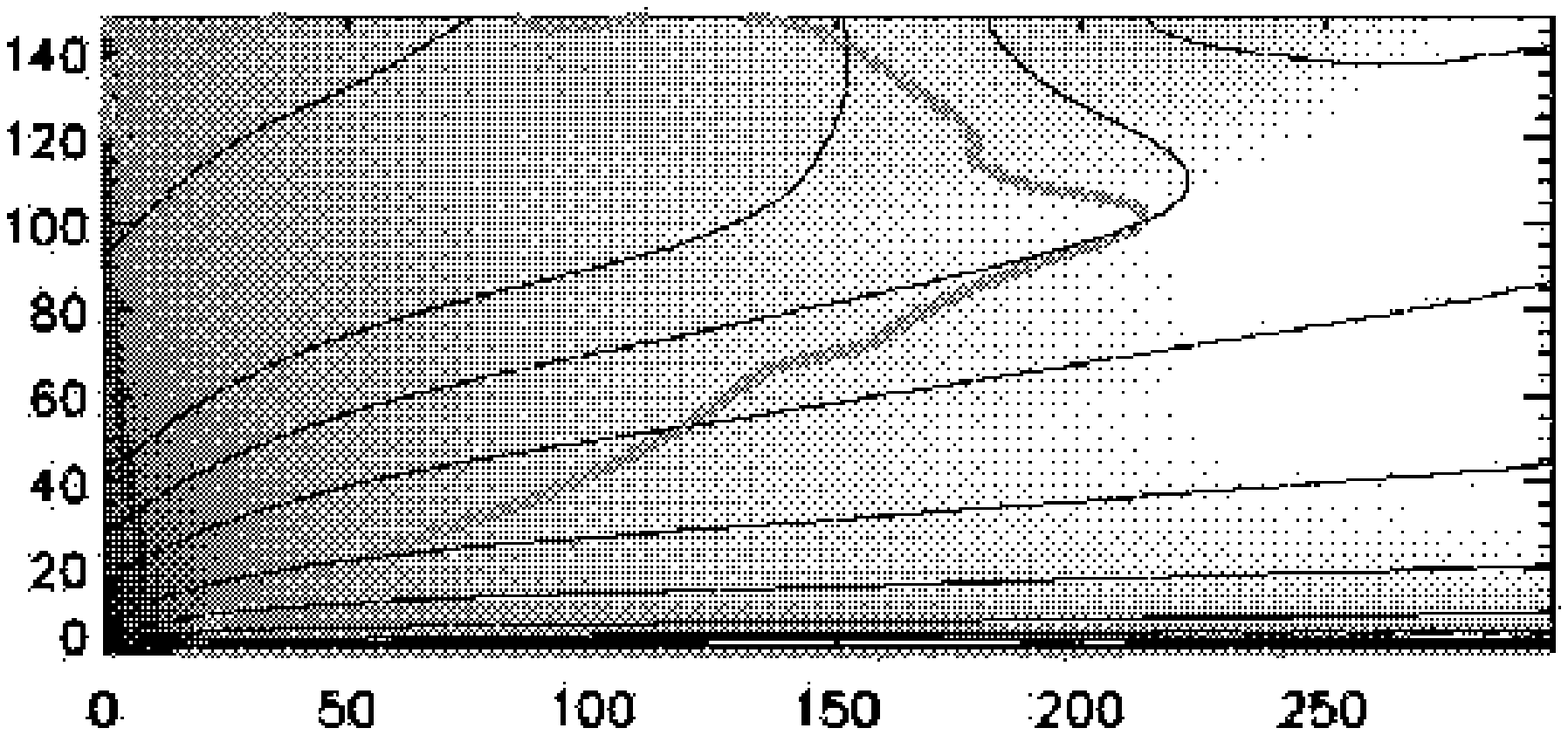}

\includegraphics[height=0.5cm]{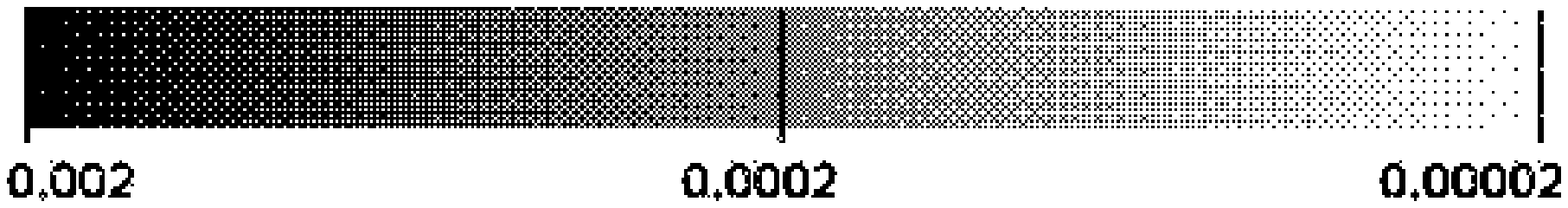}

\includegraphics[height=3.5cm]{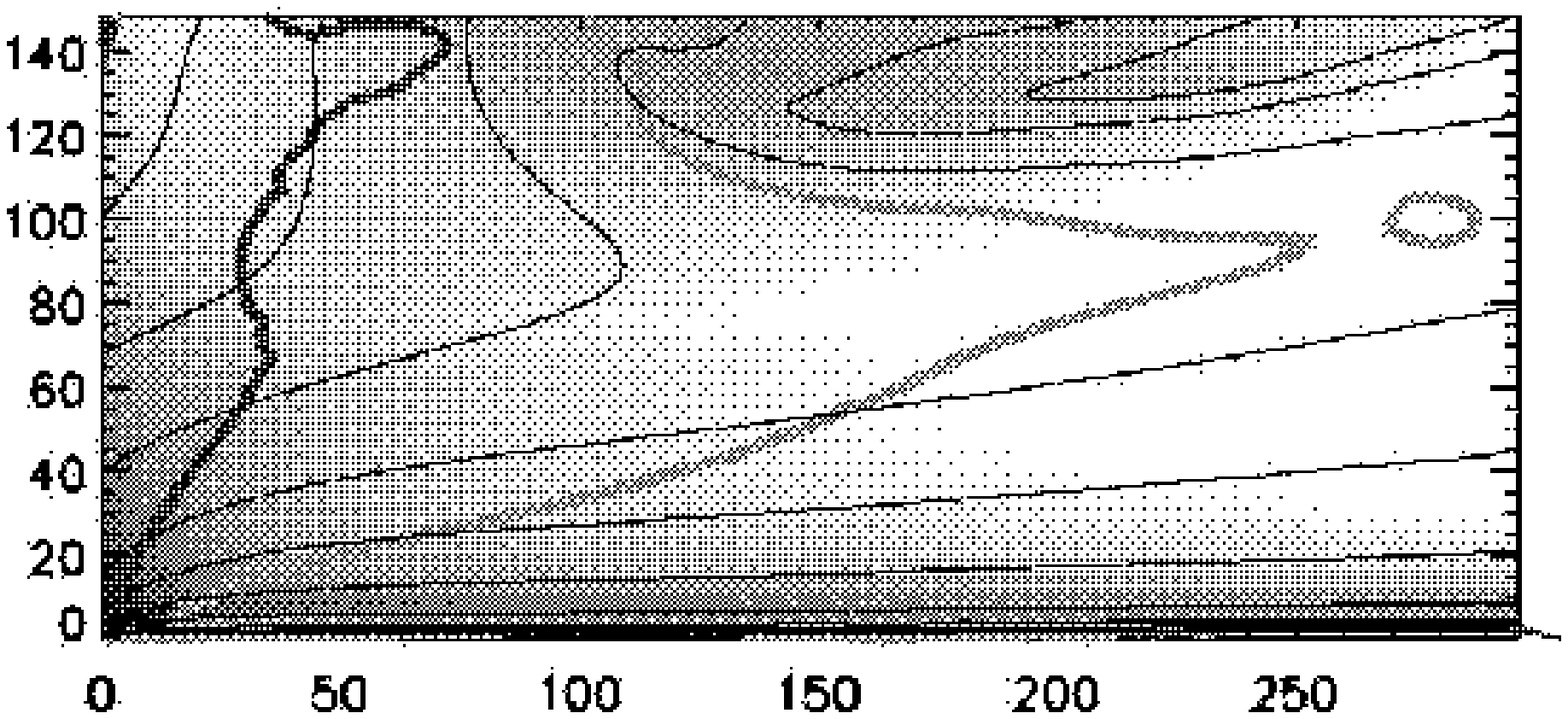}

\includegraphics[height=0.5cm]{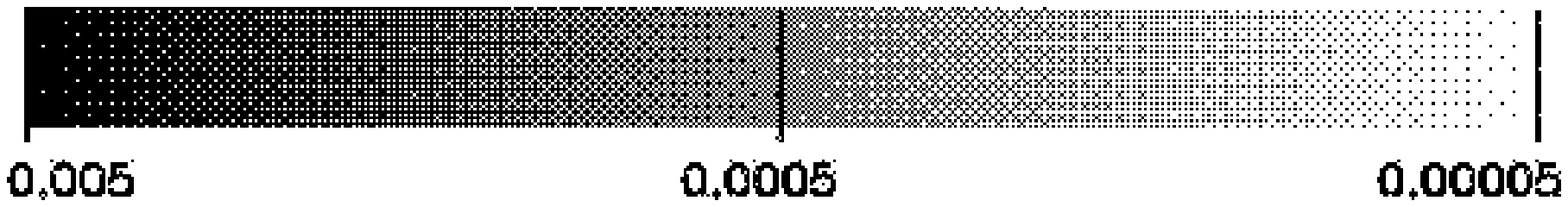}

\includegraphics[height=3.5cm]{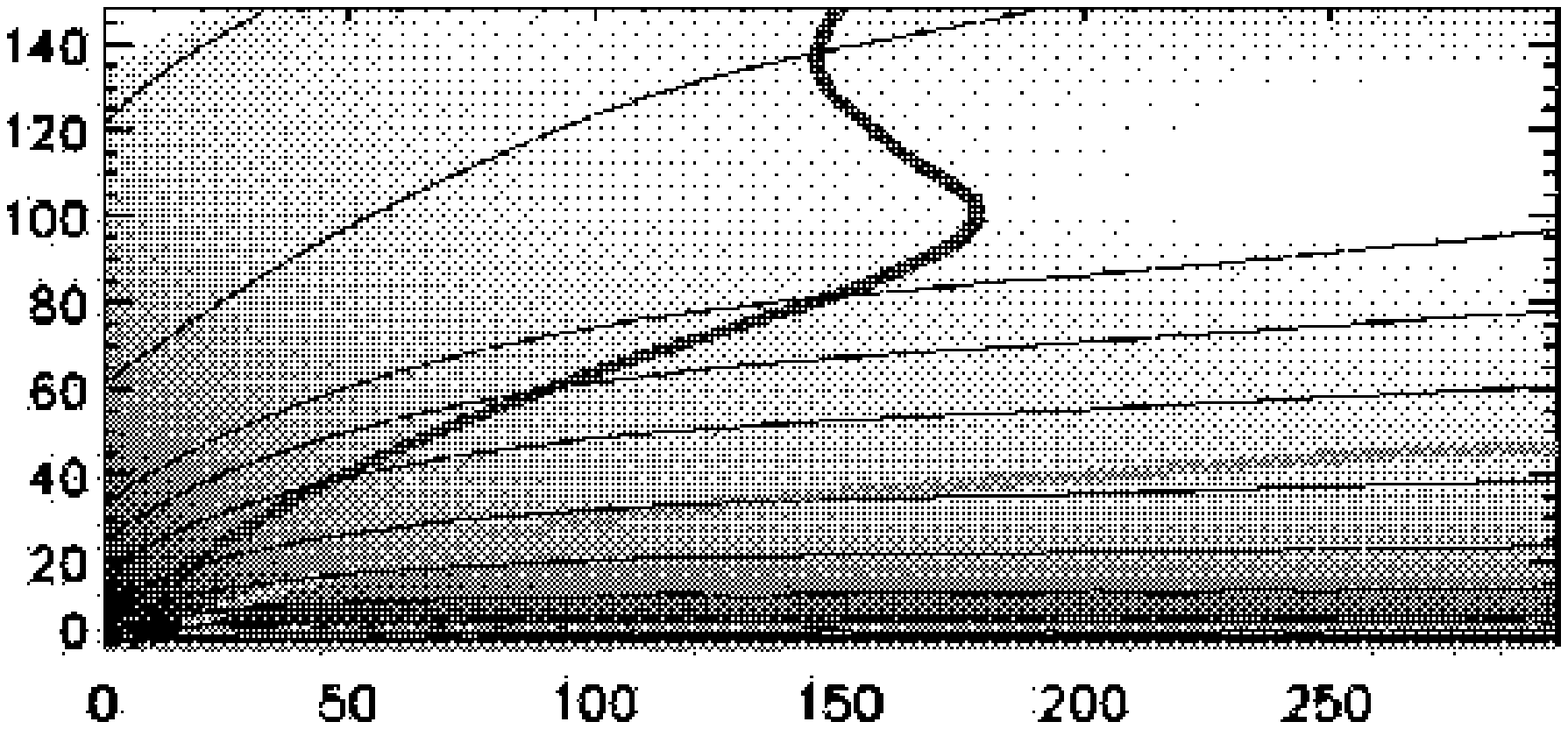}

\caption{Simulation runs with different disk 
boundary conditions (see Tab.~1).
Density and poloidal magnetic field lines for 
{\bf p4} ($t=2370$),
{\bf i15} ($t=5000$),
{\bf p20} ($t=1810$),
{\bf a2} ($t=2000$),
{\bf i10} ($t=3000$)
from {\em top {\rm to} bottom}.
See Fig.~\ref{fig_long} for further notation.
\label{fig_disk_1}
}
\end{figure}

\begin{figure}
\centering

\includegraphics[height=0.5cm]{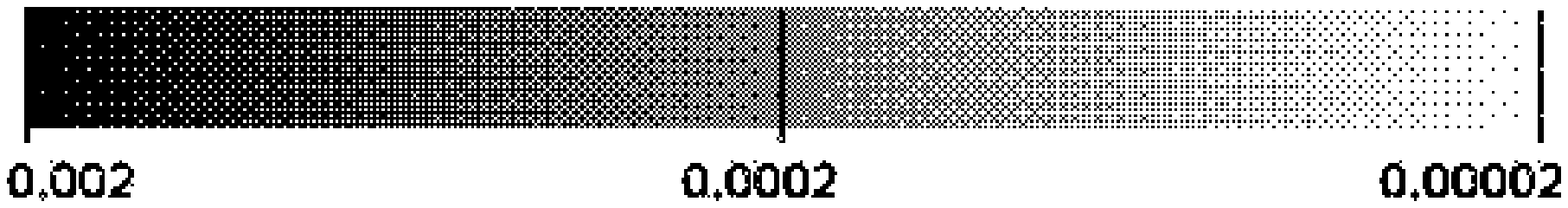}

\includegraphics[height=3.5cm]{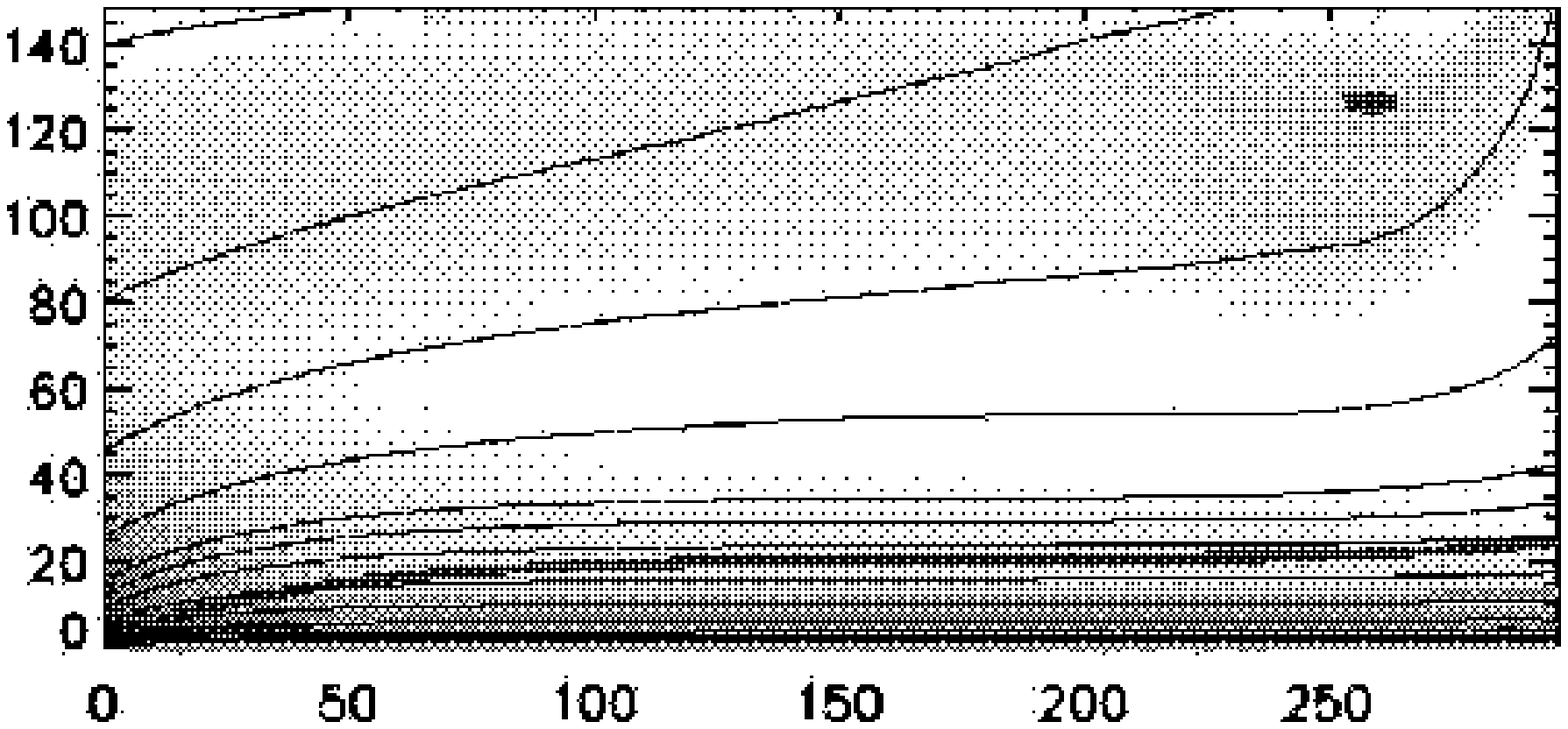}

\includegraphics[height=0.5cm]{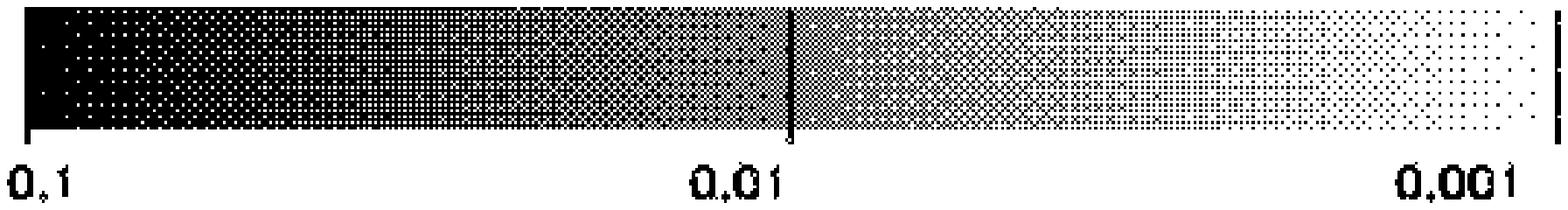}

\includegraphics[height=3.5cm]{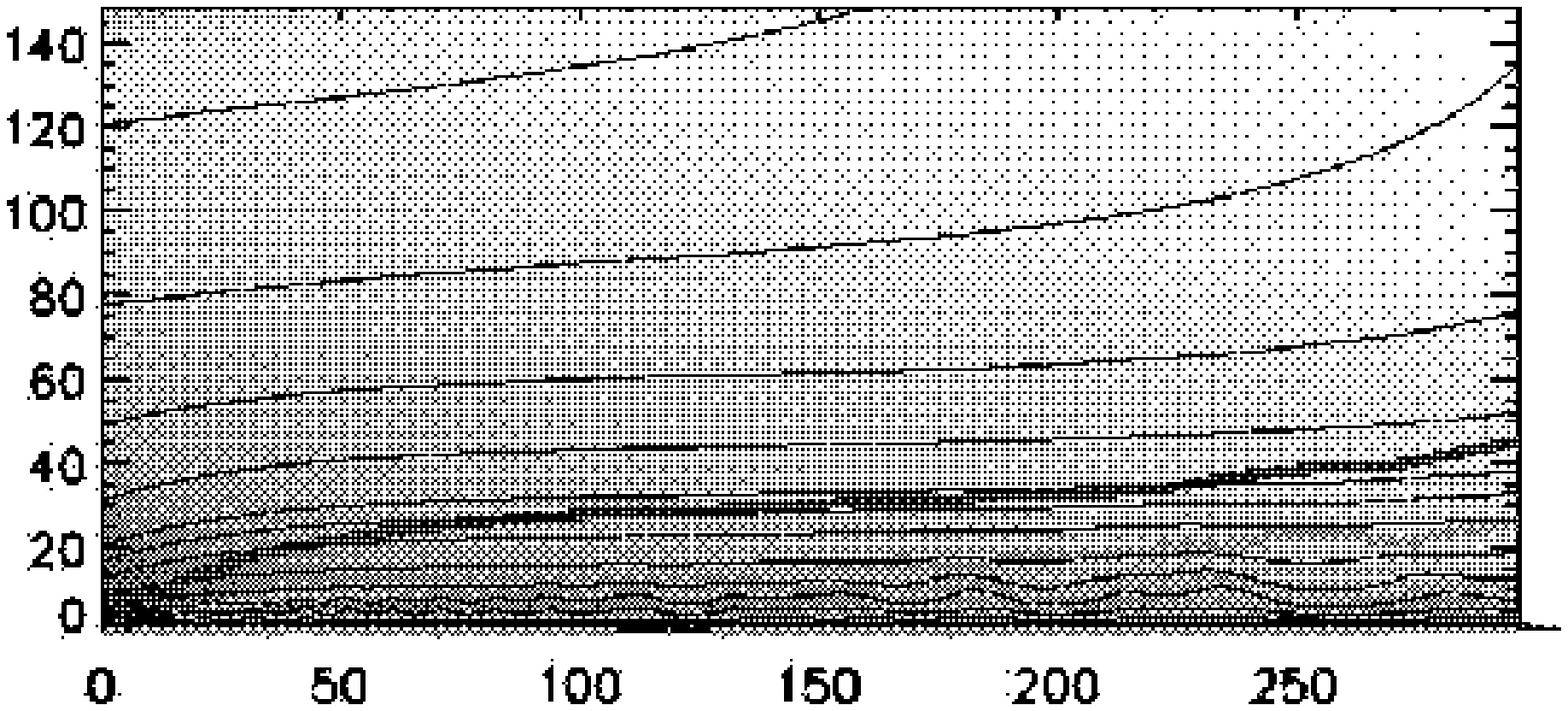}

\includegraphics[height=0.5cm]{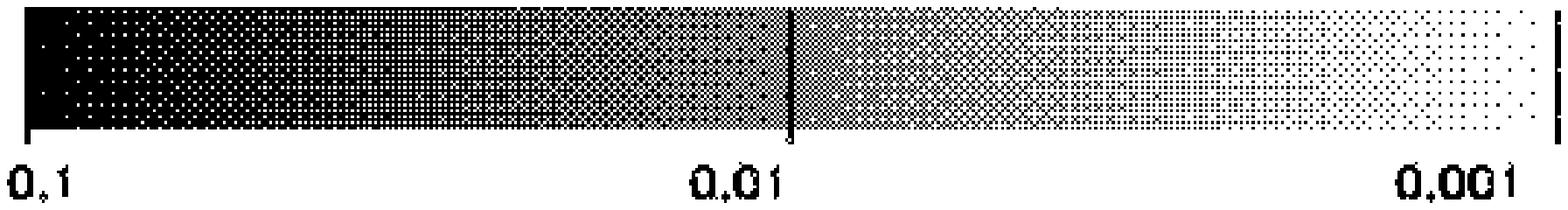}

\includegraphics[height=3.5cm]{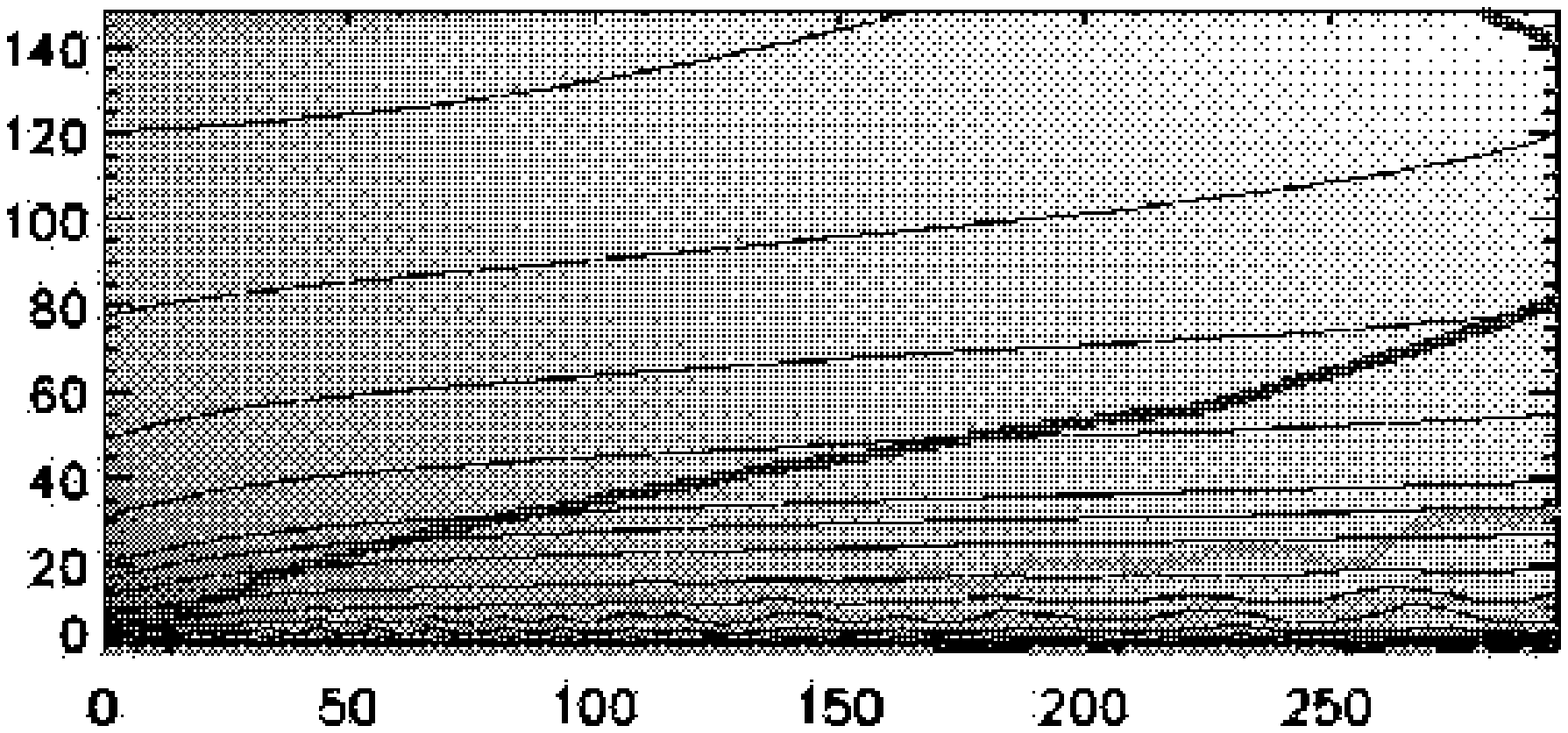}

\includegraphics[height=0.5cm]{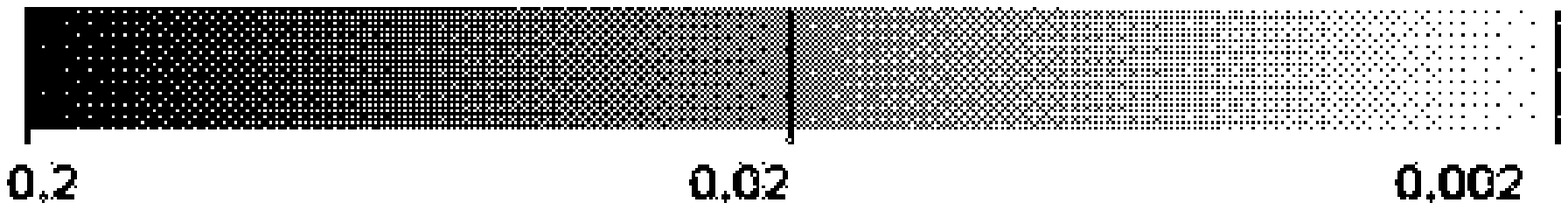}

\includegraphics[height=3.5cm]{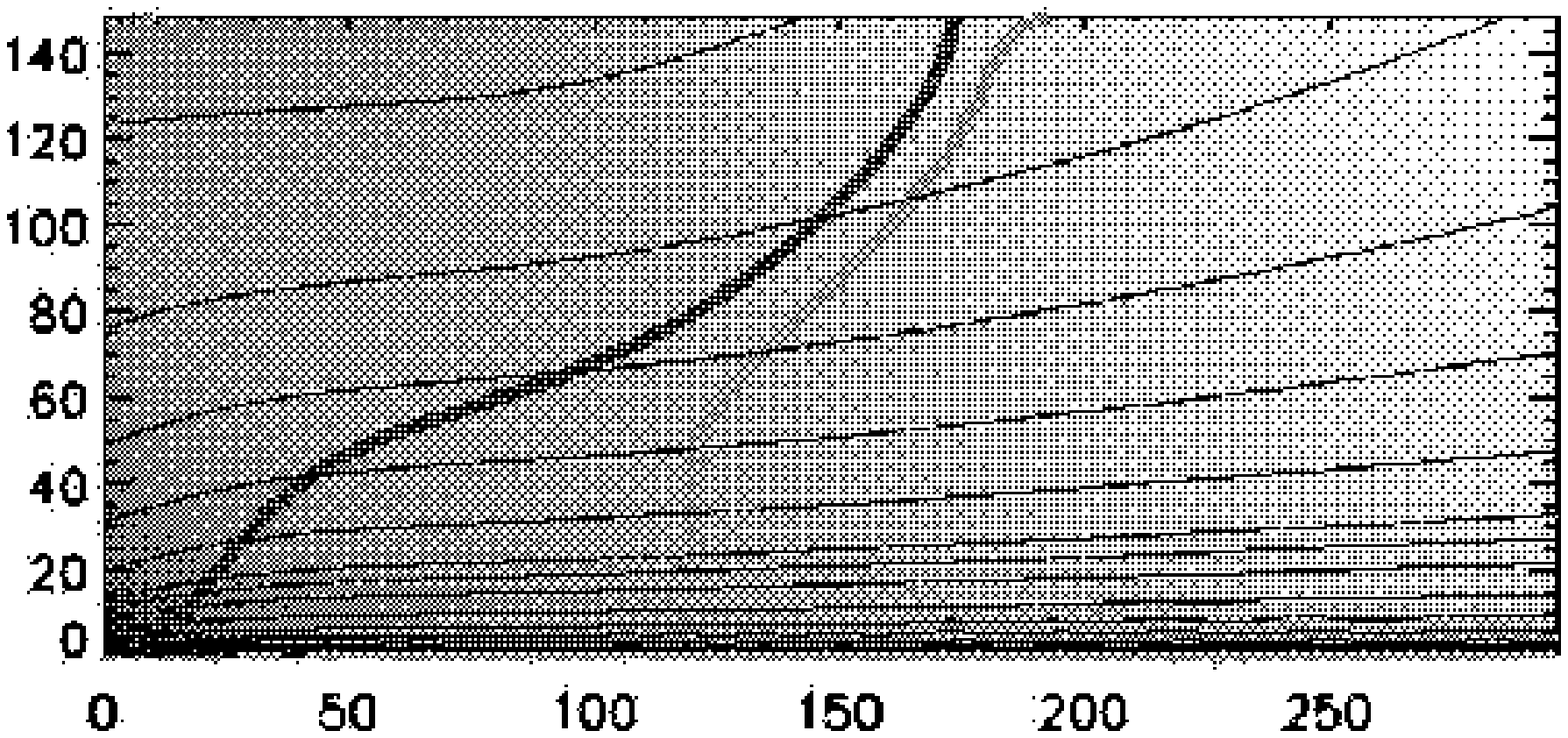}

\includegraphics[height=0.5cm]{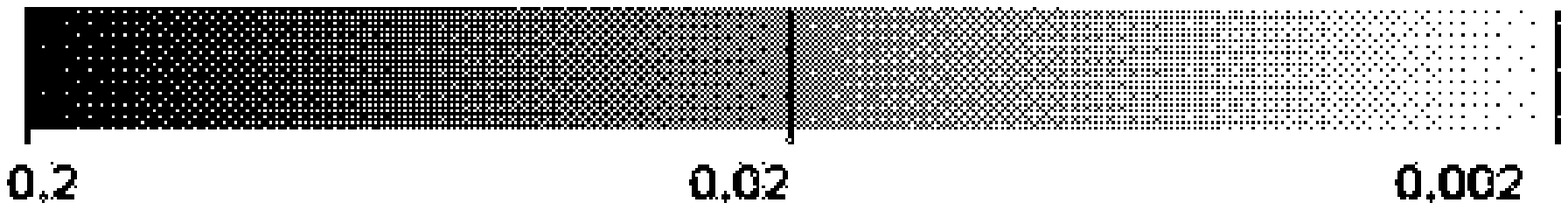}

\includegraphics[height=3.5cm]{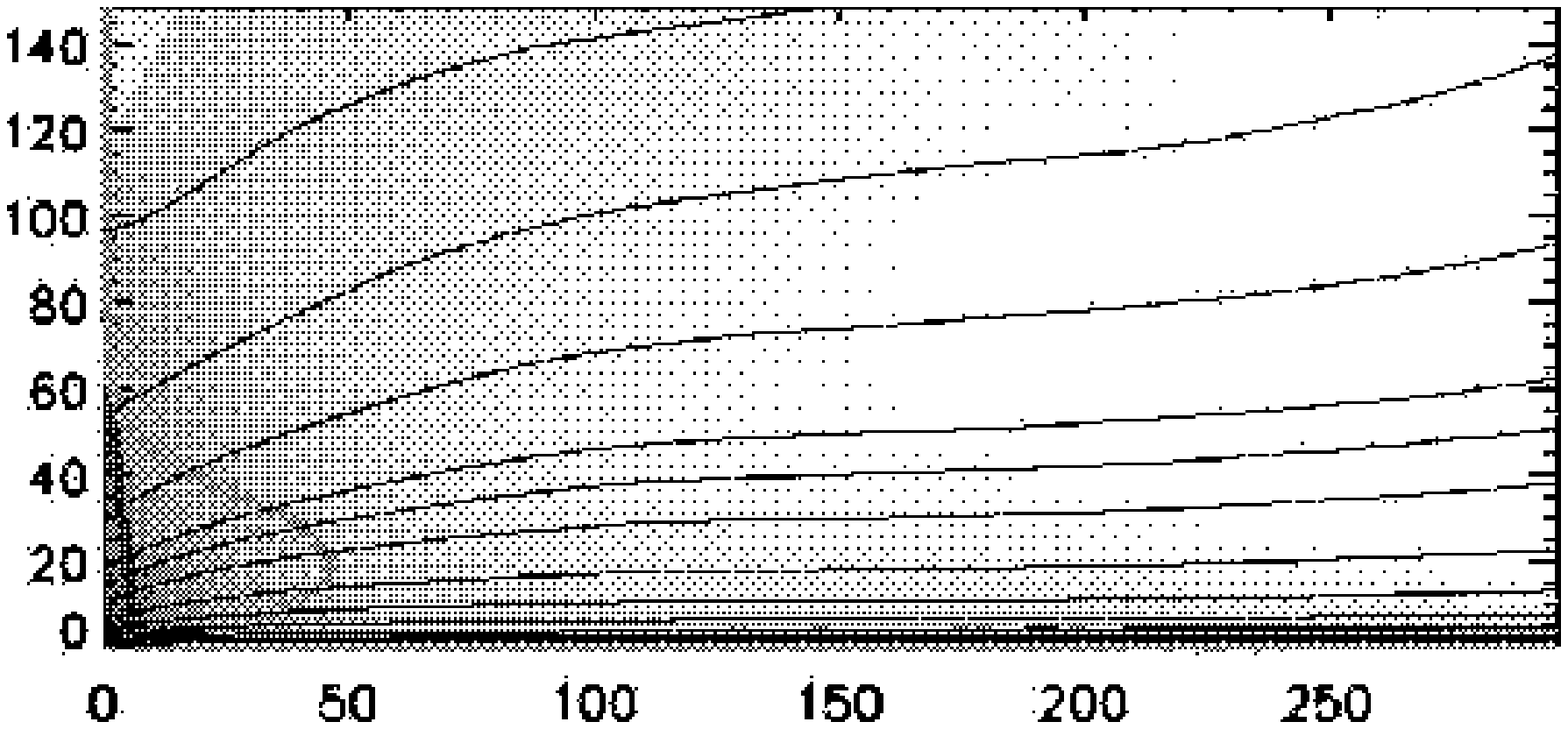}

\caption{Simulation runs with different disk
boundary conditions (see Tab.~1).
Density and poloidal magnetic field lines for 
{\bf p14} ($t=1840$),
{\bf i2} ($t=4000$),
{\bf i3} ($t=4000$),
{\bf p16} ($t=2790$),
{\bf c10} ($t=1400$)
from {\em top {\rm to} bottom}.
See Fig.~\ref{fig_long} for further notation.
\label{fig_disk_2}
}
\end{figure}
\clearpage
\subsection{The general flow evolution}

The main features of jet formation in the model scenario applied have been
described in the previous literature \citep{ouye97a,ouye99,fend02} and will
not repeated here in detail.
In summary,
the early evolution ($t< 100$) is characterized by the propagation of torsional
Alfv\'{e}n waves launched by the differential rotation between disk and corona
(see Fig.~\ref{fig_long}).
The Alfv\'{e}n waves slightly distort the initial field structure into a new
state of hydrostatic equilibrium which will remain stationary until reached by
the MHD outflow launched from the disk.
The region beyond the wave front remains undisturbed.
Torsional Alfv\'{e}n waves are still propagating across the grid, when the MHD
wind begins to accelerate from the disk surface and becomes collimated beyond
the Alfv\'{e}n surface.
The outflow rams into surrounding magnetohydro{\em static} corona developing a
bow shock which propagates with velocities comparable to the material speed.
The flow structure which is left when the bow shock has passed the computational
grid is a pure disk wind/jet, generated and energetically maintained by the disk
rotation, the continuous mass injection from the disk and the disk magnetic flux.
We note that at these later evolutionary stages the outflow has completely 
swept out of the grid the initial condition and is governed solely by the boundary
conditions.

As a general result, outflows launched from a flat disk magnetic field profile 
($\mu < 0.5$) do not reach a steady state.
Instead, a wavy pattern evolves close to the symmetry axis with almost constant
mass flow rate in axial direction but variable mass flow rate in lateral
direction.
This case is similar to the extreme example of an initially {\em vertical}
magnetic field examined by \citet{ouye97b}.
\clearpage
%
\begin{table*}
\scriptsize
\begin{center}
\caption{Summary table of our simulation runs.
Shown are parameters for the disk poloidal magnetic field profile ${\mu}$, 
the disk wind density profile $\mu_{\rho}$,
and the disk wind magnetization profile $\mu_{\sigma}$.
The inner disk radius poloidal field strength is $B_{\rm p,i}$.
The density at this radius is $\rho_{,i}=1$, but is sometimes lowered by a
factor of two (denoted by a $\star$ in column 1).
Mass flow rates in $r$- and $z$-direction along the three sub-grids
$(r_{{\rm max},i} \times z_{{\rm max},i}) = 
(20.0 \times 60.0),(40.0 \times 80.0), (60.0 \times 150.0)$
are denoted as $\dot{M}_{{\rm z},i}, \dot{M}_{{\rm r},i}$, respectively.
The average degree of collimation measured by the mass flow rates in 
$z$ and $r$-direction, and normalized by the area threaded, is $<\!\!\zeta\!\!>$.
The maximum flow velocity obtained outside the axial spine is 
$v_{\rm mx}$.
\label{tab_all}
}
\begin{tabular}{lllllllllllll}
\tableline\tableline
\noalign{\smallskip}  run & 
 $\mu$ & $\mu_{\rho}$ & $\mu_{\sigma}$ & $B_{\rm p,i}$ & $\tau $ & 
 $\dot{M}_{z,1},\dot{M}_{r,1}$ & $\dot{M}_{z,2},\dot{M}_{r,2}$ & $\dot{M}_{z,3},\dot{M}_{r,3}$ &
 $\hat{\zeta}_1, \hat{\zeta}_2, \hat{\zeta}_3$ & $v_{\rm mx}$ & $<\!\!\zeta\!\!>$ \\
\noalign{\smallskip}
\tableline
\noalign{\medskip}
p21 & 1.5 & 2.0 & -1.5 & 0.214 & 2010 & 
 0.25, 0.16 & 0.40, 0.07 & 0.42, 0.16 & 9.6, 22.8, 13.0 & 1.25 &  ~15 \\
i9 & 1.5 & 2.0 & -1.5 & 0.667 & 840 &
 0.14, 0.28 & 0.37, 0.10 & 0.42, 0.16 & 3.0, 14.8, 13.0 & 3.08 & ~15  \\
c3 & 1.5 & 1.5 & -2.0 & 0.214 & 1940 & 
 0.49, 0.36& 0.78, 0.29 & 0.85, 0.30 & 8.4,10.8,14.0 & 1.07 & ~15 \\
p4 & 1.5 & 1.5 & -2.0 & 0.321 & 2370 & 
 0.40,0.46 & 0.72,0.33 & 0.87, 0.41 & 5.2, 8.8, 11.0 & 1.74 &  ~10  \\
i16 & 1.5 & 1.0 & -2.5 & 0.321 & 2040 & 
 0.56,1.51 & 1.21,1.84 & 1.73,2.12 & 2.2, 2.6, 4.0 & 2.50 & ~5 \\
i4,$\star$ & 1.5  & 0.5 & -3.0 & 0.40 & 550 & 
 0.93, 2.10 & 1.88, 4.07 & 1.30, 7.07 &  2.7, 1.9, 0.92 & 1.47 & ~2  \\
 i8 & 1.5 & 0.5 & -3.0 & 2.67 & 290 & 
 0.11,5.78 &           &           & sup.-Alf.~inflow & &  ?  \\
i14 & 1.25  & 1.5 & -1.5 & 0.355 & 610 & 
 0.63,0.49 & 1.01,0.46 & 1.09,1.40 & 7.8, 8.8, 4.0 & 1.53 & 10  \\
i15,$\star$  & 1.25 & 1.5 & -1.5 & 0.177 & 5000 & 
 0.25,0.28 & 0.42,0.24 & 0.52,0.22 & 5.4, 7.2, 12.0 & 1.22 & 10  \\
p20 & 1.25 & 1.0 & -2.0 & 0.355 & 1810 & 
 1.07,1.58 & 1.98,2.00 & 2.31,2.69 & 4.2, 4.0, 4.5 & 1.47 & 5  \\
a2 & 1.25 & 0.5 & -2.5 &  0.177 & 2350 & 
 0.56, 0.51 & 0.89, 0.69 & 1.03, 0.58 & 6.6, 5.1, 8.8 & 0.91 & 8  \\
p2 & 1.0 & 1.5 & -1.0 & 0.155 & 2550 &
 0.79,0.53 & 1.21,0.40 & 1.43,0.38 & 9.0, 12.0, 18.5 & 0.87 & 15  \\
i10 & 1.0 & 1.5 & -1.0 & 0.155 & 3330 &
 0.75,0.49 & 1.11,0.49 & 1.32,0.52 & 9.0, 9.2, 12.5 & 0.89 & 10  \\
i11 & 1.0 & 1.0 & -1.5 & 0.155 & 2090 & 
 1.60,1.66 & 2.75, 2.13 & 3.18, 2.97 & 5.8, 5.2, 5.5 & 0.84 & 5  \\ 
c9 & 1.0 &  0.5 & -2.0 & 0.358 & 1730 &
 3.14, 6.14 & 6.64, 11.5 & 8.30, 18.5 & 3.1, 2.3, 2.3 & & 2  \\ 
p13 & 0.8 & 1.5 & -0.6 & 0.922 & 290 &
 0.93, 0.53 & 1.87,1.28 & 6.00, 1.90 & 10.5, 5.8, 15.8 & 2.33 & 15  \\
p14 & 0.8 & 1.5 & -0.6 & 0.307 & 1840 &
 0.90, 0.46 & 1.34, 0.39 & 1.41, 0.43 & 12.3, 18.2, 16.5 & 1.18 & 20  \\
a1  & 0.8 & 1.5 & -0.6 & 0.247 & 2830 &
 1.02, 0.35 & 1.39, 0.26 & 1.55, 0.34 & 17.5, 20.5, 22.8  & 1.08 & 20    \\
i17 & 0.8 & 1.0 & -1.1 & 0.247 & 2790 &
 1.88, 1.49 & 3.44, 1.89 & 4.58, 1.59 & 7.6, 7.3, 14.5 & 1.05 &  $\simeq$ 15   \\
p15 & 0.8 & 0.6 & -1.5 & 0.310 & 990 & 
 3.34, 4.60 & 6.95, 7.81 & 9.13, 11.2 & 4.4, 3.6, 4.1 & 1.05 & 4  \\
c10 & 0.8 & 0.5 & -1.6 & 0.247 & 1400  & 
 3.85, 6.08 & 8.05, 11.8  & 11.3, 18.8 & 3.8, 2.7, 3.0 & 0.93 & 3   \\
i19 & 0.65 & 1.5 & -0.3 & 0.279 & 1420 &  
 0.53, 0.17 & 0.73, 0.16 & 0.82, 0.27 & 18.4, 18.1, 15.5 & 1.01 & 20  \\
i18 & 0.65 & 0.5 & -1.3 & 0.279 & 1600 & 
 4.93, 5.58 & 10.8, 10.1 & 14.2, 16.6 & 5.3, 4.3, 4.3 & 1.01 & 4  \\
a5  & 0.65 & 1.0 & -0.8 & 0.279 & 1700 &
 2.14, 1.51 & 3.79, 1.75 & 4.59, 2.22 & 8.5, 8.6, 10.1 & 1.05  & 10  \\
p6 & 0.5 & 1.5 & 0.0 & 0.112 & 1800 &
  1.18,0.33 & 1.65,0.23 & 1.82, 0.25 & 21.5, 28.7, 36.4 & & 30  \\
i1 & 0.5 & 1.5 & 0.0 & 0.112 & 3600 & 
 1.18, 0.04 & 1.35, -.048 & 0.92, 0.18 & 161, (112), 25.6 & 0.7 & 30 \\
p7 & 0.5 &  1.5 & 0.0 & 0.225 & 1000 & 
 1.19, 0.30 & 1.63, 0.26 & 2.07, 0.51 & 23.8, 25.1, 20.3 & 0.86 & 20  \\
c5 & 0.5 & 1.5 & 0.0 & 0.225 & 430 &
 1.23, 0.39 & 2.44, 0.35 & 4.89,-0.22 & 18.9, 27.9, (111) &  0.78 & 30  \\
c8 & 0.5 & 0.8 & -0.7 & 0.112 & 3800 &
 3.66, 2.08 & 7.38, 2.40 & 10.1, 1.77 & 10.6, 12.3, 28.5 & 0.7 & 30 \\
i2 & 0.5 & 0.8 & -0.7 & 0.112 & 3800 & 
 3.69, 2.07 & 7.37, 2.38 & 10.3, 1.75 & 10.7, 12.4, 29.4  & 0.5 & 30   \\
p8,$\star$ & 0.5 & 0.5 & -1.0 & 0.112 & 3500 &
 3.18,2.35 & 7.19, 3.86 & 10.5, 5.10 & 8.1, 7.5, 10.3 &  0.65 & 10  \\
i3,$\star$ & 0.5 & 0.5 & -1.0 & 0.112 & 3900 & 
 3.13, 2.15 & 7.74, 3.10 & 11.3, 4.07 &  8.7, 10.0, 13.9 & 0.65 & 15  \\
p18 & 0.5 & 0.5 & -1.0 & 0.112 & 2210 &
 5.60, 5.48 & 12.5, 9.57 & 17.6, 15.3 & 6.1, 5.2, 5.8 & 0.51 & 5  \\
p19 & 0.5 & 0.5 & -1.0 & 0.225 & 1640 & 
 5.88, 5.14 & 12.9, 9.09 & 17.2, 15.1 & 6.9, 5.7, 5.7 &  0.95 & 6  \\
p16 & 0.5 & 0.3 & -1.2 & 0.225 & 2790 &
 9.60, 7.87 & 24.2, 15.2 & 36.8, 27.0 & 7.3, 6.4, 6.8 & 0.58 & 7  \\
c11 & 0.35  & 1.5 & 0.3  & 0.366 & 1250 & 
1.14, 0.25 & 2.23, 0.55 & 2.39, 1.20 & 27.4, 16.3, 10.0 &    & $\simeq$ 10 \\
a4  & 0.35  & 1.0 & -0.2 & 0.255 & 2000 & 
3.02, 0.70 & 5.05, 1.04 & 7.75, -.39 & 25.9, 19.4, (99) &    & $\simeq$ 25  \\
a6  & 0.35 & 0.5 & -0.7 & 0.255 & 1900 &  
 8.51, 2.44 & 18.8, 1.38 & 23.9, 3.47 & 20.9, 54.3, 34.3 & 0.8 & $> 30$  \\
a8  & 0.2  & 1.0 & 0.1  & 0.296 & 1950 & 
 2.62, 0.67 & 4.93, 0.61 & 7.64, -2.5 & 23.5, 32.3, (-19) &     & $\simeq$ 30  \\
a7 & 0.2 &  0.5 & -0.4 & 0.296 & 2100 & 
  8.89, 2.05 & 15.7, 3.67 & 29.9, 5.72 & 26.0, 17.1, 26.1 &  0.68  & $\simeq$ 25    \\
\noalign{\medskip}
\tableline 
\end{tabular}
\end{center}
\end{table*}
\clearpage
\begin{figure}

\includegraphics[height=5cm, width=7.3cm]{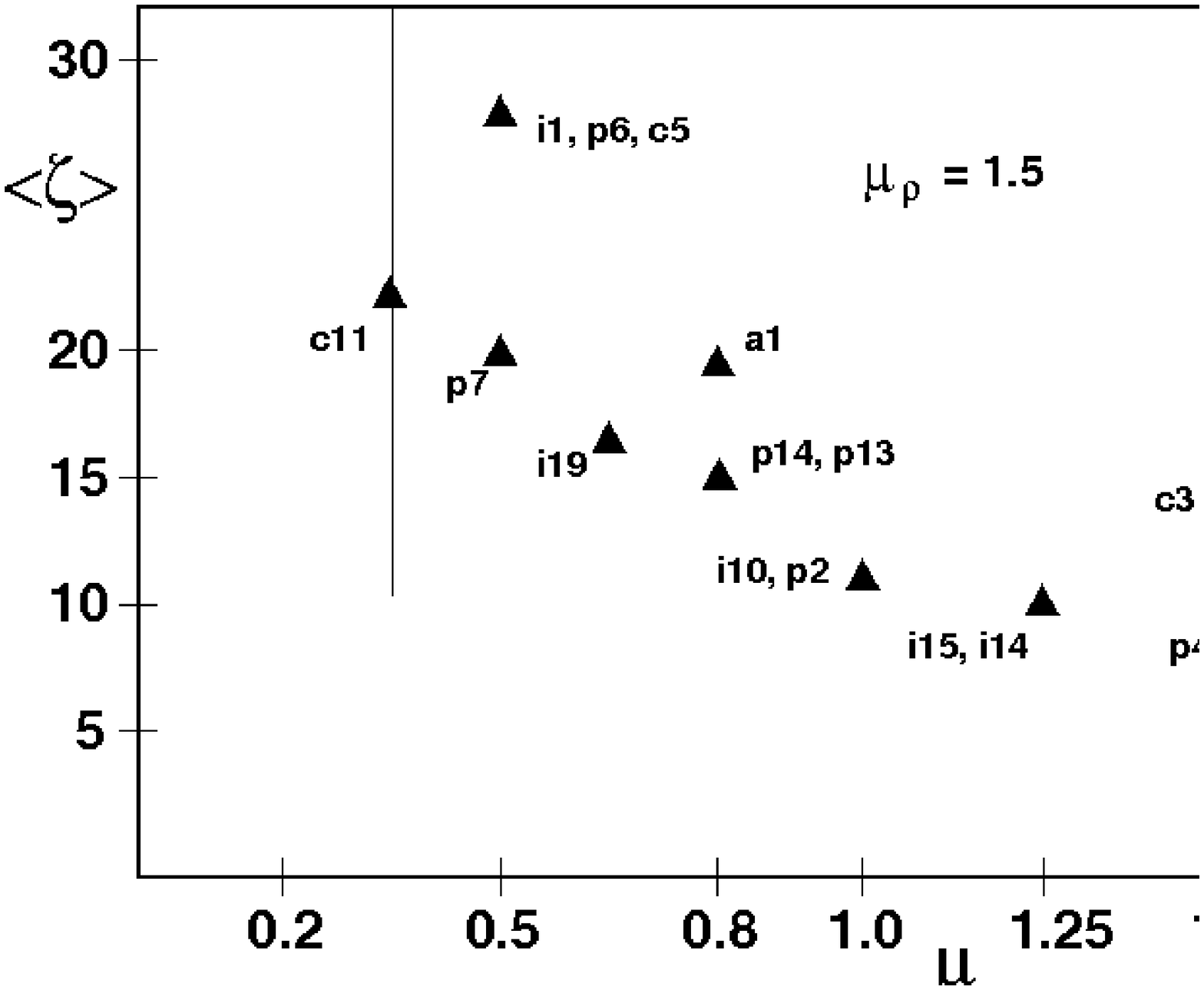}

\bigskip
\includegraphics[height=5cm, width=7.3cm]{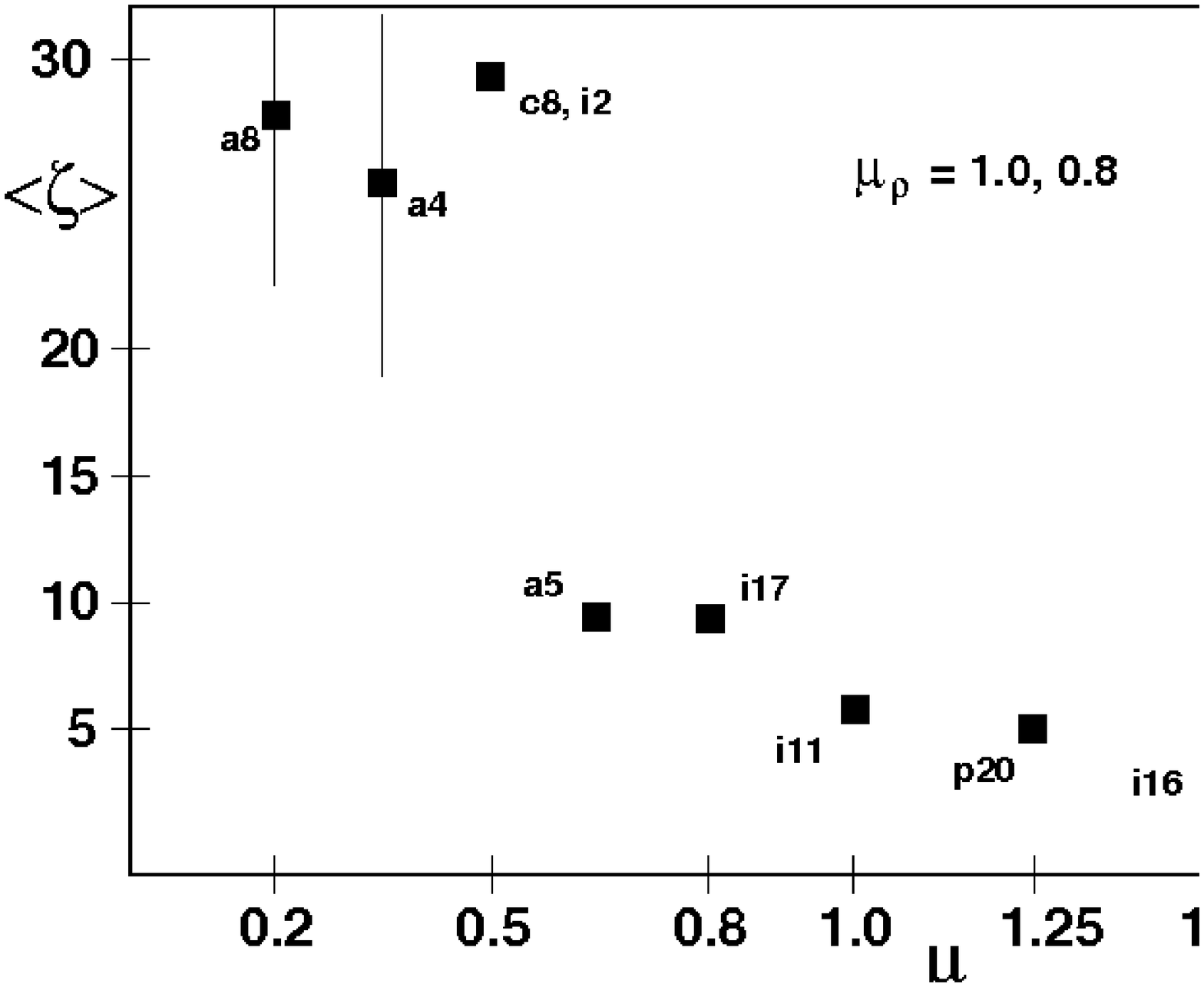}

\bigskip
\includegraphics[height=5cm, width=7.3cm]{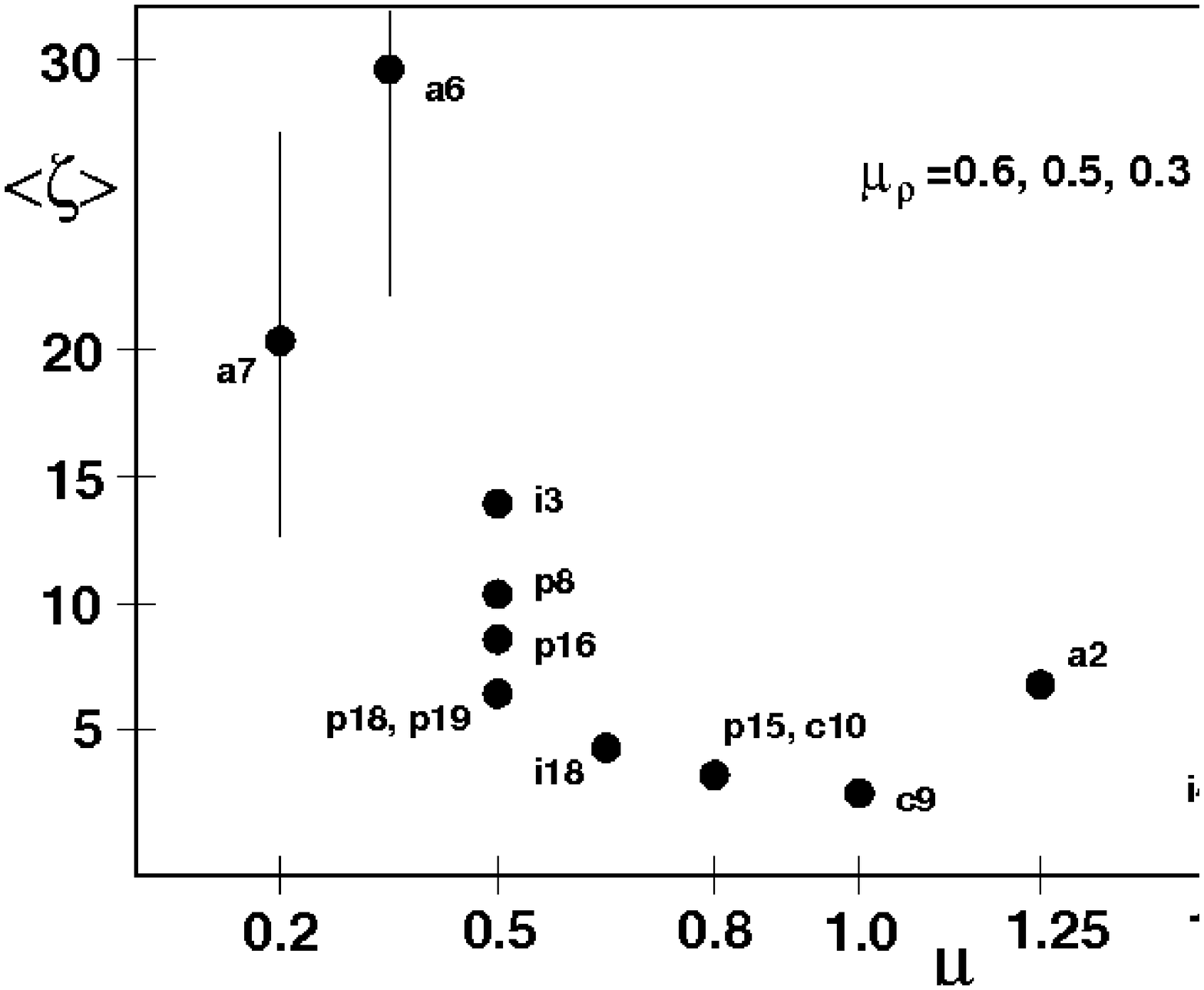}

\caption{Degree of collimation measure (denoted by $<\!\!\zeta\!\!>$) as 
defined in Sect.~\ref{sec_measure} and power law index of the disk poloidal
magnetic field profile $\mu$ (see Tab.~\ref{tab_all}).
The three sub-figures refer to simulations applying inflow density profiles,
$\mu_{\rho} = -1.5$ (top), 
$\mu_{\rho} = -1.0$ (middle), 
$\mu_{\rho} = -0.5$ (bottom).
Data points with bars indicate a variable collimation degree resulting in most
cases from axial instabilities of highly collimated outflows.
\label{fig_dg_mu}
}
\end{figure}
\clearpage
\subsection{How to measure the degree of collimation?}  \label{sec_measure}

The main goal of this study is to investigate the relation between
the accretion disk magnetic flux profile and the collimation of the outflow.
In order to {\em quantify} such an relation we need to define a measure
for the degree of outflow collimation.
As a matter of fact, some ambiguity exists about such a definition and a few
words of clarification are needed.
What is usually observed as jet flow, is an elongated feature of radiation
emitted in combination with a signature of high velocity.
This is then associated with
(i) an elongated mass distribution ("collimation") emitting the radiation and
(ii) a strong axial velocity component ("acceleration").
Therefore, from the theoretical point of view, it is natural to apply the
{\em directed mass flux as measure of jet collimation} (see below).

In this respect, we also note the different meaning of the 
"degree of collimation" and the "ability to collimate". 
The first term refers to the status of the established - collimated - outflow
as directed mass flux which is probably equivalent to the observed jet.
The second term refers to the question of how well an initially weakly
collimated disk wind can be turned into a collimated stream by means of
internal MHD forces.
It is straight forward to expect that for the same disk magnetic field
profile, 
a more concentrated disk wind density profile (i.e. large $\mu_{\rho}$) 
will result in a more concentrated density profile of the asymptotic jet
flow as well (presumably leading to a smaller jet radius observed).
However, both a wide and a narrow jet, i.e. a jet with a flat or steep 
asymptotic density distribution, 
may have the same asymptotic opening angle (e.g. a cylindrical shape).

For ideal MHD jets in stationary state, collimation can be measured
by the opening angle of the magnetic field lines.
In the case of resistive jets, or if the outflow has not yet reached a
steady state, this is not appropriate, as mass flux direction and 
magnetic field direction are not aligned.

In the present paper, we consider the {\em directed mass flux} as 
appropriate quantitative measure of outflow collimation.
For simplicity, we measure the {\em degree of collimation} by the ratio 
$\zeta$ of mass flux in axial versus lateral direction,
\begin{equation}
\zeta \equiv \frac{\dot{M}_{\rm z}}{\dot{M}_{\rm r}}
= {
{  2\pi \int_{0}^{r_{\rm max}} r \rho v_z dr }
 \over 
{ 2\pi r_{\rm max} \int_{0}^{z_{\rm max}} \rho v_r dz }.
}
\label{eqn_zeta}
\end{equation}
This parameter depends on the size of the integrated volume.
We determine the mass flow rates applying three differently sized volumes,
$\dot{M}_{{\rm z},i}, \dot{M}_{{\rm r},i}$,
considering different sub-grids of the whole computational domain of 
$150\times300\,\ri$ corresponding to cylinders of radius and height
$(r_{\rm max} \times z_{\rm max}) =
(20.0\!\times\!60.0), (40.0\!\times\!80.0), (60.0\!\times\!150.0)$
(see Tab.~\ref{tab_all}, Tab.~\ref{tab_michel}, Tab.~\ref{tab_diff}).
We then consider the mass flux {\em ratio} $\zeta$ normalized to
the size of the areas passed through by the mass flow in r- and z-direction,
$\hat{\zeta}_1, \hat{\zeta}_2, \hat{\zeta}_3$.
The ratio of grid size $(r_{\rm max}/ z_{\rm max})$ as displayed
in our figures converts into a ratio of cylinder {\em surface areas} of 
$ A_{\rm z}/A_{\rm r} \sim 0.5 (r_{\rm max}/ z_{\rm max})$.
Thus, a mass flux ratio $\zeta_2 = 2$ as defined by Eq.~(\ref{eqn_zeta})
corresponds to a mass flux ratio normalized to the surface area threaded of 
$\hat{\zeta}_2 = 8$ which could be considered as ``very good'' collimation.

The difference in degree of collimation calculated for differently sized
volumes, $\hat{\zeta}_1, \hat{\zeta}_2, \hat{\zeta}_3$,
may also indicate different evolutionary states in different regions of
the outflow.
At a certain time,
the outflow may already be well collimated close to its origin and also 
have reached a stationary state in this region.
However, 
on larger spatial scales the same outflow may still interact violently with 
the ambient initial corona (see e.g. simulation runs c3, p4).
Some simulations could be performed until an almost stationary flow on the 
whole numerical grid has developed (e.g. c8, p8).
In other cases a quasi-stationary state of the outflow within the inner
region (say $50\%$ of the grid) has been reached.
In most cases the initial corona has completely swept out of the grid.
From the mass flow rates on the sub-grid we derive an
{\em average degree of collimation} $<\!\!\zeta\!\!>$
considered to be typical for that simulation, 
but depending on the evolutionary state of the 
flow.

We note another interesting consequence of the large grid applied
compared to previous studies on smaller grid size 
\citep{ouye97a,pro06}.
In some of our simulations which reaches later evolutionary stages
of the outflow the Alfv\'{e}n surface leaves the grid in axial direction
(see e.g. run i3).
Thus, these outflows are sub-Alfv\'{e}nic in their outer layers and the jet
structure will consist of three nested layers (cylinders) of different
magnetosonic properties -- an outer sub-Alfv\'{e}nic layer, 
a super-Alfv\'{e}nic / sub-fast magnetosonic
layer at intermediate radii, and a super-fast magnetosonic jet in the 
inner part 
(here we neglect the very inner axial spine where there is no mass 
inflow from the disk but potentially from a stellar wind).
The extension of these layers depends on the parameters 
$\mu$ and $\mu_{\rho}$ and will be discussed in more detail below. 
For example, for simulation run i10, we estimate the asymptotic radius of
the Alfv\'{e}n surface to about $120 \ri$, and
the asymptotic radius of the fast magnetosonic surface to about $50 \ri$.
For comparison, in  the original approach \citep{ouye97a} passes the radial
grid boundary at a height of about $z=60$.
A multi-layer structure of jets in respect to the magnetosonic surfaces
has been proposed in stationary studies by \citet{fend96},
however, then suggesting layering inverted to what we find now
(ie. a super fast-magnetosonic flow in the outer part respectively).
We expect that such a stratification in lateral direction should be
essential for the jet stability. 
This particular issue has not yet been treated in the literature.
\clearpage
\begin{figure}
\centering

\includegraphics[width=7.5cm]{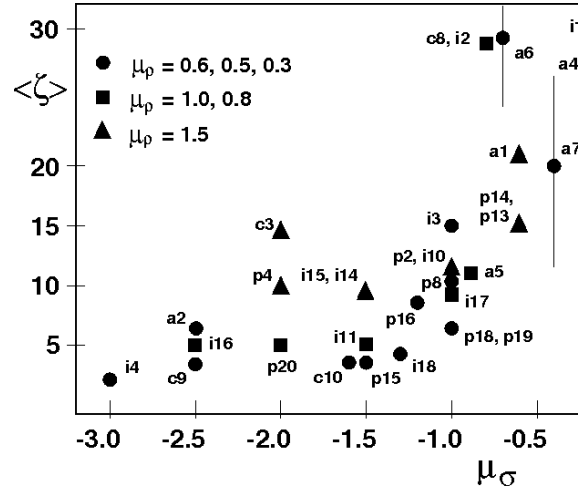}

\caption{Degree of collimation (denoted $<\!\!\zeta\!\!>$) as defined in  
Sect.~\ref{sec_measure} and power law index of the disk wind magnetization profile 
$\mu_{\sigma}$ (see Tab.~\ref{tab_all}).
This figure combines all three sub-figures of Fig.~\ref{fig_dg_mu} applying
$\mu_{\sigma} = \mu_{\rho} -2 \mu -1/2$.
Data points with bars indicate a time variable collimation degree resulting in 
most cases from axial instabilities of highly collimated outflows.
\label{fig_dg_sig}
}
\end{figure}
\clearpage
\subsection{Degree of collimation - MHD simulations}

Simulation run i10 with $\mu = 1.0$, $\mu_{\rho} = 1.5$ serves as our ``reference run''
resembling just the same parameter set as applied in the exemplary solution
by \citet{ouye97a},
but now performed on a substantially larger physical grid.
After $t=3000$ (i.e. 3000 inner disk orbits)                        
a stationary outflow is obtained over a large fraction of the grid.
The run of the Alfv\'{e}n surface at large radii indicates that the 
outermost layers of the outflow have not yet reached a stationary 
state (see Fig.\ref{fig_disk_1}).
We obtain degrees of collimation of $\hat{\zeta}_{1,2,3} = 9.0,9.2,12.5$
resulting in an "average" degree of collimation of $<\!\!\zeta\!\!> = 10$
(see Tab.~\ref{tab_all}).
As the outflow has not yet settled in a stationary state over the whole grid, 
these values will vary {\em slightly} for later times. 
Note that even 3000 orbits of the inner disk correspond to about only two
orbits at the outer disk radius at 150 $\ri$.
The sequence $\hat{\zeta}_{1,2,3}$ also indicates an {\em increasing degree of collimation}
along the outflow.

\clearpage
\begin{figure}
\centering

\includegraphics[height=0.5cm]{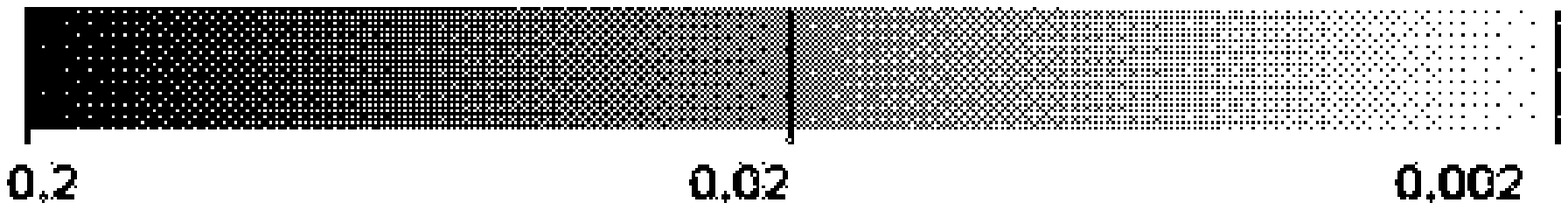}

\includegraphics[height=4.0cm]{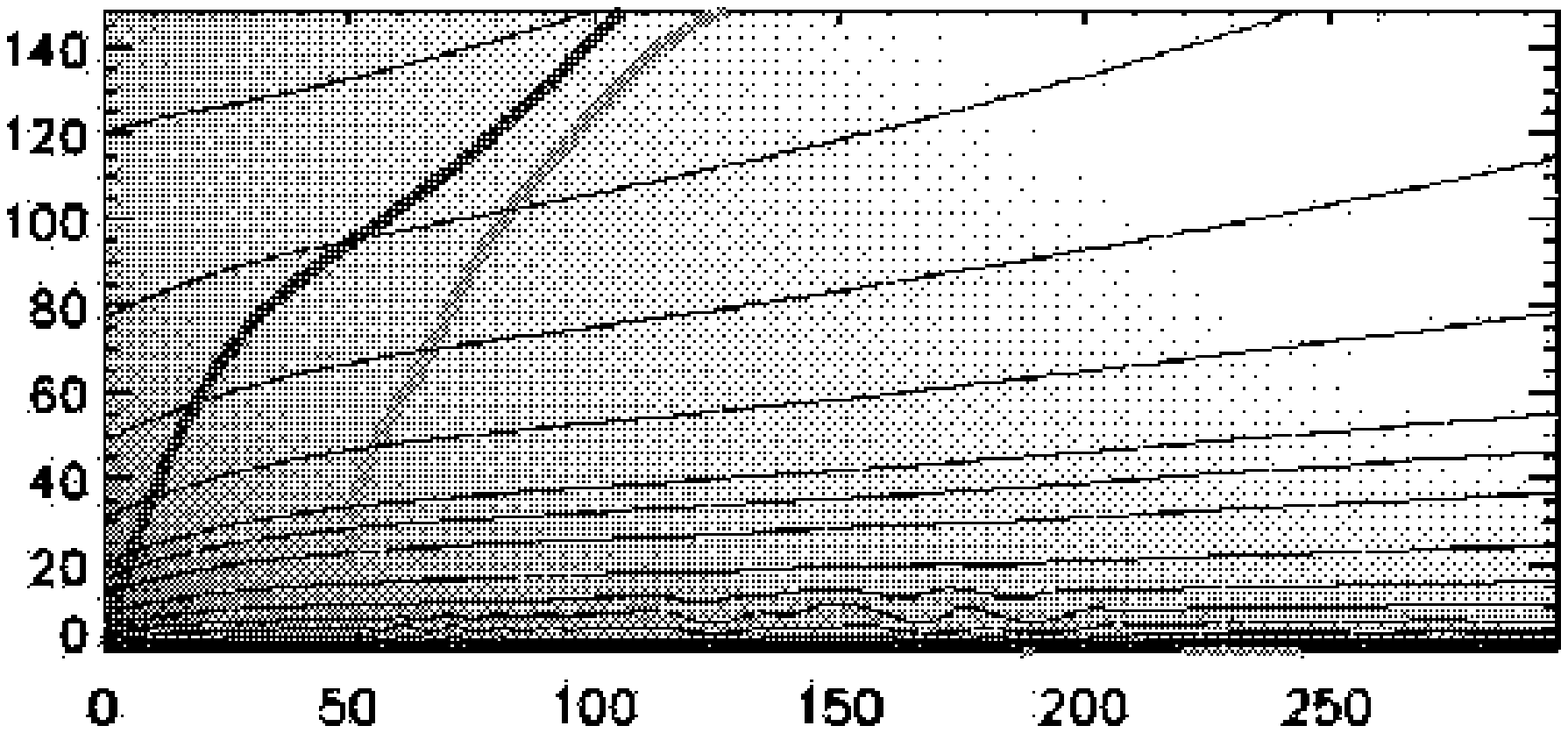}

\includegraphics[height=4.0cm]{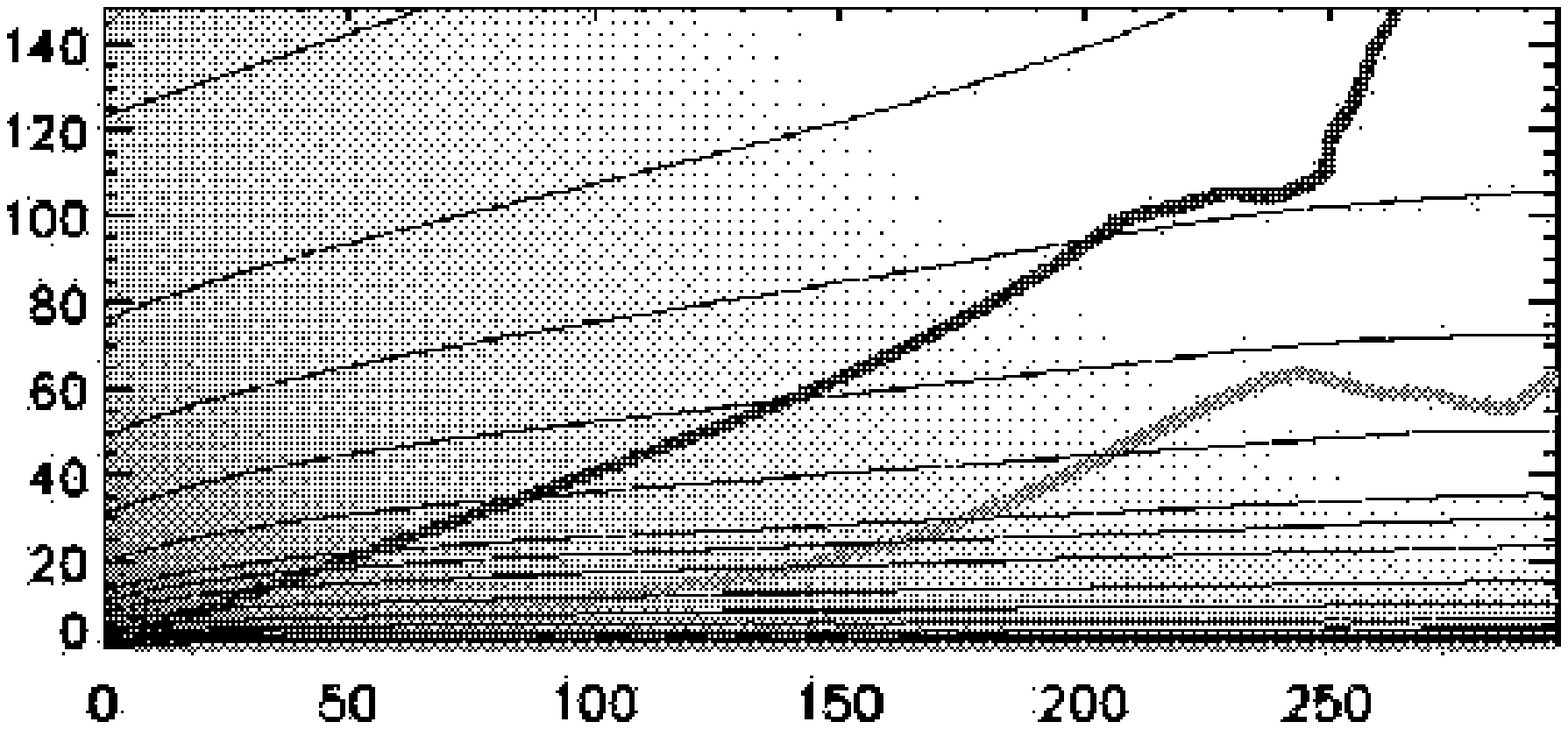}

\caption{Simulation runs with different 
disk boundary conditions (see Tab.~1).
Impact of a total disk magnetic flux variation (factor 4).
Density and poloidal magnetic field lines for
{\bf p18} ($t=2210$),
{\bf p19} ($t=1640$)
from {\em top {\rm to} bottom}.
See Fig.~\ref{fig_long} for further notation.
\label{fig_disk_3}
}
\end{figure}
\clearpage

\subsubsection{Disk magnetic field profile and collimation}

We now compare simulation runs with different disk magnetic flux profile $\mu, \mu_{\Psi}$,
but similar inflow density profile $\mu_{\rho}$.
We have run about 40 different simulations which in total required about one year of CPU 
time on various work stations.

Table \ref{tab_all} shows the main input parameters for the each simulation run,
in particular the power law index of the magnetic field and density distribution.
Table \ref{tab_all} further shows the main parameters of the resulting outflow, 
namely the mass flow rates in lateral and axial direction defining the collimation 
degree.
What is also indicated is the physical time step $\tau$ at which the mass flow rates 
are calculated.
This is in most cases, but not always, the final time step of the simulation.
When comparing different simulation runs, one has to keep in mind that they 
evolve with different dynamical speed.
Thus, if possible, only outflows at late (or equivalent) evolutionary stages will be 
compared, which may have been reached after different physical time
(as e.g. stronger magnetized outflows evolve faster, see below).

In general, we find an 
increasing degree of collimation with decreasing slope of the disk magnetic field 
profile\footnote{The mass flow rates in simulations
with flat magnetic field profile $\mu <0.4$ are difficult to determine as these outflows
do not reach a stationary state and show a time-dependent variation in the lateral
mass flux close to the axis}
(see Fig.~\ref{fig_dg_mu}).
This holds for all of the disk wind density profiles investigated so far.
A steep density profile generally leads to a higher degree of jet collimation
which is, however, not surprising as the mass flux is more concentrated along
the axis just by definition of the boundary condition.

For the simulations with relatively flat density profile 
($\mu_{\rho} =0.3, ..., 1.0$), 
a strong increase in collimation degree can be observed just below $\mu=0.6$.
However, simulations with a flat disk magnetic field profile (below $\mu =0.5$) 
do not reach a stationary state within the time scale of our simulations. 
In general, the degree of collimation for these outflows is high. 
The axial mass flow rate stays more or less constant whereas the radial mass flux
varies in time, resulting in a "wavy" pattern evolving along the outflow axis.
This has been discussed by \citet{ouye97b} for the case of an initially
vertical coronal magnetic field and in the context of knot generation in
protostellar jets.

Now we compare our results obtained for the different density profiles
{\em quantitatively}.
We find that a physically meaningful classification can be achieved by shifting and
spreading the three figures (Fig.~\ref{fig_dg_mu}) not in vertical but in
horizontal direction.
Essentially, this corresponds to a comparison {\em collimation degree versus magnetization}
by applying $\mu_{\sigma} = \mu_{\rho} - 2\mu - 1/2$.
Indeed, the resulting diagram (Fig.~\ref{fig_dg_sig}) shows a convincing correlation 
between the magnetization profile power law index $\mu_{\sigma}$ and the degree of
collimation obtained $<\!\!\zeta\!\!>$.
As we naturally expect that the inflow parameters will change during the
life time of an accretion disk, our results might explain why not all accretion 
disks do launch a jet and why disks do not launch jets for all their life time.

The above mentioned steep rise in collimation degree for $\mu <0.6$ is now
indicated for a magnetization profile $\mu_{\sigma} > -1.0$.
Obviously, the slope of this rise depends in particular on the reliability of simulation
runs a6, c8 and i2. 
We note that simulation i2 and c8 follow a relatively flat density profile, 
and did evolve up to several 1000 disk orbital periods into an almost stationary state on a
global scale.
Simulation a6 remains sub-Alfv\'{e}nic over most of the grid size and has developed an
instable flow pattern along the jet axis.

The width of the $(\mu_{\sigma}$ - $<\!\!\zeta\!\!>)$-correlation shown in Fig.~\ref{fig_dg_sig}
is due to detailed differences in the parameter setup.
The data points shown for a certain simulation which is deviating from the general trend
of the correlation indeed have a particular origin. 
For example, simulation run c3 is strongly magnetized with the Alfv\'{e}n surface very close
to the $r$-axis. The same holds for simulation run i4.
Therefore, both simulations do not consider a typical Blandford-Payne magneto-centrifugally
launched jet, but have the Alfv\'{e}n points close to the foot points of the field lines
(hence a short lever arm).
However, in summary we conclude that the numerical data give a clear correlation between
the power law index of the disk wind magnetization profile and the degree of collimation.

\clearpage
%
\begin{table*}
\scriptsize
\begin{center}
\caption{Summary of simulation runs obeying an artificial
decay of the toroidal field component.
The parameter $f_{B3}$ gives the factor by which the toroidal 
magnetic field is artificially decreased each time step
(see Tab.~\ref{tab_all} for other notations).
\label{tab_diff}
}
\begin{tabular}{lllllllllllll}
\tableline \tableline 
\noalign{\smallskip} run & 
 $\mu$ & $\mu_{\rho}$ & 
 $\mu_{\sigma}$ & $B_{\rm p,i}$ & $\tau $ & 
 $f_{B3}$ &
 $\dot{M}_{z,1},\dot{M}_{r,1}$ & $\dot{M}_{z,2},\dot{M}_{r,2}$ & $\dot{M}_{z,3},\dot{M}_{r,3}$ &
 $\hat{\zeta}_1, \hat{\zeta}_2, \hat{\zeta}_3$ & $v_{\rm mx}$ & $<\!\!\zeta\!\!>$ \\

\tableline 
\noalign{\smallskip}
\noalign{\medskip}
p10 & 1.5 & 0.8 & -2.7 & 0.40 & 1800 & 
 0.1 &
 -.04, 1.58 & 0.07, 2.45 & 0.11, 3.19 & (0.2), 0.1, 0.2 & 0.18 & 0 \\
\noalign{\medskip}
p11 & 1.5 & 0.8 & -2.7 & 1.20 & 3100 & 
0.1  &
 -.051,1.59 & 0.057, 2.47 & 0.064, 3.40 & (0.2), 0.1, 0.1 & 0.17 & 0  \\
\noalign{\medskip}
p17 & 1.0 & 1.5 & -1.0 & 0.232 & 1890 & 
 0.9 & 
 0.30,1.08 & 0.82,0.97 & 0.96,1.39 & 1.7, 3.4, 3.4 & 0.17 & 3 \\
\noalign{\medskip}
i12 & 1.0 & 1.0 & -1.5 & 0.387 & 1520 & 
 0.9 & 
 0.60,2.65 & 1.73,3.20 & 2.59,5.66 & 1.4, 2.2, 2.3 & 0.29 & 2 \\
\noalign{\medskip}
i13 & 1.0 & 1.0 & -1.5 & 0.246 & 1020 & 
 0.9 &
 .092,3.19 & 1.33,3.73 & 3.48,7.38 & 0.17, 1.4, 2.4 & 0.32 & 2 \\
\noalign{\medskip}
p12 & 0.8 & 1.5 & -0.6 & 0.922 & 870 & 
0.9 &  
 0.87, 0.69 & 1.55, 1.19 & 1.26, 1.75 & 7.6, 5.2, 3.6 & 0.27 & 4 \\
\tableline \tableline 
\noalign{\medskip}
\end{tabular}
\end{center}

\end{table*}
\clearpage
\subsubsection{Field strength and collimation}

While the slope of the disk magnetic flux / magnetization profile clearly governs
the asymptotic collimation of the outflow, 
the amount of total disk magnetic flux is seemingly less important.
We have investigated different absolute magnetic fluxes for the same flux profile 
and did not detect any relation with the degree of collimation.
Simulations with high magnetic field strength are CPU time consuming due to
the Alfv\'{e}n speed time stepping.
On the other hand, the jet evolves faster and also propagates across the numerical 
grid within a shorter period of time (physical units).
We thus reach similar evolutionary stages at earlier physical time (but after
more numerical time steps).

For reference we compare simulation runs 
i14 and i15 with field strength $B_{p,i}(i15) = 0.5 B_{p,i}(i14)$
and initial flow density $\rho_{inj}(i15) = 0.5 \rho_{inj}(i14)$;
runs p13, p14 and a1 with $B_{p,i}(p13) = 3.0 B_{p,i}(p14)= 3.73 B_{p,i}(a1)$
and $\rho_{inj}(i15) = 0.5 \rho_{inj}(i14)$;
and 
runs p18, and p19 with $B_{p,i}(p19) = 2.0 B_{p,i}(p18)$ 
(see Fig.~\ref{fig_disk_3}).
Table \ref{tab_all} shows that the collimation degree is indeed similar for each 
example, as it is the evolving jet structure 
(see examples p18 and p19, Fig.~\ref{fig_disk_3}).
On the other hand, the total magnetic flux governs the asymptotic velocity gained
by the jet (see below).

\subsection{Jet collimation for highly diffusive toroidal field}

As mentioned above (Sect.\ref{sec_tor_pol}), studies in the literature have
proposed a "poloidal collimation" of the jet, 
i.e. jet collimation due to the pressure of an ambient poloidal magnetic 
field \citep{spru97,matt03}.
In order to investigate the feasibility of such a process, 
we have performed a set of simulations applying an artificially strong decay
of the toroidal field component, 
invented to mimic the toroidal field decay due to the kink-instability \citep{spru97}.

Our simulations (see Tab.~\ref{tab_diff}) follow a parameter setup equivalent
to those in Tab.~\ref{tab_all} with the single exception that the toroidal
field is artificially decreased by either a factor
$f_{B3} = 0.9$ or $0.1$ at each numerical 
time step\footnote{This is done in the ZEUS subroutine "ct" treating the
constrained transport and updating the magnetic field components}.
Thus, this part of our paper considers a "toy model" mimicking an artificially
enhanced magnetic diffusivity, 
only affecting the toroidal field component\footnote{We note that the
artificial toroidal field decay sometimes causes problems for the usual
outflow boundary condition in axial direction.  However, so far the internal
structure of the outflow has not been not affected}.
A fully self consistent approach would involve a tensorial description
of magnetic diffusivity with strong diffusivity for those components affecting
the toroidal field component. This is beyond the scope of the present paper.

During the long-term evolution of these outflows the resulting toroidal magnetic
field is decreased by 5 orders of magnitude compared to the simulations presented 
before (independent of $f_{B3}$).
The simulation runs we have to compare are 
p17 with p2 and i10,
runs i12 and i13 with i11,
run p12 with p14 and p13,
and runs p10 and p11 with i16.
All outflows with decaying toroidal magnetic field are substantially less
collimated (Tab.~\ref{tab_diff}).
This is not surprising as the collimating tension force by the toroidal field 
has just been switched off.
The exception is simulation run p12 which, due to its steep disk wind density
profile, 
shows relatively good collimation comparable to the previous simulations runs 
without toroidal field decay.
\clearpage
\begin{figure}
\centering

\includegraphics[height=4cm, width=7cm]{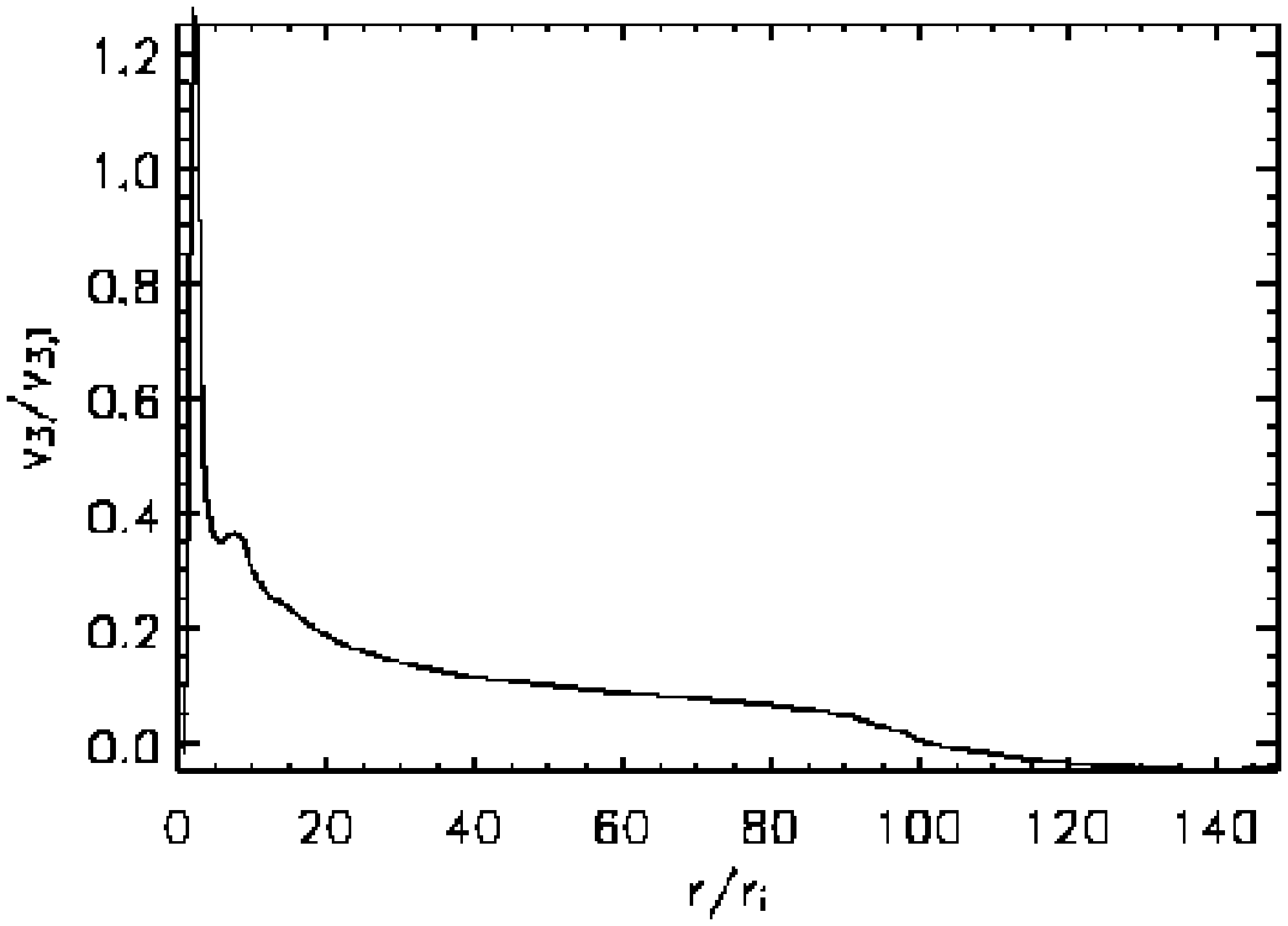}

\includegraphics[height=4cm, width=7cm]{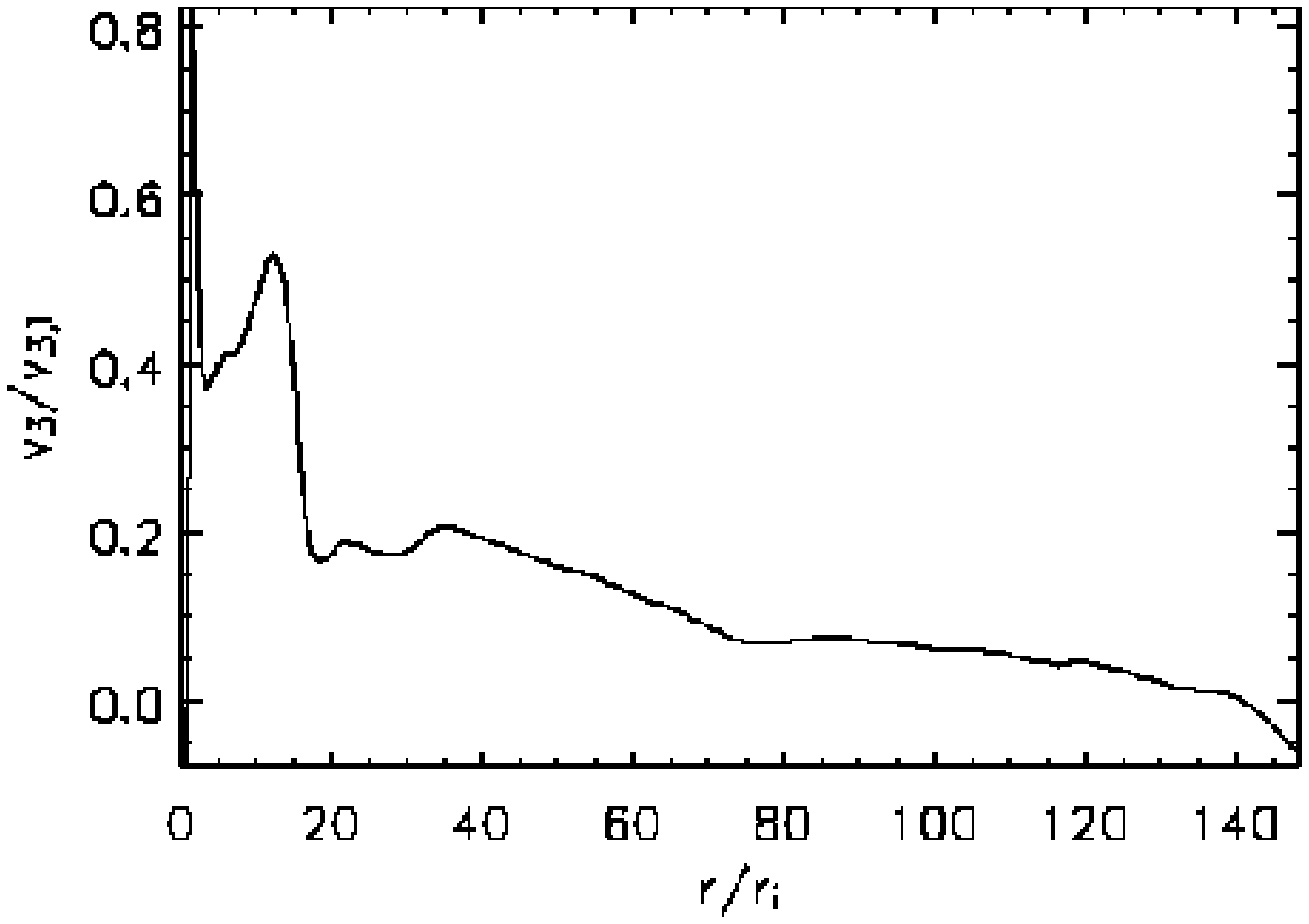}

\includegraphics[height=4cm, width=7cm]{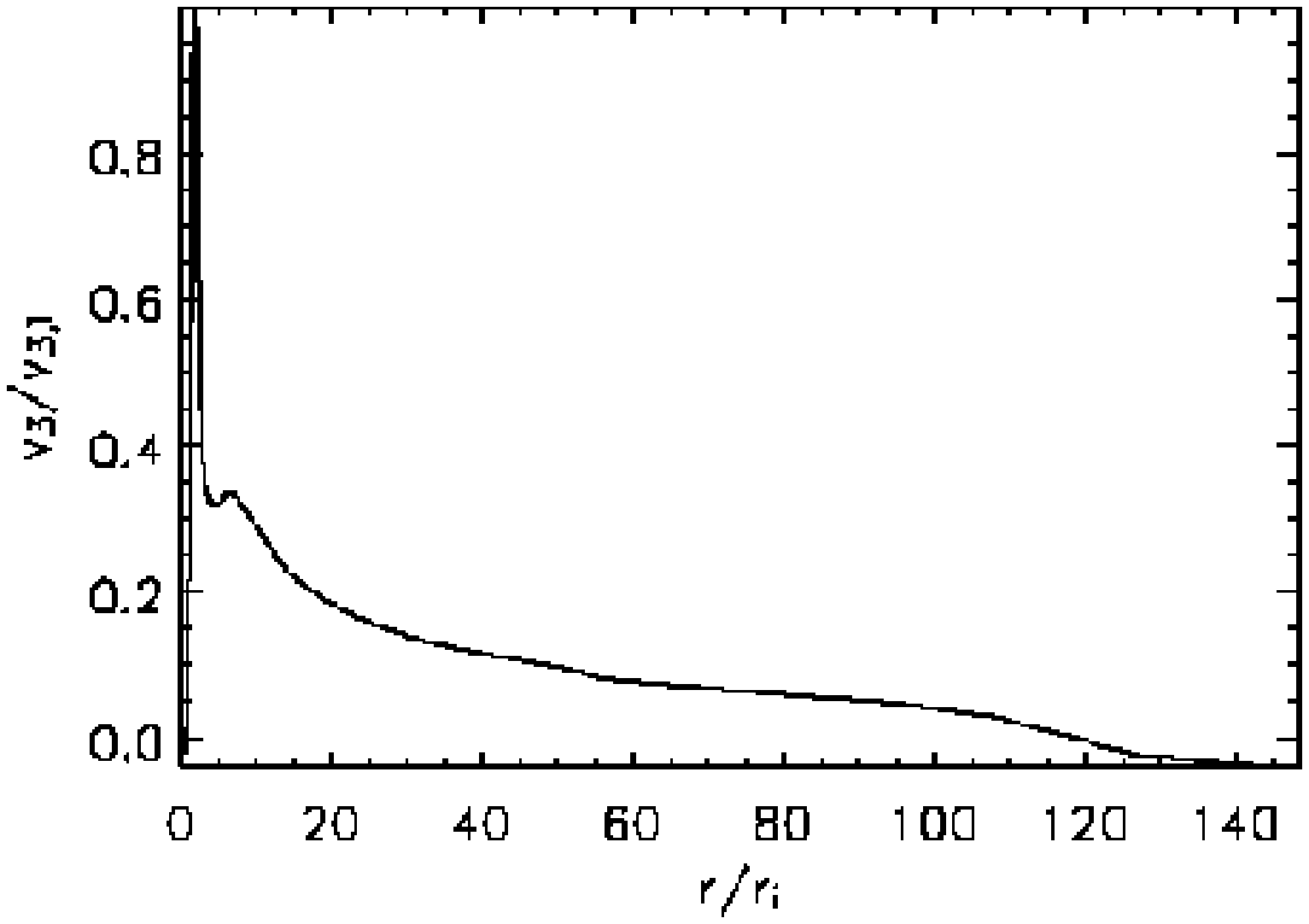}

\includegraphics[height=4cm, width=7cm]{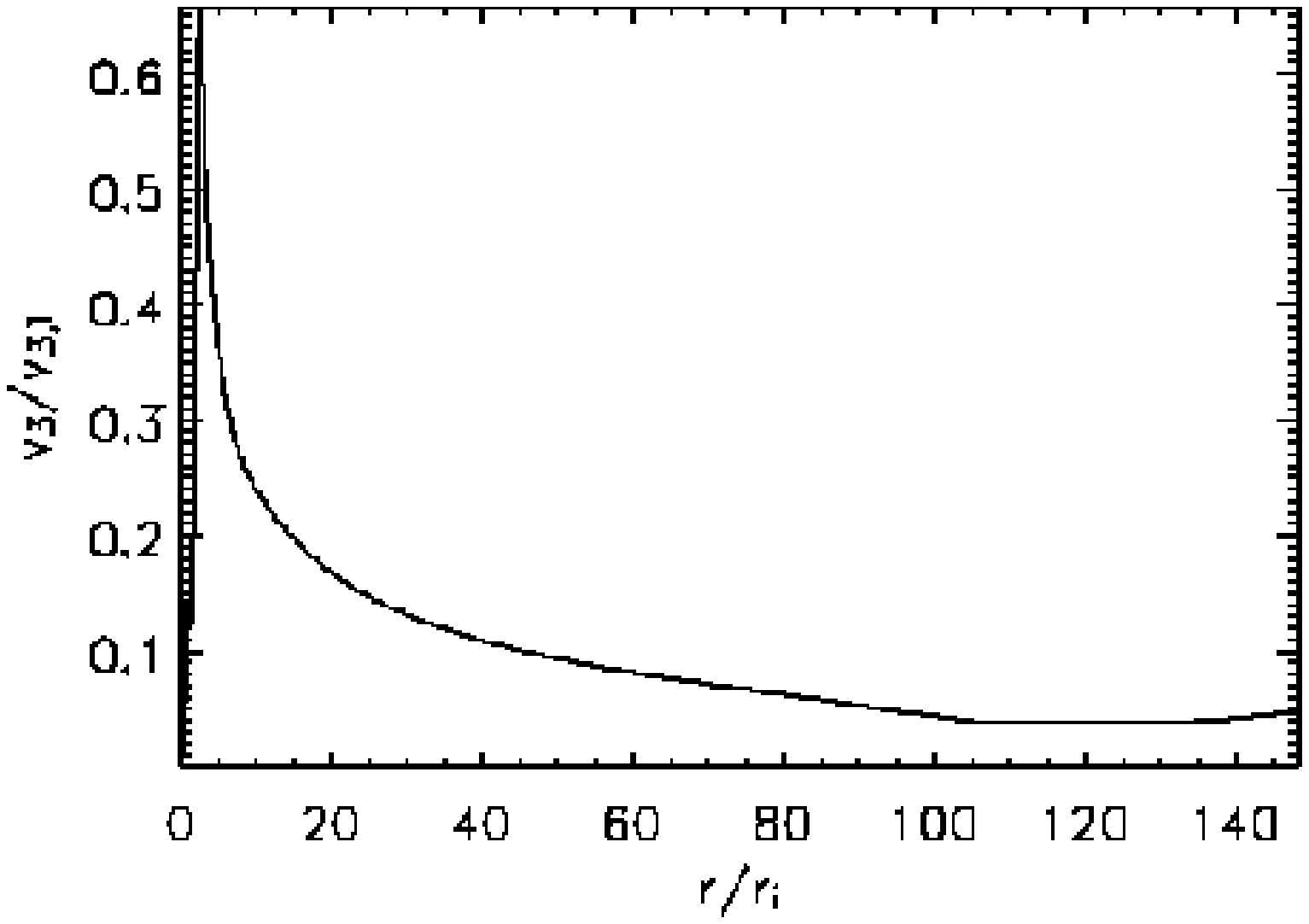}

\caption{Toroidal velocity profile across the jet at $z= 160$ for 
selected parameter runs at the final stages of simulation.
{\bf i3} ($t=4000$),
{\bf i1} ($t=3000$),
{\bf p8} ($t=3200$),
{\bf i11} ($t=2000$)
from {\em top {\rm to} bottom}.
\label{fig_v_phi}
}
\end{figure}
\clearpage
\subsection{Jet velocity \& Michel scaling}

In general we obtain a narrow range for the maximum velocity of the 
super-Alfv\'{e}nic flow at large distance from the origin.
The maximal jet velocity is typically in the range of the Keplerian 
speed at the inner disk radius, 
in agreement with the literature \citep{ouye97a,cass02,kudo02}.
However, in our simulations we noticed a slight trend in the jet velocity in
relation to the magnetization $\sigma$.
In order to further investigate the Michel scaling for collimated jets,
we have performed simulation runs with identical initial field and mass 
flow {\em profile}, but different absolute number values for mass flow
{\em rate} and magnetic field {\em strength} - either varying the initial
magnetic field strength or the inflow density by a certain factor.

These simulations are summarized in Tab.~\ref{tab_michel}.
In particular, we may compare simulations p21 to i9, c3 to p4,
i14 to i15, p13 to p14 and a1, i1 to p7 and c5, and p18 to p19 
(see Fig.\ref{fig_disk_3}).
Table \ref{tab_michel} shows the ratio of the maximum poloidal 
velocity for either of the simulations (labeled as 'A' and 'B')
as expected from the original Michel scaling
\begin{equation}
{\cal{R}}_{\rm v} \equiv \frac{v_{\infty,A}}{v_{\infty,B}} =
 \left(\frac{\sigma_{A}}{\sigma_{B}}\right)^{1/3} =
\left(\frac{B_{\rm p, A}}{B_{\rm p, B}}\right)^{2/3}
\left(\frac{\sigma_{\rm i, A}}{\sigma_{\rm i, B}}\right)^{1/3},
\end{equation}
and each of the numerical values involved
(iso-rotation $\Omega_{\rm F}$ and injection velocity are the same 
for all examples).
The jet velocity has been measured along a slice across the 
jet at $z=160$.  
The poloidal velocity profile has a maximum close to the axis.
For each sample, the velocity measure of the asymptotic flow has been taken 
at similar spatial location and comparable evolutionary time step.
This is an important point as we compare simulations which different
evolutionary time scales where in some cases the large scale
flow has not yet settled in a steady state.

The situation for the toroidal velocity is less pronounced.
The typical rotation velocity at $(z=160,r=80)$ is 1/10 of the
Keplerian speed at the inner disk for most of the cases
investigated.
Examples for toroidal velocity profiles across the jet are shown in 
Fig.~\ref{fig_v_phi}.

Our large-scale numerical grid allows for direct comparison of the numerically
derived velocity structure with recent observations indicating rotation in the
jet from the young stellar object DG\,Tau \citep{bacc02}.
The numerical grid covers about $7\times14$\,AU which is about the size of the 
innermost slit position shown in Fig.\~7 of \citet{bacc02}.
These authors measure radial velocities up to 400 km/s and find two
major outflow components - a low velocity and a high velocity component.
It is the low velocity component of
de-projected\footnote{Assuming an inclination of $38\degr$,
see \citep{bacc02} and references therein}
jet velocity of about 90\,km/s which shows radial velocity
shifts of up to 12\,km/s.
The radial velocity differences across the jet have been widely
interpreted as indication of rotation although this is not yet
completely confirmed.
The observed velocity data corresponding to our numerical grid
(equivalent to "Slit S3/S5" and "Region I" in \citet{bacc02}) 
are 64\,km/s in poloidal velocity (in [SII]) and 8\,km/s
as total radial velocity shift.
The Keplerian speed at the inner disk radius for DG\,Tau is 
\begin{equation}
v_{\rm K, i} = 113 {\rm km/s} 
\left(\frac{M}{0.67\,\msun}\right)^{1/2}
\left(\frac{R_{\rm i}}{10\,\rsun}\right)^{-1/2},
\end{equation}
assuming a central mass of $0.67\,\msun$ (see \citet{bacc02} and references 
therein).
According to such a normalization, our simulations which fit best to
the low velocity jet of DG\,Tau are those with a flat magnetic field
profile $\mu \simeq 0.5$ and with moderate field strength,
namely i1, ..., p8 (see Tab.~\ref{tab_all})
These runs also show a high degree of collimation which is observed
as well.
The numerically derived rotational velocity for these examples is 
about $0.1 v_{\rm K, i} \sim 10\,{\rm km/s}$ which is surprisingly
close to the observed value (see Fig.~\ref{fig_v_phi}).
The launching radius can then derived just by following back the
streamlines along the collimating jet flow. 
In the runs i1, ..., p8 the launching region is distributed over
a relatively large disk area of $r$\lax$80$ corresponding to
about 3\,AU.
Note that in these simulations only the inner part of the jet is
super-Alfv\'{e}nic.
Note also that the toroidal velocity in the numerical models {\em decreases} 
with radius in apparent contradiction with the observations of DG\,Tau
(which, however, do not reolve the region close to the axis).
We like to emphasize that in a self-consistent picture also the high
velocity component has to be explained by the same simulation.
As another note of caution we mention that all MHD simulations of jet
formation done so far have shown a terminal jet speed of the order of
the Keplerian speed at the launching region 
\citep{ouye97a,ouye99,fend02,vito03}. 
It is therefore difficult to interpret the very high velocity
450\,km/s features observed for DG\,Tau in such a framework unless
we assume very strong magnetic flux respectively a very low mass
loading.

We finally discuss briefly the outflow velocities obtained in simulations
with strongly diffusive toroidal field.
As seen from Tab.~\ref{tab_diff}, 
the typical velocities in such outflows are about a factor 5 lower compared
to the standard approach.
In particular, this follows from comparing simulations 
p2 to p17, i11 to i12 and i13, and p13 to p12.
This is not surprising as we have artificially reduced the accelerating
Lorentz force component (along the field), 
$\vec{F}_{\rm L,||} \sim \vec{j}_{\perp}\times \vec{B}_{\phi}$,
where $\vec{j}_{\perp}$ is the poloidal electric current density component 
$\vec{j}_{\rm p} \sim \nabla \times \vec{B}_{\phi} $
perpendicular to the magnetic field.
\clearpage
%
\begin{table}
\begin{center}
\caption{Subset of Tab.~\ref{tab_all}, summarizing simulation runs with similar
field and density profiles but different relative magnetic flux. 
For the notation of the jet velocity ratio 
${\cal{R}}_{\rm v} \equiv 
v_{\infty,A}/v_{\infty,B}$ and the magnetization ratio
${\cal{R}}_{\sigma} \equiv \sigma_{\rm A}/\sigma_{\rm B}$ we refer to the text and Eq.\,12.
\label{tab_michel}
}
\small
\begin{tabular}{lllllclll}
\tableline \tableline 
\noalign{\smallskip} 
 run & $\mu$ & $B_{\rm p,i}$ & $\rho_{\rm i}$ & $v_{\rm mx}$ & $<\!\!\zeta\!\!>$ & 
 ${\cal{R}}_{\sigma}^{1/3}$ & 
 ${\cal{R}}_{\rm v} $   \\
\noalign{\smallskip} 
\tableline 
\noalign{\medskip}
c3 & 1.5 & 0.214 & 1.0 & 1.07 & 15 & 1.31 & 1.63 \\
\noalign{\medskip}
p4 & 1.5 & 0.321 & 1.0 & 1.74 & 10 &  &  \\
\noalign{\smallskip}
\tableline 
\noalign{\smallskip}
i14 & 1.25 & 0.355 & 1.0 & 1.53 & 10 & 1.26 & 1.25 \\
\noalign{\smallskip}
i15 & 1.25 & 0.177 & 0.5 & 1.22 & 10  \\
\noalign{\smallskip}
\tableline 
\noalign{\smallskip}
p13 & 0.8 & 0.922 & 1.0 & 2.33 & 15 & 2.08 & 1.98 \\
\noalign{\medskip}
p14 & 0.8 & 0.307 & 1.0 & 1.18 & 15  & 1.17 & 1.09 \\
\noalign{\medskip}
a1  & 0.8 & 0.247 & 1.0 & 1.08 & 15    \\
\noalign{\smallskip}
\tableline 
\noalign{\medskip}
i1 & 0.5 & 0.112 & 1.0 & 0.7 & 30 & 1.59 & 1.23 \\
\noalign{\medskip}
p7 & 0.5 & 0.225 & 1.0 & 0.86 & 20  \\
\noalign{\medskip}
c5 & 0.5 & 0.225 & 1.0 & 0.78 & 30  \\
\noalign{\smallskip}
\tableline 
\noalign{\medskip}
p18 & 0.5 & 0.112 & 1.0 & 0.5 & 5 & 2.13 & 2.46 \\
\noalign{\medskip}
p19 & 0.5 & 0.225 & 1.0 & 0.95 & 6  \\
\noalign{\medskip}
p16 & 0.5 & 0.225 & 1.0 & 1.1 & 7  \\
\noalign{\medskip}
\tableline \tableline 
\end{tabular}
\end{center}
\end{table}
\clearpage
\subsection{The run of the Alfv\'{e}n surface}

Another parameter which measures the relative strength of the magnetic field
is the Alfv\'{e}n Mach number $M_{\rm A}$,
\begin{equation}
M_{\rm A}^2 \equiv \frac{4 \pi \rho v_{\rm p}^2}{B_{\rm p}^2} 
\end{equation}
At the inflow boundary (= disk surface) the Alfv\'{e}n Mach number varies as
\begin{equation}
M_{\rm A,0}^2(r) \sim r^{2\mu - \mu_{\rho} - 1} 
\sim r^{-\mu_{\sigma}-3/2} \equiv r^{-\mu_{\rm A}},
\end{equation}
with $\mu_{\rm A} = \mu_{\sigma}+3/2$.
A classical Blandford-Payne type MHD jet is launched as slow sub-Alfv\'{e}nic disk wind, 
i.e.  $M_{\rm A,0}(r) < 1 $ for all radii along the disk surface.
In turn, this condition constrains the disk wind poloidal magnetic 
field and density profile, respectively, $\mu_{\rho} > 2\mu + 1 $.
For smaller $\mu$, a situation may occur that at a certain radius the disk wind
will be launched as {\em super-Alfv\'{e}nic} flow.
The location of this radius depends on the actual values of
$B_{\rm P,i}$ and $\rho_{\rm inj}$.
Therefore, in general sub-Alfv\'{e}nic disk winds require a positive exponent
for the $M_{\rm A}(r)$ profile.
In the self-similar Blandford \& Payne approach we have $\mu_{\sigma} = -3/2$,
while in the \citet{ouye97a} simulations $\mu_{\sigma} = -1$
A dipolar (stellar) magnetic field distribution with $\mu_{\rho} \sim 3/2$ 
implies $\mu_{\sigma} = -1$ and,
thus, $M_{\rm A,0}(r) \sim r^{-1/4}$, that means a critically negative exponent.
The disk wind launched will be super-Alfv\'{e}nic \citep{fend00}. 

Note that the relation between the magnetization $\sigma$ and the
collimation measure $<\!\!\zeta\!\!>$ shown in Fig.~\ref{fig_dg_sig} 
holds similarly for the Alfv\'{e}n Mach number as 
$\mu_{\rm A} = \mu_{\sigma} + 3/2$.

\begin{figure}
\centering

\includegraphics[width=8cm]{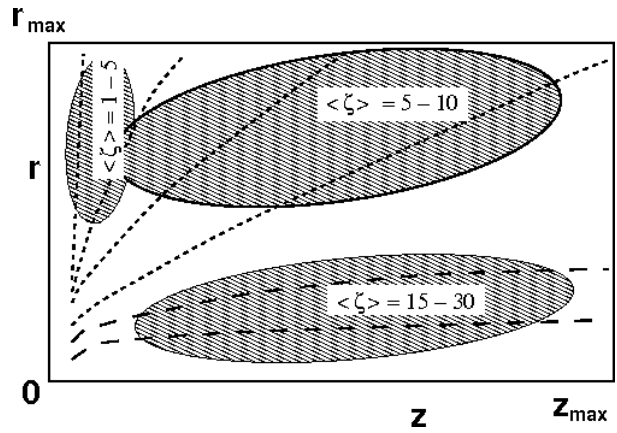}

\caption{Showcase run of Alfv\'{e}n surfaces for different simulation runs
(see Figs.~\ref{fig_disk_1}, ~\ref{fig_disk_2},~\ref{fig_disk_3},
~\ref{fig_disk_4}).
Selected are simulations: p20, i16; a2, i11, p4; p19, p16, p15, i15;
i3, i10; i2, i9; p7, i1, p14 (clockwise from $r$-axis).
The shape of the Alfv\'{e}n surfaces s governed by the parameters $\mu$ and
$\mu_{\rho}$. 
Dashed/dotted lines highlight Alfv\'{e}n surfaces leaving the grid through the
$z$-boundary / $r$-boundary.
As indicated by the ellipsoids, the latter ones are typical for highly
collimated outflows, the first ones for less collimated outflows.
Only examples with a sub-Alfv\'{e}nic disk wind profile were chosen for this
figure.
\label{fig_alfven}
}
\end{figure}

Figure \ref{fig_alfven} summarizes our results concerning the shape of
the Alfv\'{e}n surfaces in respect of outflow collimation degree.
The showcase plot indicates the run of the Alfv\'{e}n surface of different
simulations, overlaid by the numerically observed degree of collimation for
these outflows (indicated by ellipsoidal areas).
Only examples with a sub-Alfv\'{e}nic disk wind profile were chosen for this
figure.
In general, a "flat" Alfv\'{e}n surface (i.e. curving towards the $r$-direction)
indicates a less collimated jet.
Some of these outflows are not fully evolved dynamically on a global scale, 
but as we measure the collimation degree only within $r = 150$, the main
results derived are nevertheless reliable.
For highly collimated outflows the Alfv\'{e}n surface curves parallel to the
jet axis already at rather low altitudes above the disk.
\clearpage
\clearpage
We have also extreme examples (i5, i8) for which the Alfv\'{e}n surface curves
towards the equatorial plane (see Fig.~\ref{fig_alfven2} for i8) such that 
the outer part of the disk wind is launched super-Alfv\'{e}nic.
Those winds cannot not {\em magneto-centrifugally} accelerated but are
{\em magnetically} driven by the Lorentz force in vertical direction,
$\vec{F}_{\rm L,z} \sim \vec{j}_{r}\times \vec{B}_{\phi}+ \vec{j}_{\phi}\times \vec{B}_{r}$.
Note that simulations i15 and p20 seemingly look similar to the case
of a super-Alfv\'{e}nic disk wind,  
but actually start as sub-Alfv\'{e}nic disk wind everywhere as 
resolved by our scaled grid.

The kink in the Alfv\'{e}n surface in simulation i10 (Fig.~\ref{fig_disk_1}, {\em bottom})
demonstrates that the outer part of the outflow has not yet reached a stationary state.
The foot point radius of that flux surface intersecting the kink is at
about $50\,r_{\rm i}$.
Similarly, this holds for most of the other simulation runs.

\section{Summary}

We have performed numerical MHD simulations of jet formation from accretion disks.
The disk is taken as a time-independent boundary condition determining the 
disk magnetic flux profile and the density profile of the mass flux from 
the disk surface.
This implies that we implicitly assume that disk internal physical processes such
as diffusion, advection, or turbulence have worked together in generating and
maintaining a large scale disk magnetic field structure
which leading to a disk surface field profile and the mass flux profile
as prescribed in our simulations.
The underlying assumption is that a variation of internal disk parameters
can lead to a variation in the resulting magnetic and mass flux profile.

We applied the ZEUS-3D MHD code in the axisymmetric option and modified for physical
magnetic diffusivity.
Our physical grid size is $(150\times 300)$ inner disk radii corresponding to about
$(6.7\times 13.3)\,$AU for $\ri \simeq 10\,\rsun$ in the case of a protostellar jet
which is several times larger than in previous studies in the literature and comparable
to the observational resolution.
We applied a simple parameterization for physical magnetic diffusivity in which turbulent
Alfv\'{e}nic pressure is responsible for both turbulent magnetic diffusivity and pressure.
However, the magnitude of magnetic diffusivity is well below the critical value 
derived earlier beyond which the degree of jet collimation becomes affected \citep{fend02}.

The major goal of this paper was to investigate if and how the jet collimation
depends on the accretion disk magnetic flux profile.
As a quantitative measure of the collimation degree we apply the directed mass
flux (axial direction versus lateral direction).
We have run a substantial number of simulations covering a large area in the parameter
space concerning the mass flux and magnetic flux profile across the jet.
In particular, we varied 
\begin{itemize}
\item
 the disk poloidal magnetic field {\em profile} parameterized by a power law $B_p(r) ~ r^{-\mu}$, 
\item
 the disk wind density {\em profile} (i.e. the density "injected" from the disk surface into
 the jet flow) parameterized by a power law $\rho_p(r) ~ r^{-\mu_{\rho}}$, 
\item
 the disk magnetic field strength and/or the total mass flux.
\end{itemize}
This corresponds to a variation of
the {\em mass flux profile} across the jet, 
the {\em magnetization profile} across the jet, and  
the magnetosonic {\em Mach number} across the jet.

Our simulations give clear evidence that a flatter magnetic field profile in the disk
wind generally leads to a higher degree of jet collimation.
This holds for all density profiles applied, however, the collimation degree increases
with increasing slope of the density profile.
For very flat magnetic field profile ($\mu$ \lax $0.4$) the jet flows do not settle
into a stationary state. Instead, instabilities evolve along the outflow axis. 

Furthermore, we find a unique correlation between the slope of the disk wind
magnetization profile and the degree of jet collimation, potentially explaining
why not all accretion disks do launch a jet and why a disk does not 
launch jets for all its life time.

A variation of the total magnetic field strength does not seem to affect the degree
of collimation 
(at least not for a variation by factors up to 8 we investigated).
The same holds for the total mass flux in the jet flow.
However, the relative field strength affects the time scale for the dynamical evolution
of the outflow - jets with stronger magnetic flux evolve faster.

Comparing simulations with different magnetization but the same magnetic flux profile we
were able to prove the Michel scaling (Michel 1969) for the asymptotic outflow velocity 
$v_{\infty} \sim \sigma^{1/3}$ with relatively good agreement.
The small deviations may be explained by the fact that we investigate collimated
and not spherical outflows 
and also outflows which are to some part still dynamically evolving on the global scale.

In order to prove the feasibility of "poloidal collimation" which has been discussed
in the literature, we invented a toy model introducing artificial decay of the toroidal
magnetic field component
(in addition to our parameterization of physical magnetic diffusivity). 
In general, such outflows show less collimation and also reach only lower asymptotic 
speed. 
However, as a matter of fact, outflow collimation is definitely present even in the
absence of toroidal fields. 

Comparing our numerical results to observations of DG\,Tau we find indication for
a flat magnetic field profile in this source, as in this case the observed
and the numerical jet velocity (low velocity component) and jet rotation do match.

In summary, our paper clearly states a correlation between the disk wind 
{\em magnetization profile} and the {\em outflow collimation}. 
Good collimation requires a flat magnetization profile across the jet,
i.e. sufficient magnetization also for larger disk radii.
The final solution to the question of the jet mass loading and the disk-jet
magnetic field structure is expected to come from simulations
taking into account self-consistently both the internal disk dynamics and 
the outflow.

\begin{acknowledgements}
I thank the LCA team and M.~Norman for sharing the ZEUS code.
H.~Zinnecker and J.~Wambsganss are thanked for financial support during the
time in Potsdam when most of the computations for the present paper were
performed.
\end{acknowledgements}

\appendix

\section{Figures intended for electronic publication only}
\clearpage
\begin{figure}
\centering

\includegraphics[height=0.5cm]{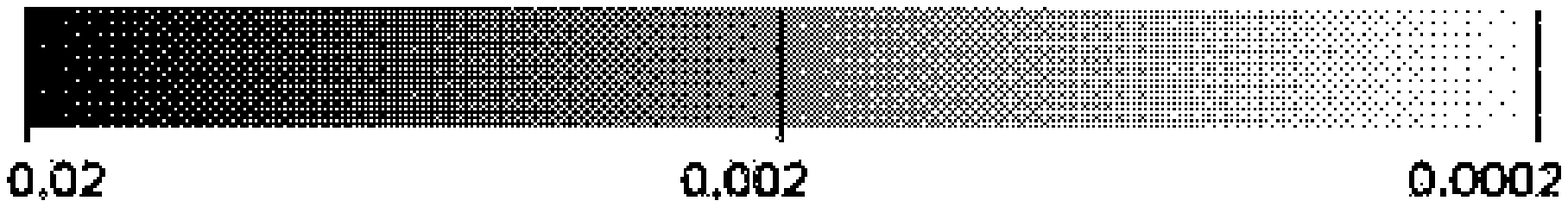}

\includegraphics[height=3.5cm]{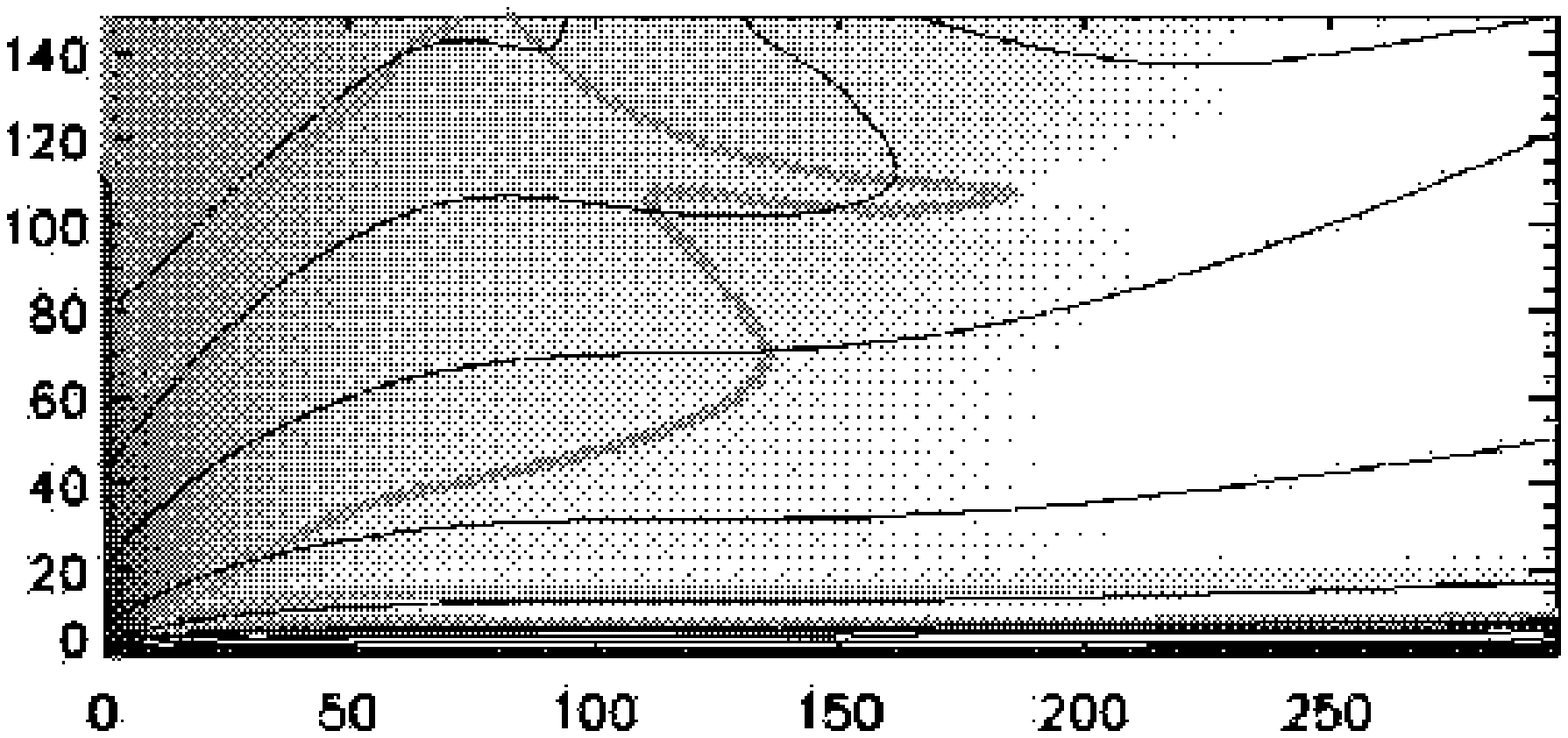}

\includegraphics[height=0.5cm]{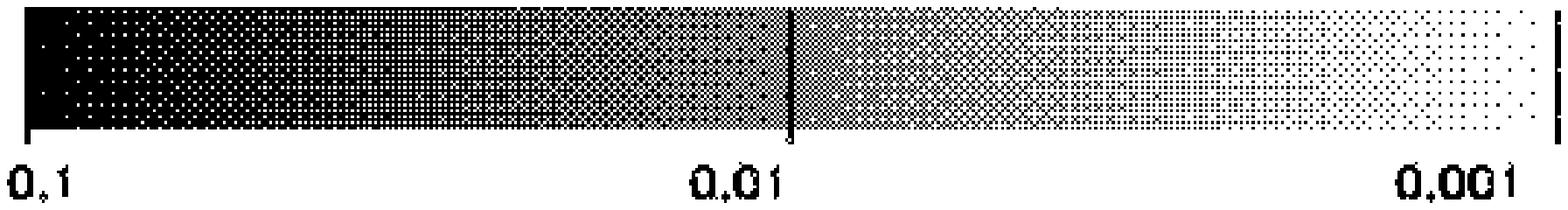}

\includegraphics[height=3.5cm]{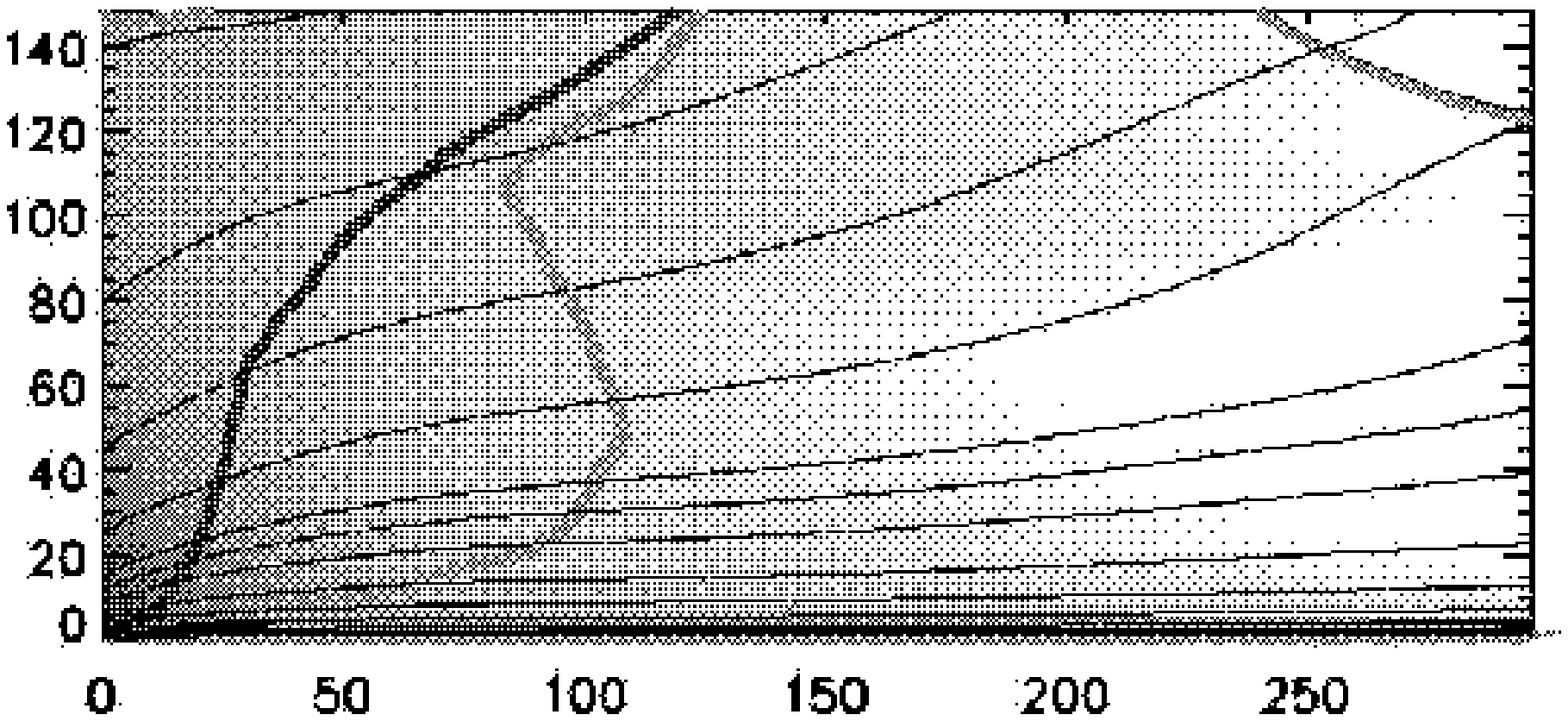}

\includegraphics[height=0.5cm]{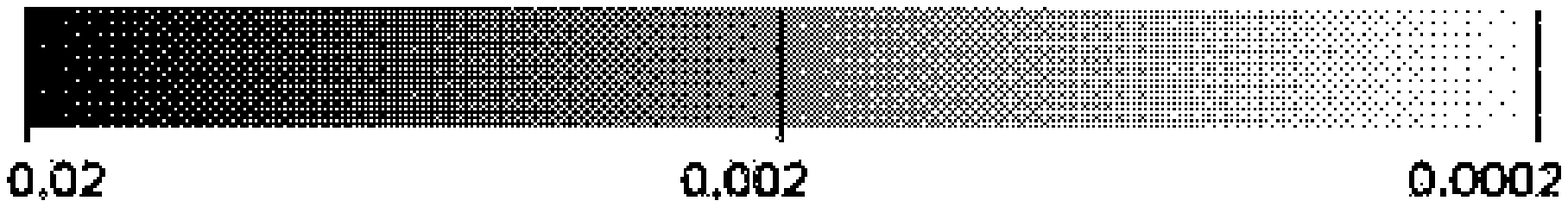}

\includegraphics[height=3.5cm]{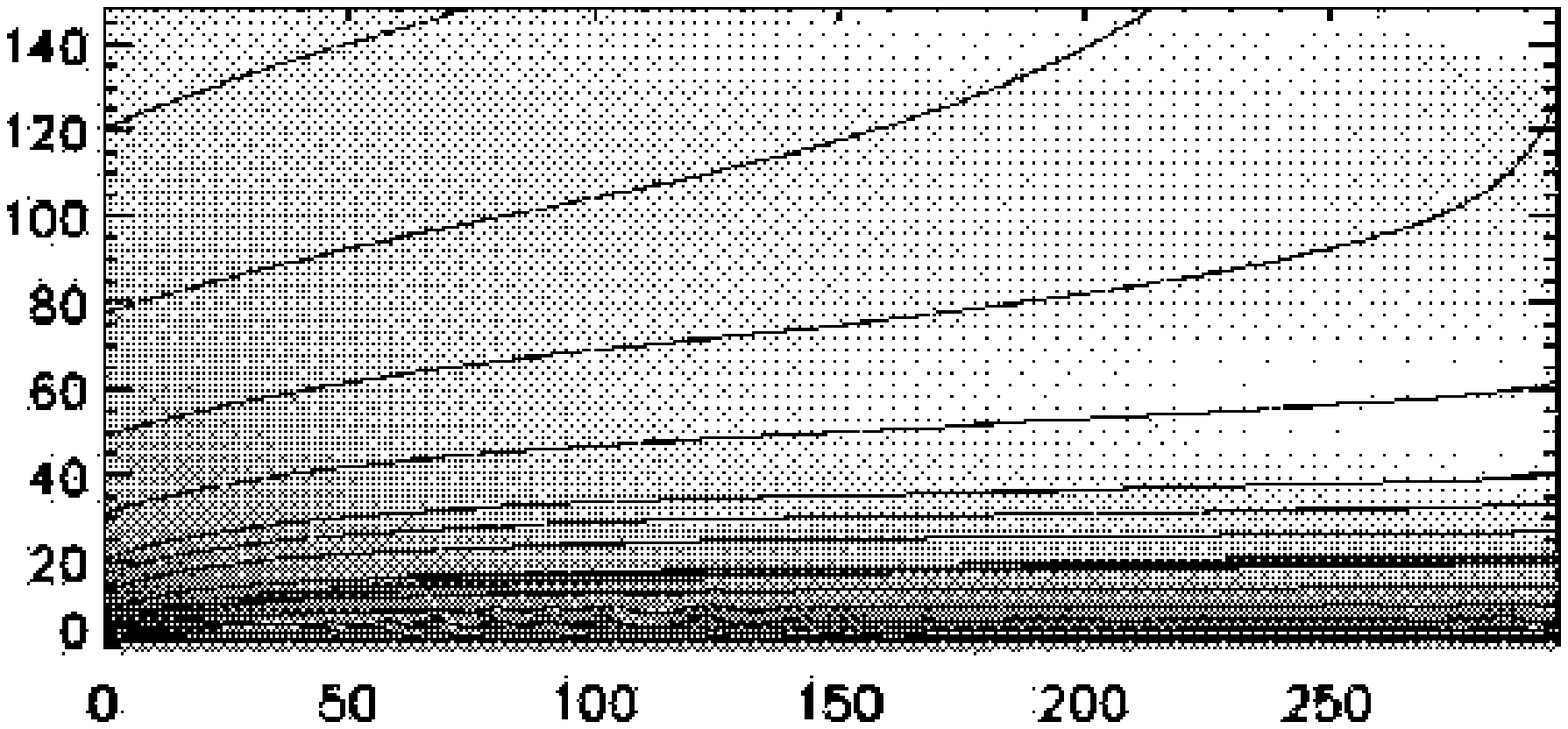}

\includegraphics[height=0.5cm]{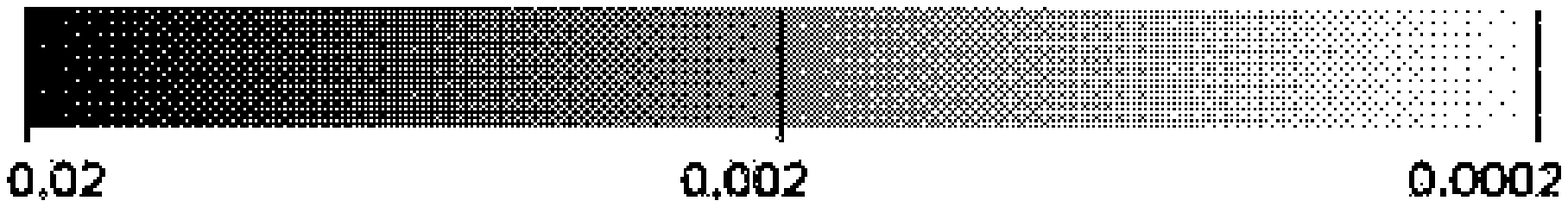}

\includegraphics[height=3.5cm]{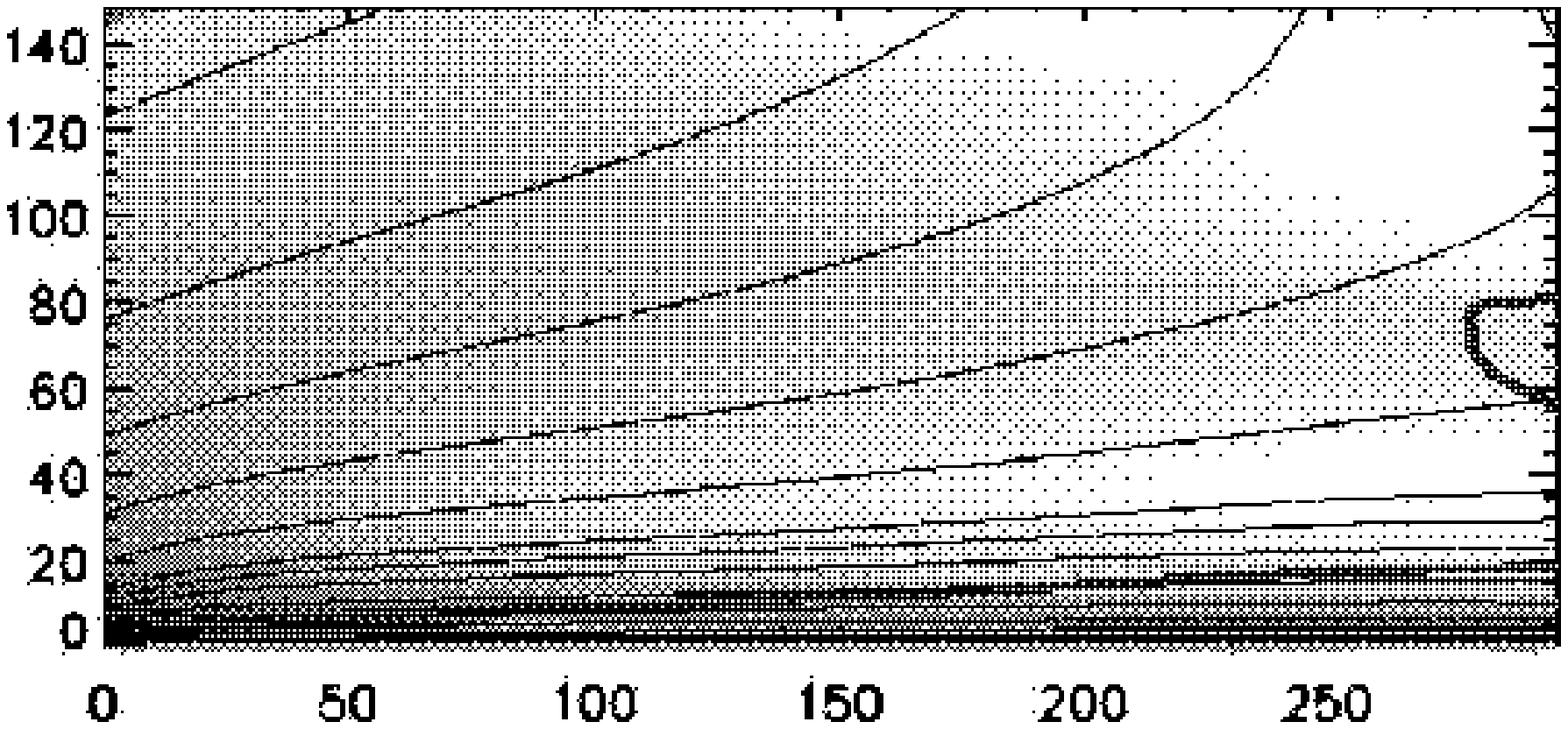}

\includegraphics[height=0.5cm]{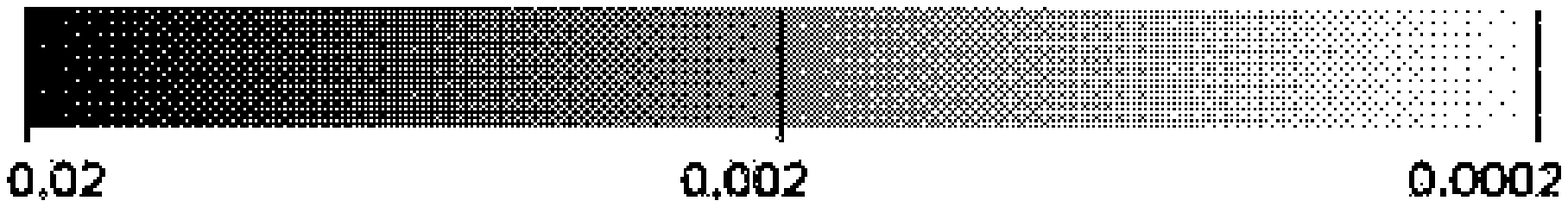}

\includegraphics[height=3.5cm]{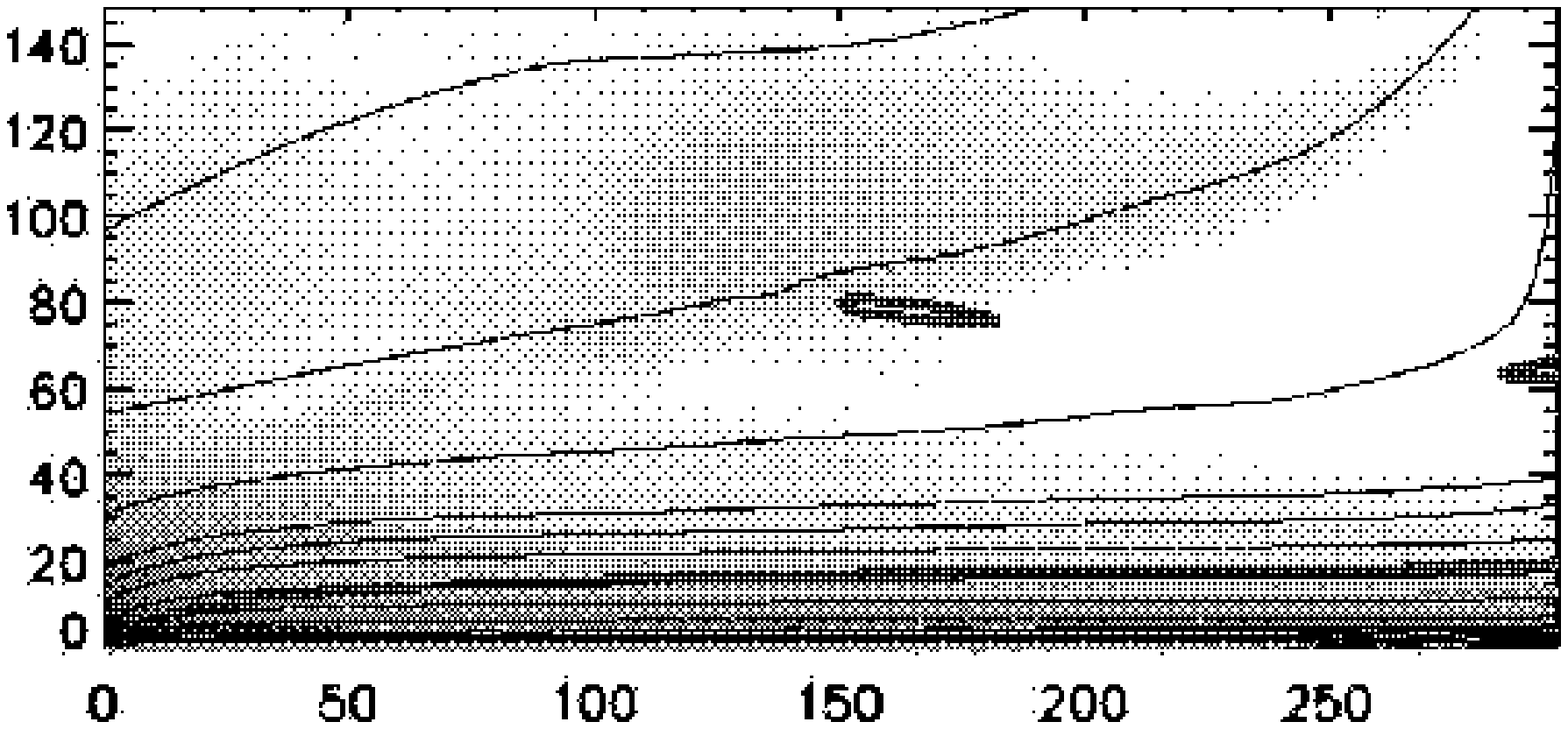}

\caption{Simulation runs with different disk boundary 
conditions (see Tab.~1).
Density and poloidal magnetic field lines for 
{\bf i16} ($t=2040$),
{\bf p15} ($t=990$) 
{\bf i1}  ($t=2000$) 
{\bf p7}  ($t=1000$), 
{\bf a1} ($t=2830$),
from {\em top {\rm to} bottom}.
See Fig.~\ref{fig_long} for further notation.
\label{fig_disk_4}
}
\end{figure}

\begin{figure}
\centering

\includegraphics[height=0.5cm]{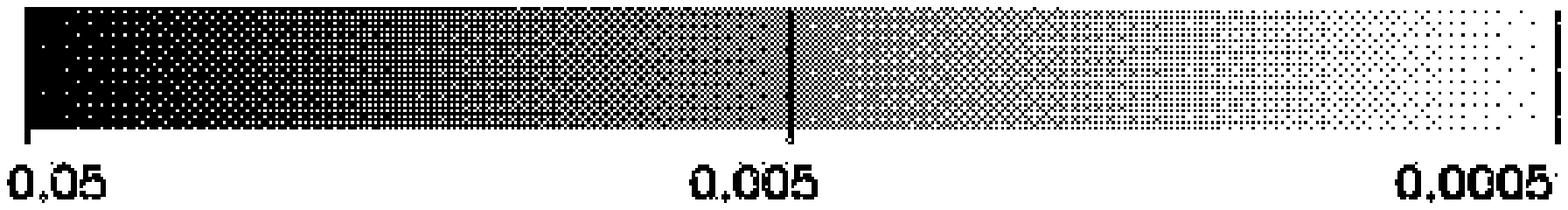}

\includegraphics[height=3.5cm]{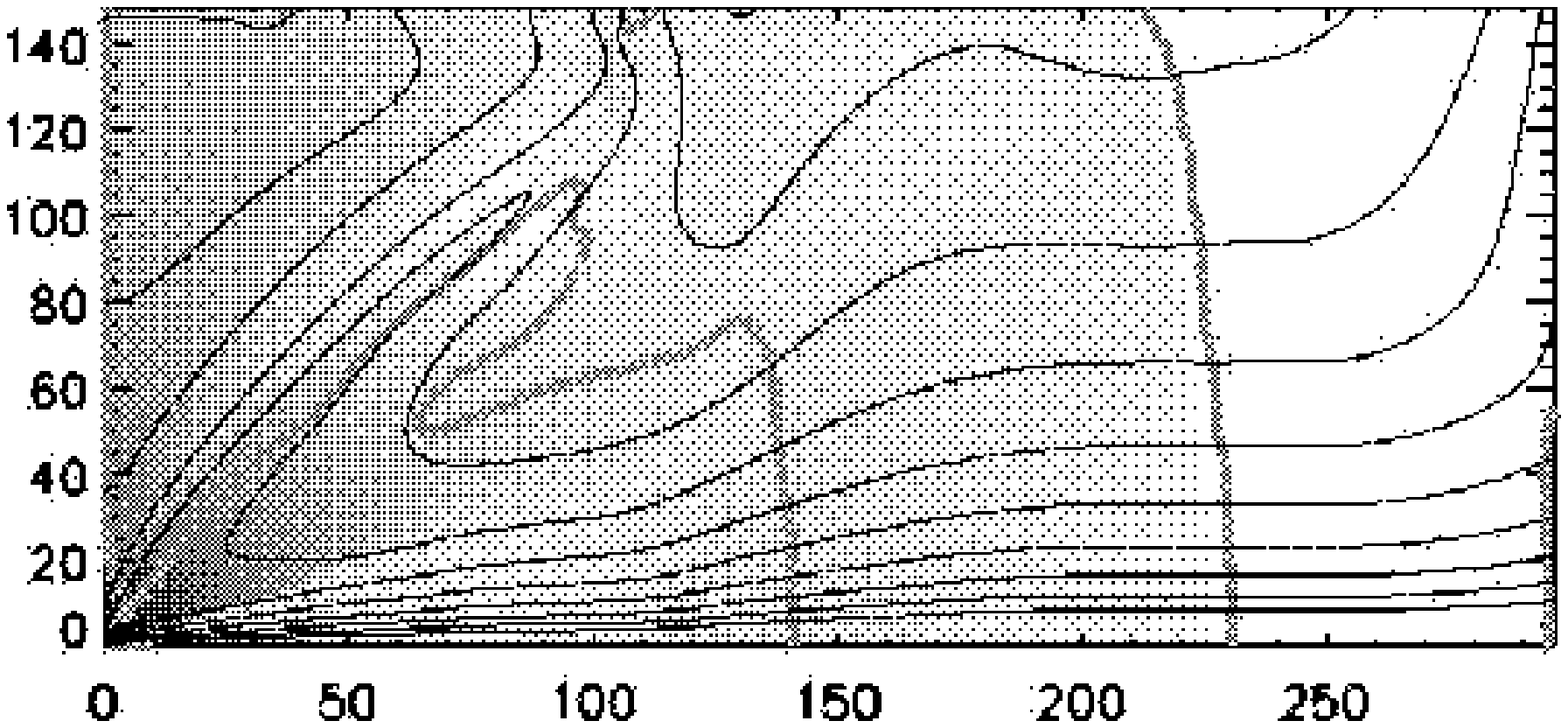}

\includegraphics[height=0.5cm]{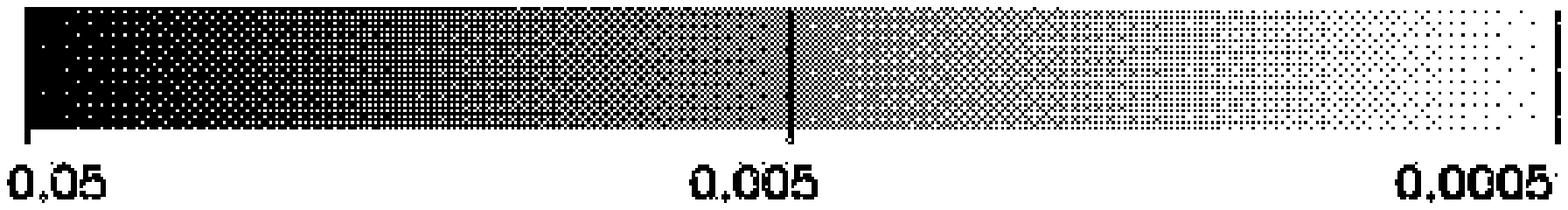}

\includegraphics[height=3.5cm]{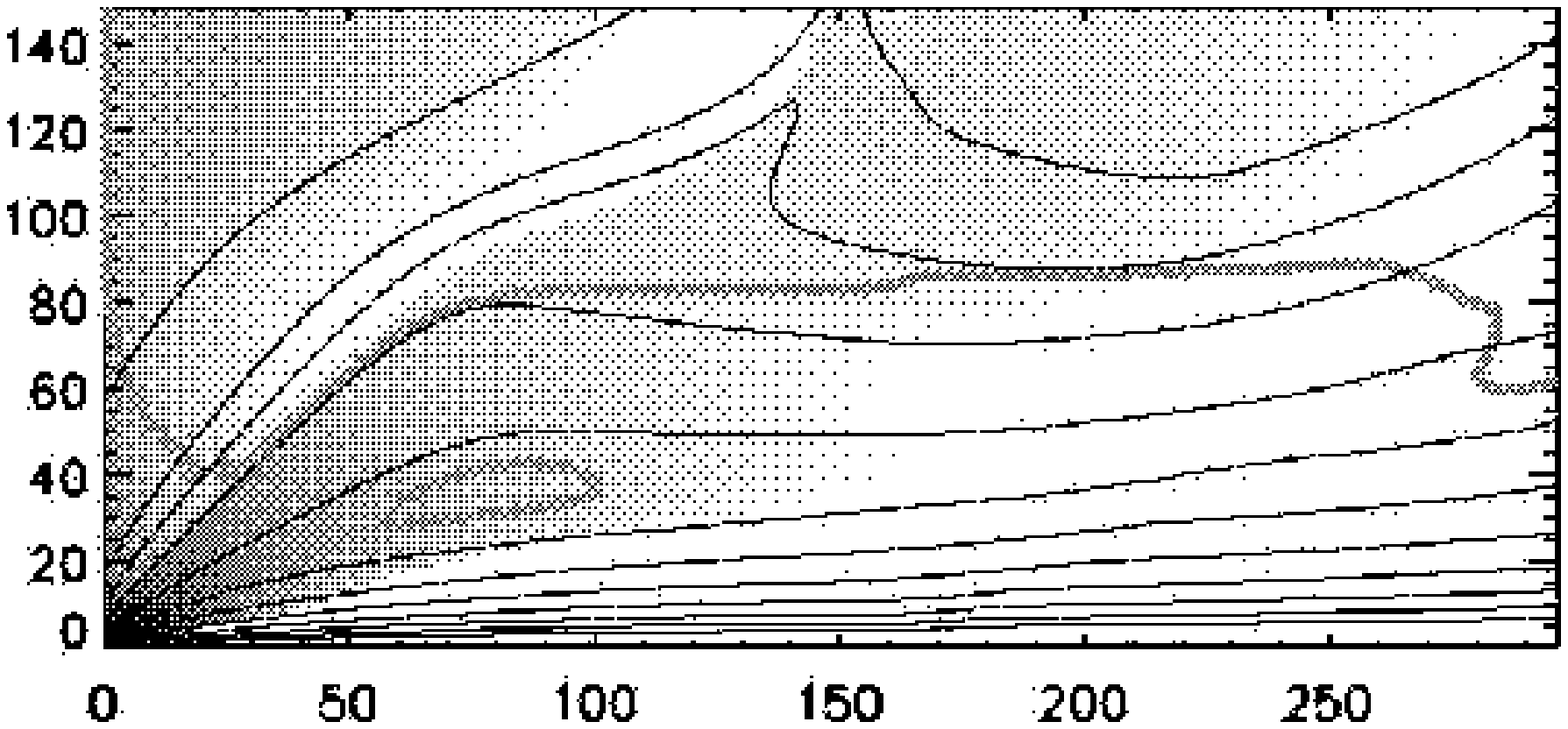}

\includegraphics[height=0.5cm]{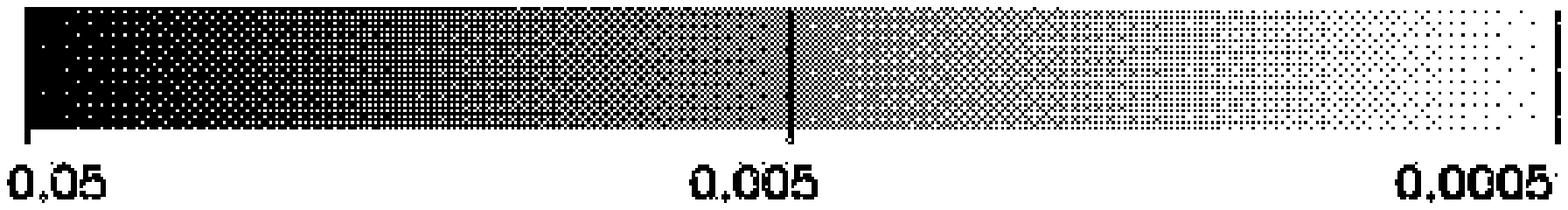}

\includegraphics[height=3.5cm]{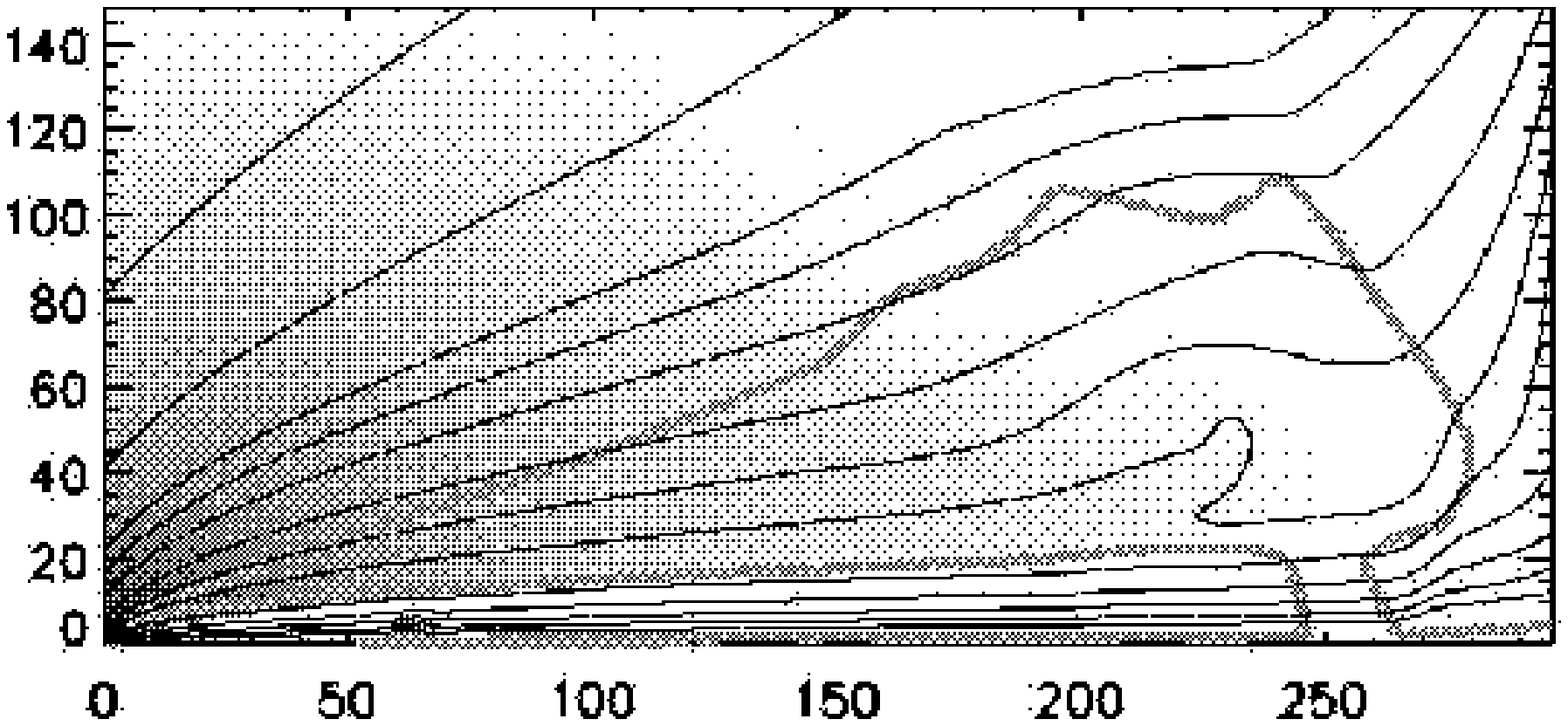}

\includegraphics[height=0.5cm]{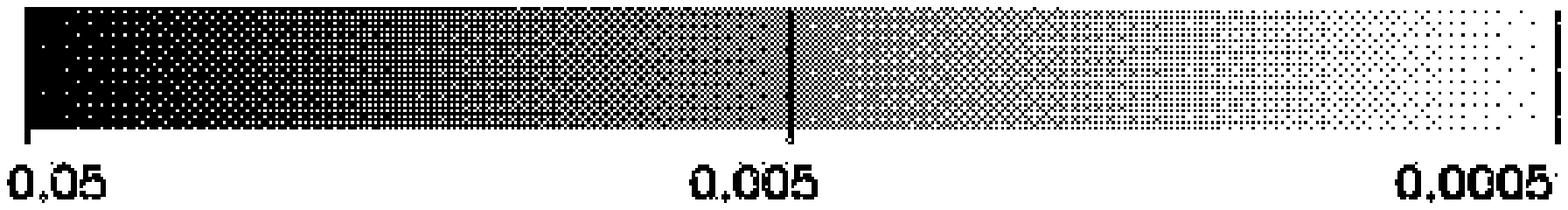}

\includegraphics[height=3.5cm]{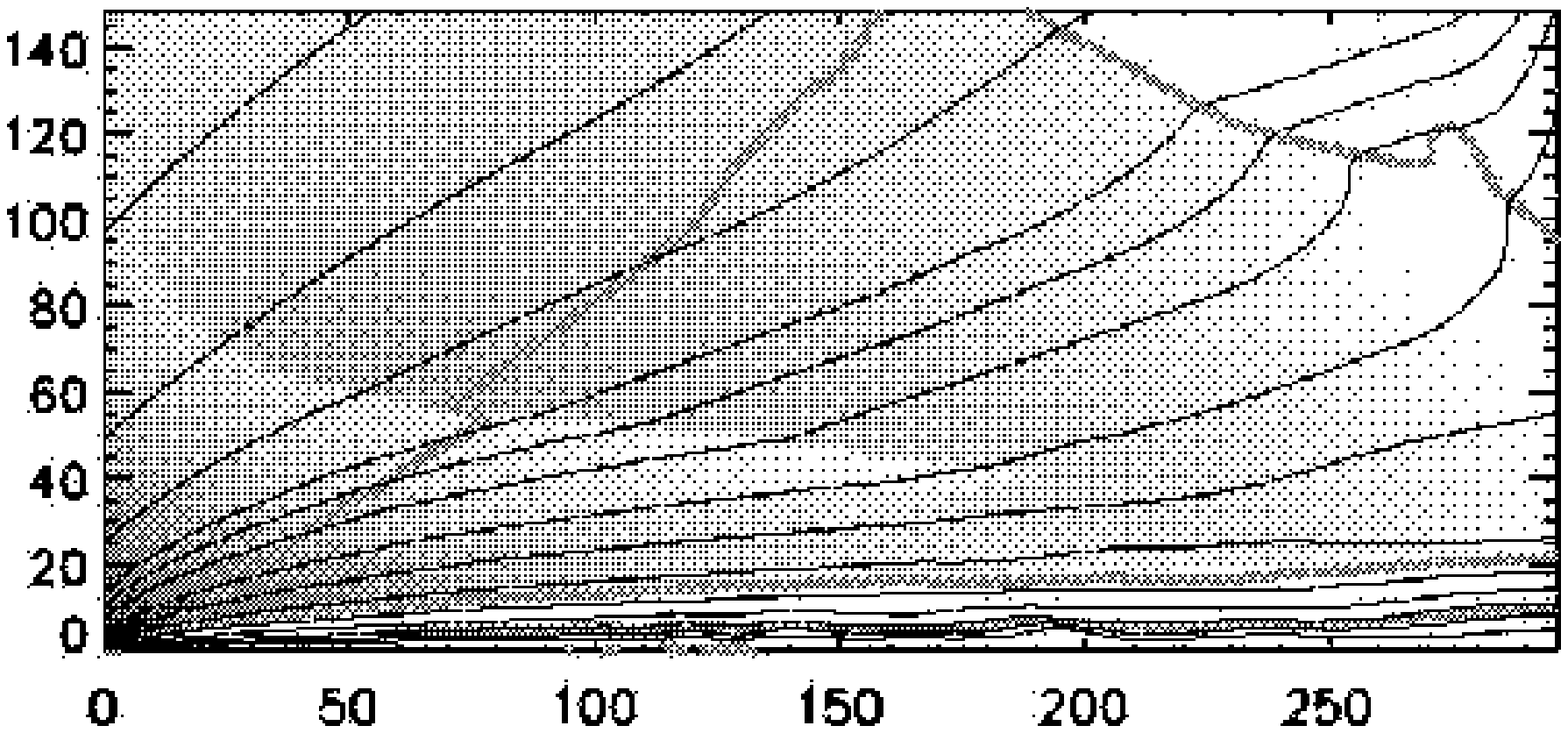}

\includegraphics[height=0.5cm]{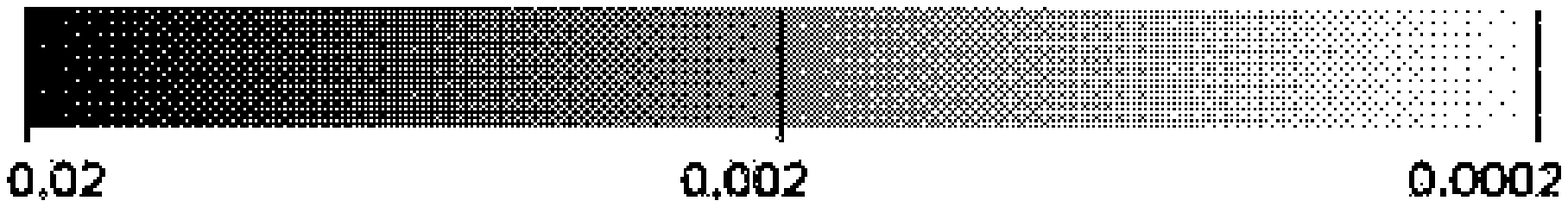}

\includegraphics[height=3.5cm]{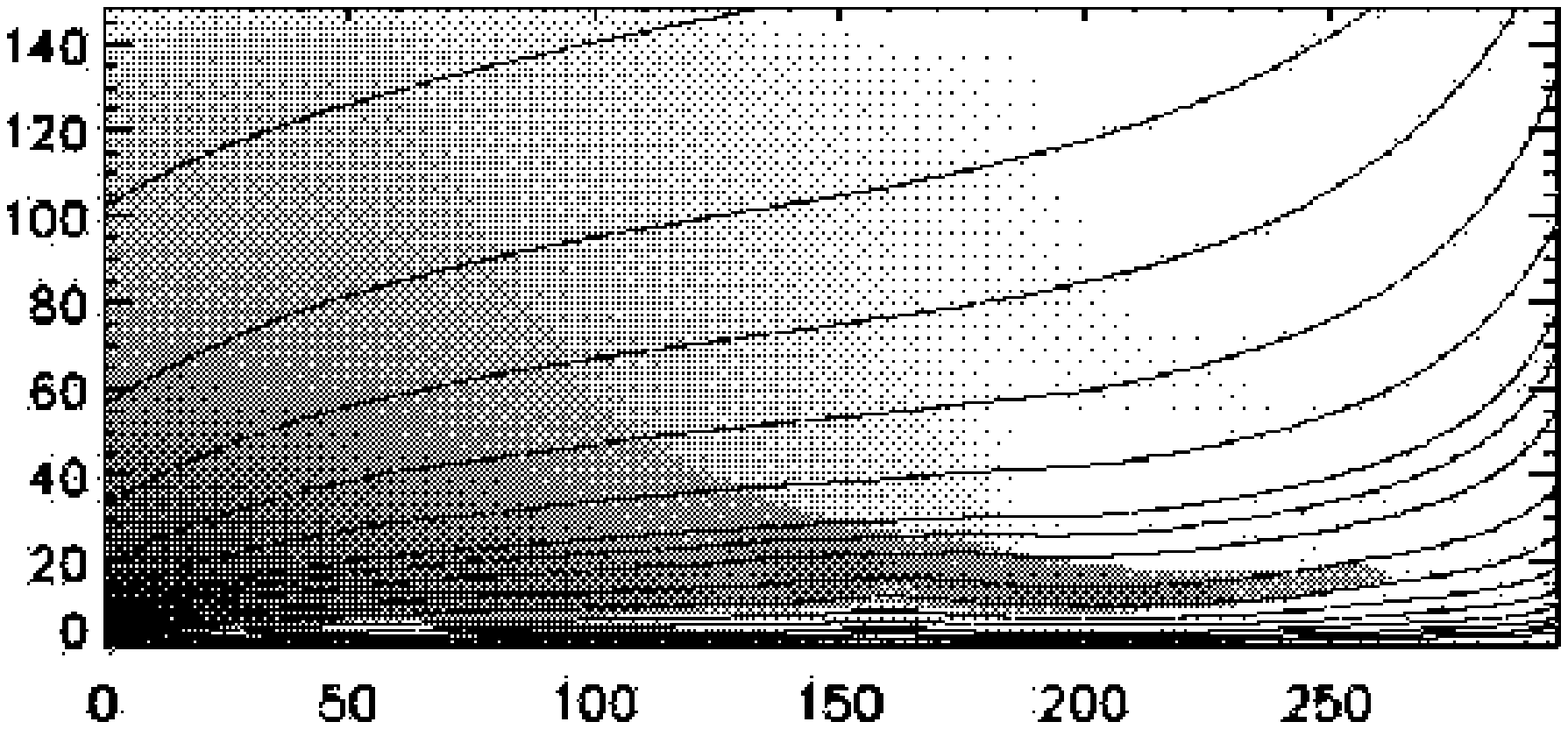}

\caption{Simulation runs with different disk 
boundary conditions (see Tab.~2).
Enhanced magnetic diffusivity for the
toroidal magnetic field component $B_{\phi}$.
Density and poloidal magnetic field lines for 
{\bf p10} ($t=1800$),
{\bf p11} ($t=3100$),
{\bf p17} ($t=1890$),
{\bf i12} ($t=1520$),
{\bf p12} ($t=870$)
from {\em top {\rm to} bottom}.
See Fig.~\ref{fig_long} for further notation.
\label{fig_b3_diff}
}
\end{figure}

\begin{figure}
\centering

\includegraphics[height=0.5cm]{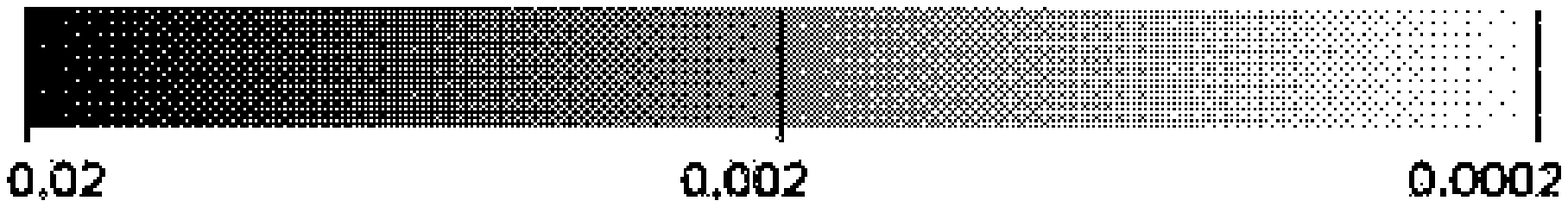}

\includegraphics[height=4.0cm]{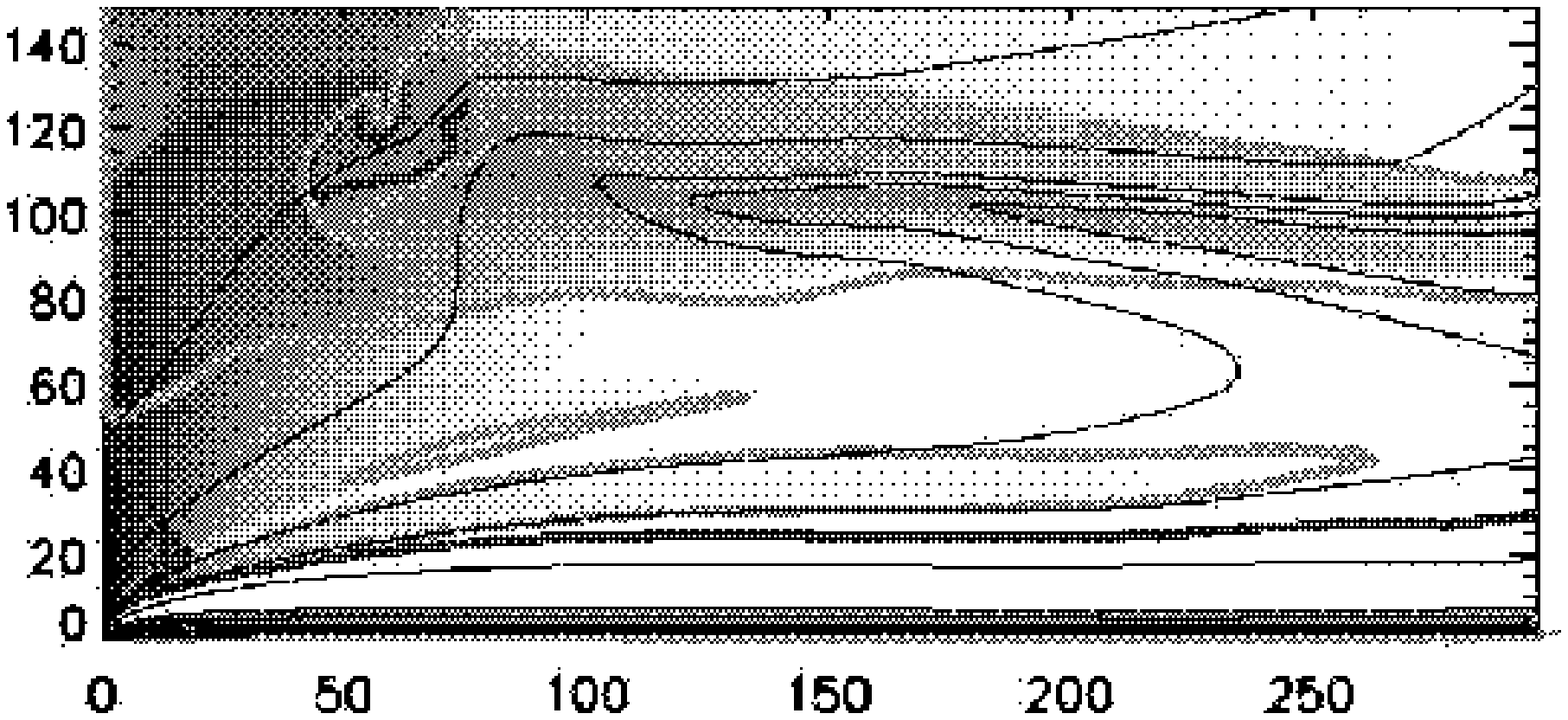}

\caption{Simulation run with different disk 
boundary conditions (see Tab.~1).
Example {\bf i8} ($t=290$) for a parameter run with the Alfv\'{e}n surface
entering the disk surface. 
Density and poloidal magnetic field lines.
See Fig.~\ref{fig_long} for further notation.
\label{fig_alfven2}
}
\end{figure}
\clearpage


\end{document}